\documentclass[11pt]{article}
\pdfoutput=1
\usepackage[centertags]{amsmath}
\usepackage[square, comma, sort&compress,numbers]{natbib}
\usepackage{array,multirow}
\numberwithin{equation}{section}
\usepackage{amssymb}
\usepackage{graphicx}
\usepackage{color}
\usepackage{relsize}

\usepackage{epsfig}

\def\Or[#1]{{\text{O}}\left({#1}\right)}
\def\dotl[#1,#2]{\left\langle #1, #2 \right\rangle}
\def\dotlb[#1,#2]{[ #1, #2 ]}
\def\dotp[#1,#2]{(#1) \cdot (#2)}
\def\aff[#1,#2]{\hat{#1}(#2)}
\def\n4sym{{\cal N}=4 SYM}
\def\>{\rangle}
\def\<{\langle}
\def\weight[#1,#2,#3]{\{(#1),#2,#3\}}
\def\ads[#1]{$\text{AdS}_{#1}$}

\newcommand{\ba}{\begin{eqnarray}}
\newcommand{\ea}{\end{eqnarray}}
\newcommand{\be}{\begin{eqnarray}}
\newcommand{\ee}{\end{eqnarray}}
\newcommand{\bq}{\begin{equation}}
\newcommand{\eq}{\end{equation}}
\newcommand{\benn}{\begin{equation*}}
\newcommand{\eenn}{\end{equation*}}
\newcommand{\bi}{\begin{itemize}}  
\newcommand{\ei}{\end{itemize}}

\newcommand{\CA}{{\cal A}}

\newcommand{\CC}{{\cal C}}
\newcommand{\CD}{{\cal D}}

\newcommand{\CI}{{\cal I}}

\newcommand{\CJ}{{\cal J}}

\newcommand{\CO}{{\cal O}}

\newcommand{\CV}{{\cal V}}
\newcommand{\nn}{\nonumber}

\newcommand\oo\infty
\newcommand\s\sigma
\newcommand\de\delta
\newcommand\De\Delta

\newcommand\f\phi
\newcommand\g\gamma
\newcommand\x\times

\usepackage{jheppub}

\makeatletter
\def\@fpheader{\vspace{-.1cm}}
\makeatother

\title{On Information Loss in AdS$_3$/CFT$_2$}

\author[a]{A.\ Liam Fitzpatrick,}
\author[b]{Jared Kaplan,}
\author[b]{Daliang Li,}
\author[b]{and Junpu Wang}
\affiliation[a]{Department of Physics, Boston University, \\
Commonwealth Avenue, Boston, MA 02215, U.S.A.}
\affiliation[b]{Department of Physics and Astronomy,  Johns Hopkins University, \\
Charles Street, Baltimore, MD 21218, U.S.A.}

\abstract{  We discuss information loss from black hole physics in AdS$_3$, focusing on two sharp signatures infecting CFT$_2$ correlators at large central charge $c$: `forbidden singularities' arising from Euclidean-time periodicity due to the effective Hawking temperature, and late-time exponential decay in the Lorentzian region.
We study an infinite class of examples where forbidden singularities can be resolved by non-perturbative effects at finite $c$, and we show that the resolution has certain universal features that also apply in the general case.    Analytically continuing to the Lorentzian regime, we find that the non-perturbative effects that resolve forbidden singularities qualitatively change the behavior of correlators at times $t \sim S_{BH}$, the black hole entropy.  This may resolve the exponential decay of correlators at late times in black hole backgrounds.  By Borel resumming the $1/c$ expansion of exact examples, we explicitly identify `information-restoring' effects from heavy states that should correspond to classical solutions in AdS$_3$. 
Our results suggest a line of inquiry towards a more precise formulation of the gravitational path integral in AdS$_3$.   
} 

\arxivnumber{}

\begin{document}  
 
\maketitle
\flushbottom

\section{Introduction}

Unitarity violation from black hole physics \cite{Hawking:1974sw} lurks within the Virasoro symmetry structure \cite{Brown:1986nw} of the AdS$_3$/CFT$_2$ correspondence \cite{Maldacena, Witten, Gubser:1998bc}.  In this paper we will identify non-perturbative effects in $G_N \equiv \frac{3}{2 c}$ that resolve this problem in an infinite class of examples.  We will argue that these results can be analytically continued to resolve information loss in the general case, and may provide clues to the correct contour of integration for the gravitational path integral.  We begin by reviewing various manifestations of information loss so that we can explain the specific problems that we will be addressing.  

\subsection{`Hard' and `Easier' Information Loss Problems}

The `hard' information loss problem is the paradox that pits local gravitational effective field theory, vis-\`a-vis the equivalence principle, against unitary quantum mechanical evolution \cite{Mathur:2009hf,Mathur:2010kx, Almheiri:2012rt,Braunstein:2009my}.  AdS/CFT has declared that unitarity must win this fight, but it does not explain how the equivalence principle can survive.    To address this question we need a general, self-consistent prescription for reconstructing local bulk observables near and across horizons using CFT data.    Since we do not expect bulk observables to be precisely defined anywhere, the prescription would need to be cognizant of its own limitations, which would presumably then answer the question of whether/when firewalls exist \cite{Almheiri:2012rt, Braunstein:2009my}.  We will have little to add to the discussion of this `hard' problem.  It seems very difficult to precisely formulate, let alone resolve, in terms of quantum mechanical observables in CFT.\footnote{For example, although bulk points outside horizons can be precisely defined in terms of the singularity structure of large central charge CFT correlators \cite{GGP, JP, Okuda:2010ym, JoaoMellin, Maldacena:2015iua}, considerations of causality show that bulk point singularities never occur behind horizons \cite{Fitzpatrick:2016thx}.}    

\begin{figure}[t!]
\begin{center}
\includegraphics[width=0.3\textwidth]{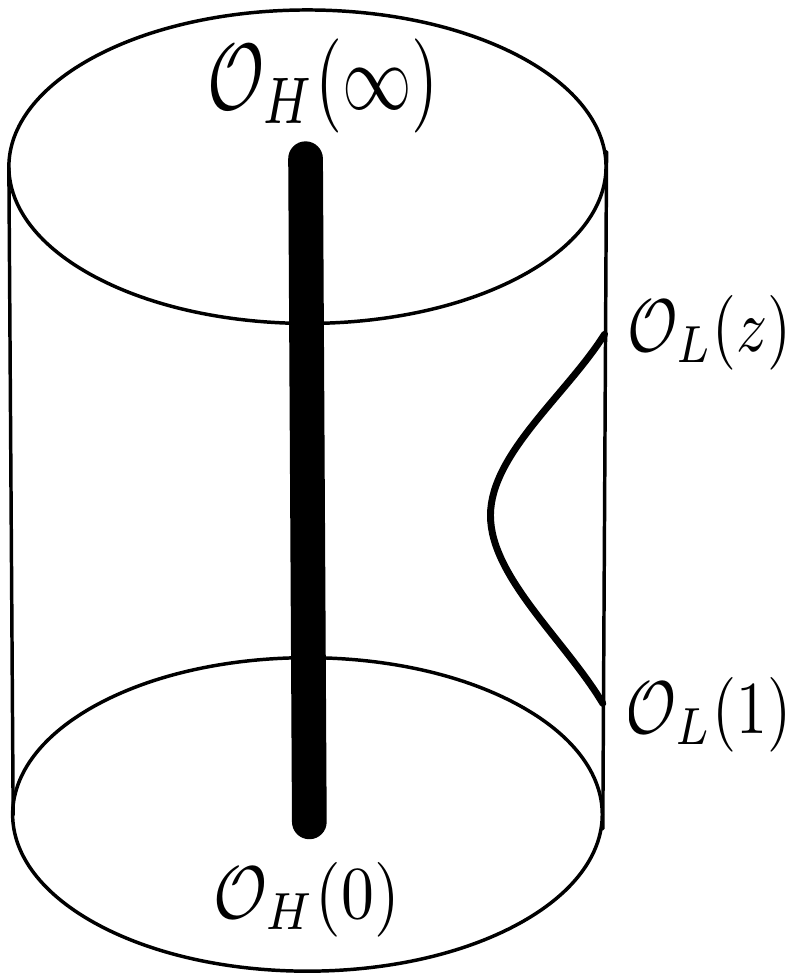}
\caption{ This figure suggests a heavy-light CFT correlator and its association with a light probe object interacting with a deficit angle or BTZ black hole background.}
\label{fig:BasicHeavyLightAdSCylinder}
\end{center}
\end{figure}

An `easier' information loss problem can be formulated directly in terms of CFT correlation functions \cite{Maldacena:2001kr}.  
A two-point CFT correlator probing a large AdS black hole will decay exponentially at late times.  This translates into the idea that all information about an object thrown into a black hole will eventually be lost.  Since field theories on compact spaces cannot forget about initial perturbations, this behavior signals a violation of unitarity.  We should emphasize that this reasoning does not apply to CFTs  on non-compact spaces, at infinite temperature, or with an infinite number of local degrees of freedom.\footnote{The first two are closely connected because as $T \to \infty$ we can measure distances in units of $1/T$, effectively decompactifying space.}  For example, the thermal 2-pt correlator of a scalar primary operator in a CFT$_2$ on an infinite line
\be
\label{eq:Thermal2dCFTCorrelator}
\< \CO(t_L) \CO(0) \>_T = \left( \frac{\pi T}{\sinh(\pi T t_L)} \right)^{2 \Delta_\CO}
\ee
decays exponentially at arbitrarily late Lorentzian time $t_L$.  The large central charge limit $c \to \infty$ can also produce thermal correlators, as we will discuss more precisely below.  

Thus the `easier' information loss problem arises because black holes are in a sense \emph{too thermal}.  To resolve it we must understand the emergence of a kind of thermodynamic limit as $G_N \to 0$, and then identify the non-perturbative `$e^{-\frac{1}{G_N}}$' corrections to this limit that restore unitarity.  

We can sharpen the problem by studying pure states, as illustrated in Figure \ref{fig:BasicHeavyLightAdSCylinder}, and by focusing on another manifestation of thermal physics.  The two-point correlator of a light probe operator in a heavy, pure-state background can be written as the four-point correlator 
\be
\mathcal{A}(z, \bar z) = \< \CO_H(\infty) \CO_L(1) \CO_L(z, \bar z) \CO_H(0) \>  \ \   \mathlarger{\overset{?}{\approx}}   \ \ \< \CO_L(1) \CO_L(z, \bar z) \>_{T_H}
\ee
using the operator/state corresponence.    In this expression $z$ is a coordinate in the plane, while $t \equiv - \log(1-z)$ will be a Euclidean time coordinate on the boundary of the global AdS cylinder.  The operator $\CO_H$ creates a black hole microstate with approximate Hawking temperature $T_H$.    If $\mathcal{A}$ is thermal then it satisfies the KMS condition, making $\CA$ \emph{periodic in Euclidean time}.  There is an intuitive but imprecise connection between this periodicity and the exponential decay in Lorentzian time discussed above.

This periodic behavior is \emph{forbidden} in the vacuum correlation functions of local CFT operators.  In the Euclidean region, CFT correlation functions can have singularities only in the OPE limit, when pairs of operators collide.  This follows because away from the OPE limit we can interpret the correlator as the inner product of normalizable states in radial quantization \cite{Pappadopulo:2012jk,Mack:1976pa,Luscher:1975js}.  If $\mathcal{A}$ were periodic in Euclidean time then it would have additional singularities at the periodic images of the OPE singularities, such as $z = 1 -e^{n/T}$ for integers $n$.  These `forbidden singularities' are a sharp manifestation of unitarity violation and information loss.  They will be a major focus in this work. 

\subsection*{A Universal Piece of the Resolution}

Information loss in black hole backgrounds appears to be a  generic feature of quantum gravity.  Therefore it would be very surprising if its resolution depended on intricate details that vary from theory to theory.  Were this the case, our task would be hopeless, since we will not be able to compute the exact heavy-light correlator in any, let alone every, holographic CFT.
Fortunately, in all CFT$_2$ there is a universal contribution to each heavy-light correlator, the Virasoro vacuum block \cite{Fitzpatrick:2014vua, Fitzpatrick:2015zha, Fitzpatrick:2015foa, Alkalaev:2015wia, Alkalaev:2015lca, KrausBlocks, Beccaria, Anous:2016kss, Besken:2016ooo}, which manifests information loss in the large central charge or $c \to \infty$ limit \cite{Fitzpatrick:2015dlt}.  

In any theory with a symmetry, it is natural to organize observables and amplitudes into irreducible representations of that symmetry group.  We encapsulate the full contribution from all states related by the Virasoro symmetry in a Virasoro conformal block.  So every theory with a vacuum state must have a Virasoro vacuum block, which contributes to a correlator like $\< \CO_H \CO_H \CO_L \CO_L \>$ because the vacuum itself contributes -- we obtain a non-vanishing result when we insert $| 0 \> \< 0 |$ between the $\CO_H$ and $\CO_L$.  Phrased in terms of Virasoro this may seem rather abstract, but via AdS/CFT we learn that
the gravitational field is related to the CFT stress tensor via
\be
g_{\mu \nu}(X) \ \ \leftrightarrow \ \ T(z)  = \sum_n z^{-2-n} L_n
\ee
and so `gravitons' in AdS are created by acting with the CFT stress-energy tensor on the vacuum.  Furthermore, the Virasoro generators $L_n$ are simply the modes of the stress tensor, so \emph{the Virasoro vacuum block includes all effects from the exchange of quantum multi-graviton states}.  The Virasoro blocks contain a great deal of exact information about quantum gravity.

We would like to study a light object probing a black hole in AdS.  This means that we should study a heavy-light correlator with $h_H \propto c$ and $h_L$ fixed at large $c$, since CFT operator dimensions correspond with AdS energies, and the Newton constant $G_N = \frac{3}{2c}$.  The corresponding Virasoro vacuum block has been computed \cite{Fitzpatrick:2014vua, Fitzpatrick:2015zha, Fitzpatrick:2015foa}, on the cylinder it is
\be
\CV_{\infty}(t) = \left( \frac{\pi T_H}{\sin(\pi T_H t)} \right)^{2 h_L}
\ee
where the Hawking temperature $T_H = \frac{1}{2 \pi} \sqrt{24 \frac{h_H}{c} -1 }$.  This pure CFT$_2$ computation clearly `knows' about black hole physics in AdS$_3$.  We  emphasize that this result is exactly periodic in Euclidean time $t$, and so it has forbidden singularites at $t = \frac{n}{T_H}$.  If we analytically continue to Lorentzian $t_L = i t$, then the vacuum block decays exponentially at late times.\footnote{By itself this does not indicate information loss in CFT$_2$ correlators, because the full correlator is an infinite sum over Virasoro blocks, and other blocks could behave differently at late Lorentzian time.  In section \ref{sec:BehaviorLargeLorentzianTime} we explain that known heavy-light blocks do all decay exponentially in $t_L$ at $c=\infty$, but we do not have explicit results when intermediate operator dimensions are of order $c$. }  Thus the $c \to \infty$ vacuum block manifests information loss.

On general grounds we expect that the Virasoro vacuum block's information loss problem must be resolved within its own structure.  In particular, the resolution  should not depend on a delicate interplay between many separate conformal blocks, since this would indicate an intricate theory-dependence.   One reason for this expectation is that at the positions of the forbidden singularities, $z=1-e^{n/T}$ is real and positive for $n<0$ and $T$ real and therefore the sum over conformal blocks is a sum over positive contributions; thus the sum over non-vacuum blocks cannot cancel the singular behavior.\footnote{Additionally, in the limit $c=\infty$, the vacuum block's forbidden singularities are sharper than those of all other Virasoro conformal blocks.}  A more general but more formal proof follows because  the vacuum block can itself be viewed as an inner product between normalizable states, and so it can only have OPE singularities at finite central charge \cite{Pappadopulo:2012jk}. 

In any case, we do not need to rely on general arguments, because we will explicitly exhibit both the Euclidean time periodicity  of large $c$ blocks and its finite $c$ resolution.  For instance,  consider the degenerate Virasoro vacuum block 
 \be
\label{eq:IntroExample}
e^{-\frac{1}{2}(1+\alpha+\epsilon)t_E } {}_2 F_1 \left(1+\epsilon,1+\alpha+ \epsilon,2+2 \epsilon , 1 - e^{-t_E} \right)
\ \   \mathlarger{\overset{c \to \infty}{\longrightarrow}}  
\ \   e^{-\frac{1}{2} t_E} \frac{\sin(\pi T_H t_E)}{\pi T_H} 
\ee
where $\alpha=\sqrt{ (1+\epsilon)^2 - 4 h_H \epsilon}$ and $\epsilon = \frac{1}{12} \left(c-13 -\sqrt{(c-1)(c-25)} \right)$.  This is a heavy-light-vacuum block, where the heavy operator dimension $h_H$ and the central charge $c$ can take any value, but the light operator dimension is pegged to the value $h_L = -\frac{1}{2} - \frac{3}{4} \epsilon$.  As $c \to \infty$ we have $\epsilon \approx \frac{6}{c}$ and the parameter $\alpha \to 2 \pi i T_H$, leading to a correlator that is periodic in Euclidean time $t_E$.  In constrast, the exact block (the hypergeometric function) is not periodic in $t_E$ for any finite $c$.  Furthermore, if we analytically continue to Lorentzian signature, the exact vacuum block does not have an exponential time-dependence.

The example of equation (\ref{eq:IntroExample}) was chosen for its simplicity, so although it is periodic in Euclidean time, it does not have any forbidden singularities.  In section \ref{sec:ExactVirasoroBlocks} we will study an infinite class of examples with degenerate external operators where the vacuum block can be computed exactly at any $c$. These special cases agree precisely with our more general results \cite{Fitzpatrick:2014vua, Fitzpatrick:2015zha, Fitzpatrick:2015dlt} as $c \to \infty$, and in particular, exhibit forbidden singularities  in the large central charge limit.  Relating the infinite discretum of degenerate vacuum blocks to the general heavy-light case requires analytic continuation, but as we review in section \ref{sec:ArgumentForAnalyticity}, the Virasoro blocks are entire functions of the external operator dimensions $h_H$ and $h_L$.  

\subsection{Borel Resummation and Classical Solutions}
\label{sec:BorelResummationReview}

It is interesting to have examples of correlators exhibiting information loss as $c \to \infty$.  But we would also like to understand the resolution of information loss from the vantage point of perturbation theory in $G_N = \frac{3}{2c}$.  In other words, we would like to expand the exact result as
\be
\mathcal{V}(z) = \mathcal{V}_{c=\infty}(z) \left(1 + \frac{f_1(z)}{c} + \cdots \right) + e^{-c s(z)} \left(g_0(z) + \frac{g_1(z)}{c} + \cdots \right) + \cdots
\ee
to explicitly identify the non-perturbative effects that restore unitarity.  The first term corresponds to perturbation theory about the AdS$_3$ vacuum.  We expect that the other terms correspond to non-perturbative corrections involving solutions to Einstein's equations incorporating the exchange of states with Planckian energy, as we will now explain.

Many series expansions in quantum mechanics have zero radius of convergence.  Given such a formal series
\be
f(g) = \sum_n a_n g^n
\ee
we can define a Borel series $B(g)$ by $a_n \to \frac{a_n}{n!}$, and in many cases $B(g)$ will then have a finite radius of convergence.  Now we can try to define a function
\be
f(g) = \int_0^\infty \frac{dy}{g} \, e^{-y / g} B (y)
\ee
as the Borel transform, which reproduces the $a_n g^n$ if we expand $B$ in $y$.  If the Borel integral converges and has no singularities on the real axis, then it can be viewed as a definition of $f(g)$.  Singularities on the real axis lead to ambiguities in $f(g)$, and more generally, singularities in the Borel plane lead to branch cuts when $f(g)$ is analytically continued \cite{Basar:2013eka}.  Relevant examples will be studied in section \ref{sec:NonPerturbativeUnitarityRestoration}.

We can connect singularities in the Borel plane to classical solutions of the field equations via an illustrative argument given by 't Hooft \cite{'tHooft:1977am}.  Simply equate the Borel transform and the  path integral description of the correlator
\be
 \int_0^\infty dy \, e^{-y / g} B (y)  \sim \int \CD \phi \, e^{-\frac{1}{g} S(\phi)}
\ee
where we use $\sim$ to denote the fact that this is a very formal relation.  It leads to 
\be
B(y) & \sim & \int \CD \phi \, \delta \left( y - S(\phi) \right)
\nn \\
& \sim & \left. \left( \frac{\partial S}{\partial \phi} \right)^{-1}  \right|_{ S(\phi) = y }
\ee
Thus we see that in order for $B(y)$ to have a singularity at some $y_*$, we expect to have
\be
\frac{\partial S}{\partial \phi_*} = 0 \ \ \  \mathrm{and}  \ \ \  S(\phi_*) = y_*
\ee
for some field configuration $\phi_*$.
Thus singularities of $B(y)$ in the $y$-plane correspond to solutions of the classical equations of motion with an action equal to $y_*$.

We will be studying the Virasoro vacuum conformal block.  In the large $c$ limit, it can be obtained from a number of direct CFT arguments \cite{Fitzpatrick:2014vua, Fitzpatrick:2015zha, Fitzpatrick:2015foa}, and also from AdS$_3$ gravity \cite{Fitzpatrick:2014vua, Asplund:2015eha, KrausBlocks, Hijano:2015qja}.  We expect that order-by-order in $1/c$ perturbation theory, the Virasoro vacuum block could be obtained, at least in principle, from AdS$_3$ calculations in a perturbative $G_N$ expansion.  The result should match with direct methods in CFT$_2$, where leading $1/c$ corrections have already been obtained.  

In section \ref{sec:BorelofVacuumBlock} we will study the exact results for the degenerate Virasoro vacuum block in $1/c$ perturbation theory and perform a Borel resummation of the result.  We will see that there are singularities in the Borel plane, and that they have a natural interpretation as specific heavy states.  In other words, when we expand the exact vacuum block in $1/c$, we will find a saddle point corresponding to the `perturbative vacuum', plus other saddles associated with the non-perturbative contributions from heavy states.

Given that we expect the $1/c$ perturbation theory to match between the gravitational path integral and direct CFT$_2$ calculations,  \emph{it is natural to conjecture that the singularities in the Borel plane must correspond to classical solutions of Einstein's equations in AdS$_3$}.  In the general case these AdS$_3$ solutions should correspond to the exchange of black holes between the light probe and the heavy background states, and should become very (numerically) important in the correlator in the vicinity of forbidden singularities.  The Virasoro vacuum block seems to know about heavy states in AdS$_3$, which emerge as `solitons' from the Virasoro `graviton' states that are created by the CFT$_2$ stress tensor.

\subsection{In Brief:  Summary and Outline}

In section \ref{sec:AdSCFTInformationLossLargeC} we provide a more complete discussion of information loss in CFT correlators.  We begin with a discussion based on AdS in section \ref{eq:SingularitiesfromAdS3}, and then in sections \ref{sec:ForbiddenSingularitiesVirasoroVacuum} and \ref{sec:BehaviorLargeLorentzianTime} we explain how the same effects arise directly from CFT$_2$ computations at large central charge.  Figure \ref{fig:ForbiddenSingularityArrangement} provides a cartoon of the locations of forbidden singularities in CFT correlators. In section \ref{sec:BehaviorLargeLorentzianTime} we arrive at a conclusion that we view as crucially important -- information loss seems to be a consequence of the behavior of the (universal and theory independent) Virasoro conformal blocks after they are expanded in the large $c$ limit.

We study degenerate external operators in section \ref{sec:ExactVirasoroBlocks} in order to obtain exact information concerning the Virasoro vacuum block.  We review Virasoro blocks and degenerate operators in section \ref{sec:DegenerateStateReview}.  For an infinite sequence of values of $h_H(r) = \frac{c}{24}(1-r^2)$, indexed by a positive integer $r$, the exact vacuum block obeys a linear differential equation of order $r$.  We provide a general argument suggesting that  these results can be analytically continued in $r$ in section \ref{sec:ArgumentForAnalyticity}, and a quick illustrative example in section \ref{sec:QuickHeavyDegenerateExample}.  Then in section \ref{sec:ConnectingDegenerateAndLargeC} we show that the exact results precisely match previous computations in the large $c$ limit.  In section \ref{sec:ResolutionofForbiddenSingularities} we show that for all values of $r$, the forbidden singularities are resolved in a universal way by non-perturbative effects at finite $c$.  More specifically,  at finite $c$ the forbidden singularities are regulated by the function
\be
S(x,c) \approx \int_0^\infty dp \, p^{2h_L - 1} e^{-p x - \frac{\sigma^2}{2 c} p^2}
\ee
where we chose $x= 0$ as the location of a singularity, and we explicitly compute $\sigma^2$ in section \ref{sec:NearAForbiddenSingularity}.  We use analytic continuation in $r$ to derive this result; this continuation passes a very non-trivial check which we describe in section \ref{sec:NearAForbiddenSingularity}.  

Finally, in section \ref{sec:LateLorentzianBehaviorDiffEq} we study late Lorentzian time behavior via an approximation motivated by the resolution of the forbidden singularities.  We show that the Virasoro blocks change qualitatively after a Lorentzian time $t_L \sim S_{BH}$, the black hole entropy.  More specifically, we derive an approximate differential equation
 \be
 -h_L g_r(t) \CV(t)  + \CV'(t) + \frac{1}{c} \Sigma(t) \CV''(t) = 0,
\ee
for the Lorentzian time behavior of the vacuum Virasoro block, which is valid for $t \ll S_{BH}$.  This equation incorporates the non-perturbative effects that resolve forbidden singularities.  The last term in the equation behaves roughly as $\frac{t}{S_{BH}} \CV''$, and it becomes as important as the other terms precisely when $| \CV | \sim e^{-S_{BH}}$.  This strongly suggests that the exponential decay of the vacuum Virasoro block ceases at precisely the timescale that is necessary to avert Maldacena's \cite{Maldacena:2001kr} information loss problem.  Thus we have found non-perturbative or `$e^{-c}$' effects that fully resolve the forbidden singularities and appear to resolve the late-time exponential decay of correlators.  

We discuss the  dependence of the exact Virasoro blocks on the central charge in section \ref{sec:NonPerturbativeUnitarityRestoration}, focusing on Borel resummation of the $G_N \propto 1/c$ expansion in section \ref{sec:BorelofVacuumBlock}.  In section \ref{sec:SaddlesStokesDegenerateBlocks} we take a different approach based on contour integral formulas for the degenerate blocks, which arise from the Coulomb gas formalism \cite{Dotsenko:1984nm, Dotsenko:1984ad}.  In both cases we identify non-perturbative contributions to the Virasoro blocks associated with heavy intermediate states.  We leave it to future work to connect our results with classical solutions of the gravitational or Chern-Simons \cite{Witten:2010cx, Gaiotto:2011nm} action in AdS$_3$.  We  provide an analysis of some more involved Coulomb gas examples in appendix \ref{app:Asymptotics31and22}; the other appendices collect various technical details.

\section{Information Loss and Forbidden Singularities in AdS/CFT}
\label{sec:AdSCFTInformationLossLargeC}

We will discuss AdS/CFT correlators  to identify signatures of information loss associated with black holes.  In section \ref{eq:SingularitiesfromAdS3} we explain how  certain singularities arise from finite temperature AdS backgrounds, and we review the explicit results in AdS$_3$.  These singularites are always present in the canonical ensemble, as a consequence of Euclidean time periodicity.  However,  as we review in section \ref{sec:ForbiddenSingularitiesVirasoroVacuum},  they also appear universally at large central charge in pure state correlators, where they represent a violation of unitarity.  These `forbidden singularities' are an avatar of information loss.  In section \ref{sec:BehaviorLargeLorentzianTime} we explain how known results on heavy-light Virasoro blocks also manifest information loss as exponential decay at late Lorentzian times.

We will be interested in exponentially small deviations from the thermodynamic limit.  In other words, we will study effects that would vanish in theories with an infinite number of local degrees of freedom, ie with the central charge $c = \frac{2}{3 G_N} \to \infty$.  
We will also need to carefully distinguish between the canonical ensemble and high-energy microstates.

\subsection{Images of OPE Singularities in AdS/CFT}
\label{eq:SingularitiesfromAdS3}

We study AdS in global coordinates, taking the curvature scale $R_{AdS} =1$ so that the pure AdS metric is
\be
ds^2 = -(r^2 + 1) dt_L^2 + \frac{dr^2}{r^2+1} + r^2 d \Omega^2,
\ee
which naturally corresponds to a CFT on the cylinder $R \times S^{d-1}$.  We can study finite temperature CFT correlators in two different phases, separated by the Hawking-Page phase transition \cite{Hawking:1982dh}.  In the thermal AdS phase we simply compactify the Euclidean time $t_E \sim t_E + \beta$.  In the AdS-Schwarzschild phase, which dominates at large temperatures, the bulk metric 
\be
ds^2 = -r^2 \left(1 - \frac{r_+^{d}+(r_+^{d-2}-r^{d-2})}{r^{d}}  \right) dt_L^2 + \frac{dr^2}{r^2 \left(1 - \frac{r_+^{d}+(r_+^{d-2}-r^{d-2})}{r^{d}}   \right) } + r^2 d \Omega^2
\ee
has a horizon at $r = r_+$.    To avoid a conical singularity at the horizon we must compactify the Euclidean time coordinate with $\beta = \frac{4 \pi r_+}{d r_+^2 +(d-2)}$.  We will always be interested in large, semi-classically stable AdS black holes with $r_+ \gtrsim 1$.

If we compute CFT correlation functions using a quantum field theory in either thermal AdS or AdS-Schwarzschild, to any order in perturbation theory we will obtain correlators satisfying the KMS condition, which requires periodicity in Euclidean time.  Perturbative corrections in $G_N$ will not alter the underlying topology of the space, or the geometry as we approach the boundary of AdS.  

This is exactly what we expect for CFT correlators in the canonical ensemble at fixed temperature. For example, the thermal two point correlator is defined by
\be
\< \CO_L(t, \Omega) \CO_L(0) \>_T \equiv \sum_{\psi}  e^{-\frac{E_\psi}{T}} \< \psi | \CO_L(t, \Omega) \CO_L(0) | \psi \>.
\ee
Notice that Euclidean-time periodicity implies that for $\Omega = 0$ there is a short-distance (OPE) singularity at $t_E = 0, \pm \beta, \pm 2 \beta, \cdots$ as a trivial consequence of the geometry.  Transforming the CFT from the cylinder to the (radially quantized) plane via $z = 1 - e^{-t + i \phi}$, these singularities occur in the Euclidean region at $z = \bar z = 1 - e^{\frac{n}{T}}$ for any integer $n$.  

Euclidean time periodicity, and the OPE image singularities that emerge as a corollary, are perfectly acceptable for a correlation function in the canonical ensemble.  However, they are impermissible in a vacuum correlation function of local operators such as
\be
\label{eq:4ptCorrelator}
\< \CO_H(\infty) \CO_L(1) \CO_L(z) \CO_H(0) \>
\ee
in a unitary CFT with a finite number of local degrees of freedom.  
This also implies that correlators in the micro-canonical ensemble cannot be exactly periodic in $t_E$, since the micro-canonical ensemble involves a finite sum over pure state correlators, ie a finite sum over heavy operators $\CO_H$ with dimensions $h_H$ in a very narrow range.

In fact, correlators such as equation (\ref{eq:4ptCorrelator}) can only have Euclidean singularities in the OPE limits $z \to 0, 1, \infty$.  The proof is an elementary consequence of the derivation of radial quantization \cite{Pappadopulo:2012jk}.  Away from the OPE limits, we can interpret the correlator as an inner product of normalizable CFT states, and so it must be finite.

Before focusing on AdS$_3$, we should note that there is another signature of information loss in CFT correlators \cite{Maldacena:2001kr}: the two-point correlator in an AdS-Schwarzschild geometry decays exponentially at late Lorentzian times.   This has an intuitive appeal, representing the fact that information tossed into a black hole never comes back out.  Of course Lorentzian-time decay has an imprecise but intuitive relationship with Euclidean periodicity, since exponential decay and periodicity are related by analytic continuation.   We expect that via the Luscher-Mack theorem \cite{Luscher:1974ez} (see \cite{Hartman:2015lfa} for a recent relevant discussion) that if correlators in the Euclidean region are non-singular and satisfy reflection positivity, then they can be continued to provide healthy Lorentzian correlators.  In what follows we will focus more on the Euclidean region, though we will discuss late time Lorentzian behavior in sections \ref{sec:BehaviorLargeLorentzianTime} and \ref{sec:LateLorentzianBehaviorDiffEq}.

Even in the semi-classical limit, there are few explicit examples (see e.g. \cite{Fidkowski:2003nf} for one) of  correlation functions in AdS-Schwarzschild backgrounds in general $d$.  However, two-point correlators in BTZ backgrounds can be easily obtained from the method of images \cite{KeskiVakkuri:1998nw}, so let us now focus on the case of AdS$_3$.  We will see that AdS$_3$ correlators in the presence of a heavy source have a nice analytic continuation in the heavy source mass, and that above the BTZ black hole \cite{BTZ} threshold, the correlators develop OPE image singularities.

For simplicity let us consider scalar probes of scalar BTZ black holes or deficit angles.  The Euclidean metric is
\be \label{eq:BTZGeometry}
ds^2 = (r^2 - r_+^2) dt^2 + \frac{dr^2}{r^2 - r_+^2} + r^2 d \phi^2
\ee
where we note that the horizon radius relates to the Hawking temperature via $r_+ = 2 \pi T_H$.  If we interpret the black hole as a CFT state, then it will have holomorphic dimension $h_H$ related to the horizon radius via  $r_+ = \sqrt{\frac{24 h_H}{c}-1}$.  Deficit angles are obtained by analytically continuing to imaginary $r_+$, which automatically occurs when $h_H < c / 24$.  In other words, all of our results can be analytically continued in $h_H$.

Since the deficit angle and BTZ geometries are orbifolds of AdS$_3$ \cite{BTZ}, we can obtain the correlator pictured in figure \ref{fig:BasicHeavyLightAdSCylinder} using the method of images \cite{KeskiVakkuri:1998nw}.  The result is
\be \label{eq:BTZDeficitCorrelator}
\< \CO_L(z, \bar z) \CO_L(1) \>_{r_+} = \sum_{n= -\infty}^\infty [V(z,n)]^{h_L} [V( \bar z,-n)]^{\bar h_L}
\ee
where
\be \label{eq:BulkVfunction}
V(z,n) = \frac{(1-z) }{ \left( \sin \left( \frac{r_+}{2} (\log(1-z) + 2 \pi i n) \right)  \right)^2}
\ee
is a function that will appear later in a different guise.    The sum over $n$ ensures that the overall correlator is single valued in the Euclidean plane, where $z$ and $\bar z$ are related by complex conjugation.  If $z$ and $\bar z$ circle the branch cut at $z = 1$ in opposite directions, then we simply have $n \to n+1$ for each summand, so that the total sum over images does not change.  Note that if  $z$ circles the branch cut while $\bar z$ remains fixed, the correlator is altered; this analytic continuation takes the correlator into the Lorentzian regime.

\begin{figure}[t!]
\begin{center}
\includegraphics[width=0.9\textwidth]{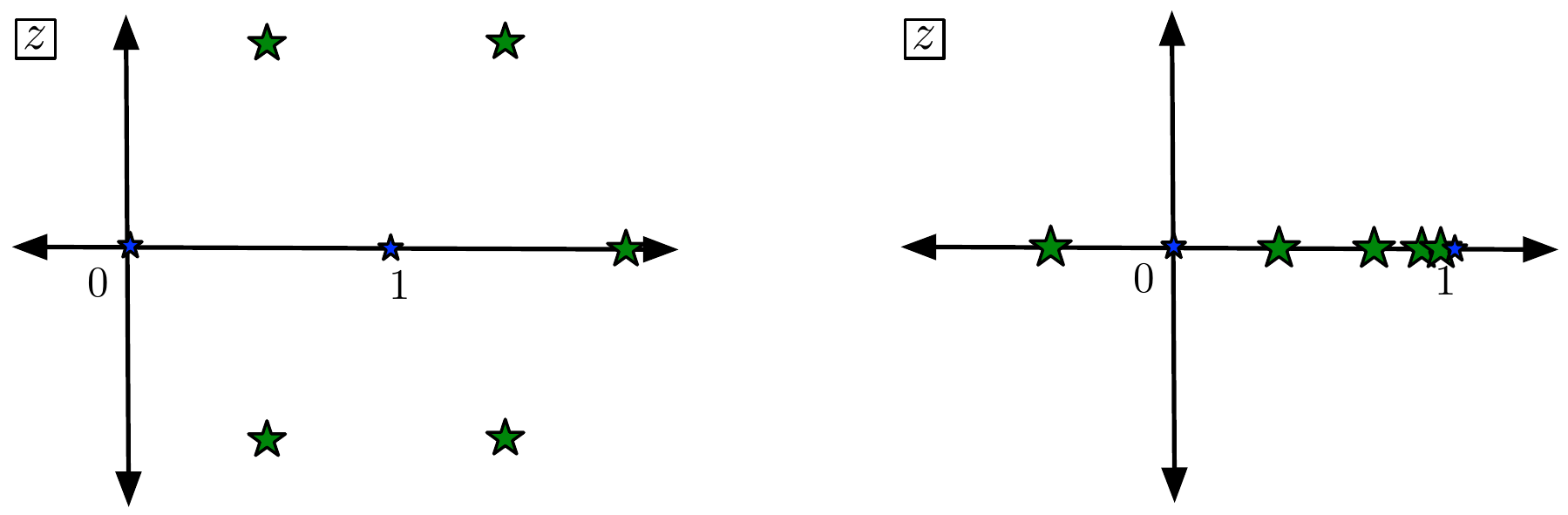}
\caption{ This figure suggests the positions of OPE image singularities of CFT correlators.   Black holes  produce the pattern on the right, while the `additional angles' discussed in section \ref{sec:ExactVirasoroBlocks} produce the pattern on the left.  These singularities are forbidden in unitary four-point correlators.  The heavy operators are located at $1$ and $\infty$, and the light probe operators are at $0$ and $z$ in the Euclidean plane. } 
\label{fig:ForbiddenSingularityArrangement}
\end{center} 
\end{figure} 

Since the full two-point correlator is single-valued in the Euclidean region, let us study its singularities on a single sheet.  In that case the imaginary part of $\log(1-z)$ varies between $0$ and $2 \pi i$, so for real $r_+$, the only term in the sum that can ever be singular is the $n=0$ term.  It is singular when
\be \label{eq:ForbiddenSingularities}
z = 1- e^{\frac{2 \pi m}{r_+} }
\ee
for all integers $m$, which always includes $z=0$ as $m=0$.  For real $r_+$ these singularities lie on the real axis, while for imaginary $r_+$ they form a unit circle around $z=1$, as pictured in figure \ref{fig:ForbiddenSingularityArrangement}.  They are simply the periodic images of the singularity $\CO_L(z) \CO_L(0) = \frac{1}{z^{2 h_L}} + \cdots$ due to the universal presence of the operator `$1$' in the light operator OPE.  Thus we see that correlators in  BTZ black hole backgrounds develop singularities that are forbidden from four-point correlators like equation (\ref{eq:4ptCorrelator}).  

In the presence of a rational deficit angle, with $r_+ = \frac{i}{k}$ and $k$ an integer, there will be no forbidden singularities.  However, if $r_+ = i k$ for an integer $k \geq 2$, then the image sum in equation (\ref{eq:BTZDeficitCorrelator}) is unnecessary (to ensure periodicity in the angular coordinate $\phi$), and the correlator develops $k-1$ extra OPE image singularities.  This case of `additional angle', pictured in figure \ref{fig:Visualization4PiAngle}, will be relevant later on; its structure of forbidden singularities is shown on the left in figure \ref{fig:ForbiddenSingularityArrangement}.

\subsection{Forbidden Singularities in the Virasoro Vacuum Block}
\label{sec:ForbiddenSingularitiesVirasoroVacuum}

A crucial feature of the AdS$_3$ correlator from equation (\ref{eq:BTZDeficitCorrelator}) is that for real $r_+$ the forbidden singularities come exclusively from the $n=0$ term in the image sum.  This fact has a natural and important interpretation in conformal field theory. 

It is not clear whether an AdS computation in a black hole background represents a thermal correlator or a correlator in the background of a heavy pure microstate, since we expect these to be indistinguishable at leading order in large $c \propto 1/G_N$.  But let us  interpret equation (\ref{eq:BTZDeficitCorrelator}) as the latter, ie as a heavy-light four-point correlator in a CFT.  All four-point correlators in CFT$_2$ can be written as a sum over Virasoro conformal blocks
\be \label{eq:HeavyLightVirasoroBlockDecomposition}
\< \CO_H(\infty) \CO_L(1) \CO_L(z) \CO_H(0) \> = \mathcal{V}_{0}(1-z)  \mathcal{V}_{0}(1-\bar z)
+ \sum_{h, \bar h} P_{h, \bar h} \mathcal{V}_{h}(1-z)  \mathcal{V}_{\bar h}(1-\bar z)
\ee
where $P_{h, \bar h}$ are products of OPE coefficients.  There is a universal contribution from the vacuum Virasoro block $\mathcal{V}_0(1-z)$ necessitated by the fact that both $\CO_L(z) \CO_L(0)$ and $\CO_H(z) \CO_H(0)$ contain the operator `$1$' in their OPE.  In fact, the vacuum block can be computed directly using the Virasoro algebra at large $c$ \cite{Fitzpatrick:2014vua, Fitzpatrick:2015zha, Fitzpatrick:2015foa, Alkalaev:2015wia, Alkalaev:2015lca, KrausBlocks, Beccaria, Fitzpatrick:2015dlt}, and it corresponds precisely with the $n=0$ term in the AdS image sum of equation (\ref{eq:BTZDeficitCorrelator}).  But before discussing this further, let us briefly review the physical content of the Virasoro blocks.

In a quantum theory with a symmetry, we can decompose the states into irreducible representations of the symmetry group.  Once we know the matrix element of a single state in the irreducible representation, we can work out matrix elements of related states using the symmetry.  This leads to a partial wave expansion for scattering amplitudes and correlation functions.  In CFTs, this conformal partial wave or conformal block decomposition can also be derived by applying the OPE expansion (see \cite{Rychkov:2016iqz, Simmons-Duffin:2016gjk} for nice reviews).  The highest weight state of the conformal algebra is called a primary state/operator, and all OPE coefficients in the theory are determined by the OPE coefficients of these primary operators.  The $P_{h, \bar h}$ in equation (\ref{eq:HeavyLightVirasoroBlockDecomposition}) are products of these OPE coefficients.

The Virasoro conformal blocks contain an immense amount of information about quantum gravity in AdS$_3$.  This follows because via AdS/CFT, the stress energy tensor $T_{\mu \nu}$ of the CFT creates gravitons in AdS.  In the case of $d=2$, the Virasoro generators $L_n$ are simply modes of the stress tensor
\be
T_{zz}(z) = \sum_n z^{-2-n} L_n
\ee
and so $L_{n}$ with $n > 2$ create states that can be naturally interpreted as `gravitons' in AdS$_3$.  Their interactions are governed by the Virasoro algebra
\be
\label{eq:VirasoroAlgebra}
[L_n, L_m ] = (n-m) L_{n+m} + \frac{c}{12} n (n^2-1) \delta_{n+m,0}
\ee
When we sum over all states related by Virasoro symmetry, we are actually including all possible effects from the exchange of gravitons.  The Virasoro vacuum block in equation (\ref{eq:HeavyLightVirasoroBlockDecomposition}) encapsulates the exchange of any number of pure graviton states between the heavy object and the light probe.

The presence of additional singularities in equation (\ref{eq:BTZDeficitCorrelator}) was rather ambiguous, since it was unclear if we should interpret the BTZ black hole background as  a pure state.  However, it has been shown \cite{Fitzpatrick:2014vua} that the function $V(z,0)$ in equation (\ref{eq:BulkVfunction}), which was obtained from a bulk computation, is in fact identical to the heavy-light Virasoro vacuum block $\mathcal{V}_0(1-z)$ in the limit $c \to \infty$ with $h_L$ and $h_H/c$ fixed.  This means that \emph{in the large $c$ limit, heavy-light CFT correlators have forbidden singularities that must be resolved at any finite $c$}.  These forbidden singularities will be present in any $c \to \infty$ limit of two-dimensional CFTs because they come from the vacuum block.\footnote{Related singularities have been noted in a few cases \cite{Maldacena:2001km, HarlowLiouville, Fateev:2011qa}.}  

Furthermore, at finite $c$ we know that the singularities must always be resolved within the structure of the vacuum block itself.  In other words, the forbidden singularities will not be resolved by a conspiratorial cancellation between the vacuum block and the sum over non-vacuum Virasoro blocks in equation (\ref{eq:HeavyLightVirasoroBlockDecomposition}).  One reason for this expectation is that vacuum block always makes the most singular contribution, proportional to $z^{-2 h_L}$ in the OPE limit $z \to 0$.  Other conformal blocks behave as $z^{h - 2h_L}$ in this limit, with $h > 0$ in unitary theories, and $h = 0$ only for conserved currents.  Since the forbidden singularities are images of the OPE singularity, other conformal blocks will be strictly less singular in both the OPE and image singularity limits.  This is borne out by the explicit formulas for general Virasoro conformal blocks \cite{Fitzpatrick:2015zha}
\be \label{eq:GeneralHLVirasoroBlock}
\mathcal{V}_{h_I}(z) \propto \left( \frac{1-w}{1-z} \right)^{h_L} w^{h_I - 2 h_L} {}_2 F_1(h_I, h_I, 2 h_I, w)
\ee
where $w \equiv 1 - (1-z)^{i r_+}$ with $r_+ = \sqrt{\frac{24 h_H}{c}-1}$ as given above.  The strength of the forbidden singularity is reduced when the intermediate dimension $h_I > 0$.  When $T$ is real, we can make an even simpler argument: the forbidden singularities at $z=1-e^{n/T}$ for $n<0$ are at real and positive $z$, and therefore the sum over the other conformal blocks is a convergent sum over positive contributions that can only add to the singularity in the vacuum block, and cannot cancel it.

There is a sharper and more formal argument  that at finite $c$, forbidden singularities must be resolved within $\CV_0$.  It is simply a restatement of the proof \cite{Pappadopulo:2012jk, Rychkov:2016iqz} that Euclidean CFT correlators only have OPE singularities.  This argument follows directly from radial quantization, whereby local operator insertions create (normalizable) states on enveloping spheres, so that correlators can be interpreted as inner products of normalizable states.  Then a basic theorem on Hilbert spaces states that when such inner products are expanded in an orthonormal basis of states, the resulting sum converges.  This argument may seem a bit formal, since it excludes singularities by presuming that local operator insertions create normalizable states.  So it is worthwhile to take a closer look at our specific setup.  The problem with the heavy-light Virasoro blocks is that as $c \to \infty$ with $h_H / c$ fixed, we must take $h_H \to \infty$, and so states created by $\CO_H$ are no longer unambiguously normalizable.    For example, the correlator $\<\CO_H(0) \CO_H(z)\> = z^{-2h_H}$ is either infinity or zero when $h_H \to \infty$.  We expect  that this underlying issue explains the presence of  forbidden singularities in the heavy-light correlators as $c \to \infty$.   In an AdS dual this occurs because perturbation theory in $G_N$ requires us to take the limit $G_N \to 0$ with the quantity $G_N M_{BH}$ fixed.

Although we are focusing on the vacuum conformal block, general blocks also have their own forbidden singularities, as can be seen directly in equation (\ref{eq:GeneralHLVirasoroBlock}) when $0 < h_I < 2 h_L$.  Even when $h_I > 2h_L$ the correlators generically have forbidden branch cuts.  We expect that these singularities must also be resolved within the structure of these more general Virasoro blocks.  We are not focusing on the general case of $h_I > 0$ because it is more complicated and less universal, but the general heavy-light Virasoro blocks certainly warrant further study.

In summary, the vacuum conformal block, a function determined purely by Virasoro symmetry, exactly matches  AdS$_3$ computations involving deficit angles and BTZ black holes \cite{Fitzpatrick:2014vua, Fitzpatrick:2015zha, Fitzpatrick:2015foa, Alkalaev:2015wia, Alkalaev:2015lca, KrausBlocks, Beccaria, Fitzpatrick:2015dlt, HartmanLargeC, Asplund:2014coa}.  In the large $c$ limit it has forbidden singularities that are indicative of unitarity violation and information loss, and the large $c$ result is analytic in the heavy state dimension $h_H$, interpolating between the deficit angle and black hole cases.  At finite $c$ the forbidden singularities must be resolved within the structure of $\CV_0(z)$  itself.  Thus we can study universal aspects of  information loss in black hole backgrounds by examining $\mathcal{V}_0(z)$ at large but finite central charge.

\subsection{Correlators at Large Lorentzian Time}
\label{sec:BehaviorLargeLorentzianTime}

Maldacena has emphasized \cite{Maldacena:2001kr}  that in a black hole background, correlators decay exponentially at late Lorentzian times.  So a small perturbation to the initial density matrix becomes arbitrarily well scrambled \cite{Bak:2007qw, Germani:2013sra, Anous:2016kss} at late times. Intuitively, this means that information thrown into a black hole never returns.   This behavior is forbidden in a theory with a finite number of local degrees of freedom on a compact space, so it provides a sharp signature of information loss when CFT correlators are obtained from AdS.  

Instead of exponential decay at arbitrarily late times, in a unitary CFT we expect \cite{Maldacena:2001kr} that correlators will have a value at least of very rough order $e^{-\kappa S_{BH}}$ for some numerical constant $\kappa$.   This expectation can be derived by imagining that the early-time correlator can be written as a coherent sum of roughly $e^{S_{BH}}$ terms, corresponding to intermediate energy eigenstates in the $\CO_H \CO_L \to \CO_H \CO_L$ channel.  If each term has a time dependence $e^{i E t}$, and if the energies  $E$ have a random distribution near the black hole mass,\footnote{See \cite{Barbon:2003aq, Barbon:2004ce, Kleban:2004rx, Barbon:2014rma} for some statistical relations between the spectrum and late time behavior.} then at late times the terms will add with incoherent phases, producing an average result suppressed by $\sim e^{-S_{BH}/2}$.

Equation (\ref{eq:HeavyLightVirasoroBlockDecomposition}) displays the decomposition of a complete CFT correlator into a sum over general Virasoro blocks, with coefficients given by products of OPE coefficients.  Furthermore, all Virasoro blocks make important contributions at large Lorentzian time, so we might not expect to be able to understand the behavior of the correlator in the large Lorentzian time regime without knowing all CFT data (the spectrum and the OPE coefficients of the theory).

However, we have computed the heavy-light Virasoro blocks  \cite{Fitzpatrick:2015zha} in the limit that the intermediate dimension $h_I$ is fixed as $h_H \propto c \to \infty$, and for all values of $h_I$, the blocks have a remarkable common feature: for $h_H > \frac{c}{24}$ they all vanish exponentially when analytically continued to large Lorentzian time.  To see this, note that these blocks have the functional form \cite{Fitzpatrick:2015zha}
\be
\CV_{h_I}(z) \propto     \left( \frac{1-w}{1-z} \right)^{h_L} w^{h_I - 2 h_L} {}_2 F_1(h_I, h_I, 2 h_I, w), \ \ \ \ \ w \equiv  1- (1-z)^{i r_+}
\ee
with $r_+ = 2 \pi T_H = \sqrt{24 \frac{h_H}{c} - 1}$, and $h_H > \frac{c}{24}$ corresponding to a BTZ black hole in AdS$_3$.  We can study the Lorentzian time $t_L$ via $z = 1 - e^{-it_L}$, in which case since $\alpha$ is imaginary, we have  $w = 1 - e^{2 \pi T_H t_L}$.  Furthermore, at large $t_L$ we have
\be
  {}_2 F_1 \left( h_I, h_I, 2 h_I, 1- e^{2 \pi T_H t_L} \right) \propto  e^{-2 \pi h_I T_H t_L}
\ee
so that overall, every  block is proportional to $e^{-2 \pi h_L T_H t_L}$ as $t_L \to \infty $, regardless of the value of $h_I \ll c$.  Notice that we have the same behavior as $t_L \to -\infty$, as we should expect since the two light operators  $\CO_L$ in the correlator are identical.  Thus all of the heavy-light, large central charge Virasoro blocks that we can explicitly compute vanish at large Lorentzian times.  Since we expect the sum over blocks to be convergent in CFT$_2$ \cite{Maldacena:2015iua}, this implies that correlators constructed from such a sum must also vanish exponentially at large $t_L$.  Since we do not have explicit expressions for the Virasoro blocks when $h_I \propto c$, a loophole remains, as it is possible that heavy-light blocks with heavy intermediate states do not vanish at late times.

Nevertheless it is interesting to ask if \emph{any} of the exact heavy-light Virasoro blocks with $h_H > \frac{c}{24}$ are non-vanishing at large $t_L$, and to study their behavior in this limit Lorentzian limit.   We will begin to address this version of information loss in section \ref{sec:LateLorentzianBehaviorDiffEq}, where in particular we show that the behavior of the vacuum block changes qualitatively at times of order $S_{BH} = \frac{\pi^2}{3} c T_H$, the black hole entropy.

\section{Exact Virasoro Blocks at Large Central Charge}
\label{sec:ExactVirasoroBlocks}

To resolve information loss, we need a method to obtain exact information about the heavy-light Virasoro blocks.
In this section we will discuss an infinite class of examples where exact information can be obtained.  First we will very briefly review degerate operators in section \ref{sec:DegenerateStateReview}. We provide an illustrative example of the general story in section \ref{sec:QuickHeavyDegenerateExample}.  Then in section \ref{sec:ConnectingDegenerateAndLargeC} we explain how the correlators of degenerate operators can be analytically continued to precisely reproduce all of our previous large $c$ results.  In section \ref{sec:ResolutionofForbiddenSingularities} we will discuss the non-perturbative resolution of the forbidden singlarities at finite $c$.  Motivated by these successes, in section \ref{sec:LateLorentzianBehaviorDiffEq} we discuss the late Lorentzian time behavior of the vacuum block.

\subsection{Brief Review of Virasoro Blocks and Degenerate States}
\label{sec:DegenerateStateReview}

Any CFT$_2$ correlator can be written as a sum over Virasoro conformal blocks
\be \label{eq:GeneralBlockDefinition}
\< \CO_1(\infty) \CO_2(1) \CO_3(z) \CO_4(0) \> = \sum_{h, \bar h} P_{h, \bar h} \mathcal{V}_{h_i, h, c}(z)  \mathcal{V}_{\bar h_i, \bar h, c}(\bar z)
\ee
where we have chosen the $12 \to 34$ channel derived from the OPE expansion of $\CO_3(z) \CO_4(0)$, and explicitly indicated the decomposition into a holomorphic and anti-holomorphic part.  The $h_i$ are dimensions of the external operators $\CO_i$ and $h, \bar h$ are intermediate operator dimensions.   These Virasoro conformal blocks, which are also known as partial waves, encapsulate the contribution of an entire irreducible representation of the Virasoro algebra to the correlator.

The holomorphic part of the blocks $\mathcal{V}_{h_i, h, c}(z)$ depends on the four external operator dimensions, the internal primary operator dimension $h$, the central charge $c$, and the kinematical variable $z$ in the plane.    Ideally we would like to have an explicit, closed-form expression for the general Virasoro conformal blocks.  Such a formula would allow us to observe how the forbidden singularities and late Lorentzian time behavior discussed in section \ref{sec:AdSCFTInformationLossLargeC} are resolved by non-perturbative effects $\sim e^{-c}$ in the large $c$ expansion.

This is probably too much to hope for.  Current tools provide  recursion relations \cite{ZamolodchikovRecursion} that efficiently compute  the series expansion \cite{Perlmutter:2015iya} of the blocks near $z=0$ with generic $h_i, h, c$; closed form results in the limit  $h \to \infty$ \cite{Zamolodchikovq}; and closed form results as $c \to \infty$ in the heavy-light limit \cite{Fitzpatrick:2014vua, Fitzpatrick:2015zha, Fitzpatrick:2015foa, Alkalaev:2015wia, Alkalaev:2015lca, KrausBlocks, Beccaria}, including general $1/c$ correction \cite{Fitzpatrick:2015dlt} to that limit.  The heavy-light limit displays the blocks' forbidden singularities at large $c$, but none of these results provide information about how those singularities are resolved at finite $c$.  The relation of the general large $c$ semi-classical blocks to the Painlev\'e VI equation \cite{Litvinov:2013sxa}, which can only be solved in terms of its own special function, does not seem to encourage those who might seek a closed form expression for $\CV$.

However, as has been known since the early days of CFT$_2$ \cite{Belavin:1984vu}, for certain special values of the parameters $h_i, h, c$ we can obtain exact information about  the Virasoro blocks.\footnote{For a thorough review see \cite{DiFrancesco:1997nk} or \cite{Ginsparg}.} 
These are cases where one of the external operators is degenerate, meaning that some of its Virasoro descendants are null states, or states with vanishing norm.  When discussing degenerate states it is useful to use a parameter $b$ so that
\be
c \equiv 1 + 6 \left(  b + \frac{1}{b} \right)^2
\ee
We can take the $c \to \infty$ limit via either $b \to 0$ or $b \to \infty$.  In this notation, the simplest example of a null state is the second level descendant
\be
\label{eq:12NullDescendant}
\left( L_{-1}^2 + b^2 L_{-2} \right) | h_{1,2} \> = 0
\ee
One can check using the Virasoro algebra of equation (\ref{eq:VirasoroAlgebra}) that the matrix of level two inner products
\be
\label{eq:Level2Det}
\left( \begin{array}{cc}
\< h | L_1^2 L_{-1}^2 | h\>  & \< h | L_1^2 L_{-2} | h\>  \\
\< h | L_2 L_{-1}^2 | h\> & \< h | L_2 L_{-2} | h\>  \end{array} \right)
\ee
has a vanishing determinant when the holomorphic dimension $h_{1,2} = -\frac{1}{2} - \frac{3}{4b^2}$; the level two descendant in equation (\ref{eq:12NullDescendant}) is the corresponding null vector.  
In general, degenerate states  can only occur for holomorphic dimensions satisfying the Kac formula
\be
\label{eq:GeneralDegenerateDimension}
h_{r,s} &=& \frac{b^2}{4}(1-r^2)  + \frac{1}{4b^2} (1-s^2) + \frac{1}{2}(1-rs)
\ee
for positive integers $r,s$.  This formula determines the values of dimension $h$ when the Kac determinant, of which equation (\ref{eq:Level2Det}) is an elementary example, vanishes.  Notice that  $r \leftrightarrow s$ simply corresponds with $b \leftrightarrow 1/b$. 

Once inserted into a correlator, the relation (\ref{eq:12NullDescendant}) becomes a very useful differential equation for the correlation functions of the primary operator $O_{1,2}(z)$ that creates $| h_{1,2} \>$.  This follows because within a correlator with operators of dimension $h_i$, a Virasoro generator $L_{-m}$ will act as the differential operator 
\be
L_{-m} \to \sum_{ \{i \}_{z_i \ne z}} \left( \frac{(m-1) h_i}{(z_i - z)^m} - \frac{1}{(z_i-z)^{m-1}} \partial_{z_i} \right) ,
\ee
as a consequence of stress energy tensor Ward identities.
For example, applying these differential operators and then performing a conformal transformation to send the operators to canonical positions, in the case of $\CO_{1,2}$ one finds 
\be
\label{eq:NullDiffEqh21}
\left( \partial_z^2 + \left(2 \frac{1+b^{-2}}{z} + \frac{b^{-2}}{1-z}\right) \partial_z  + \frac{b^{-2} h_H}{(1-z)^2} \right) \frac{\< \CO_H(\infty) \CO_H(1) \CO_{1,2}(z) \CO_{1,2}(0) \>}{\< \CO_H(\infty) \CO_H(1) \>\< \CO_{1,2}(z) \CO_{1,2}(0) \>} = 0
\ee
where $h_H$ is the dimension of $\CO_H$.
This is a version of the hypergeometric differential equation; it is an exact relation for this correlator and its conformal blocks.  One of its solutions, the vacuum conformal block, was mentioned in equation (\ref{eq:IntroExample}).

\begin{figure}[t!]
\begin{center}
\includegraphics[width=0.6\textwidth]{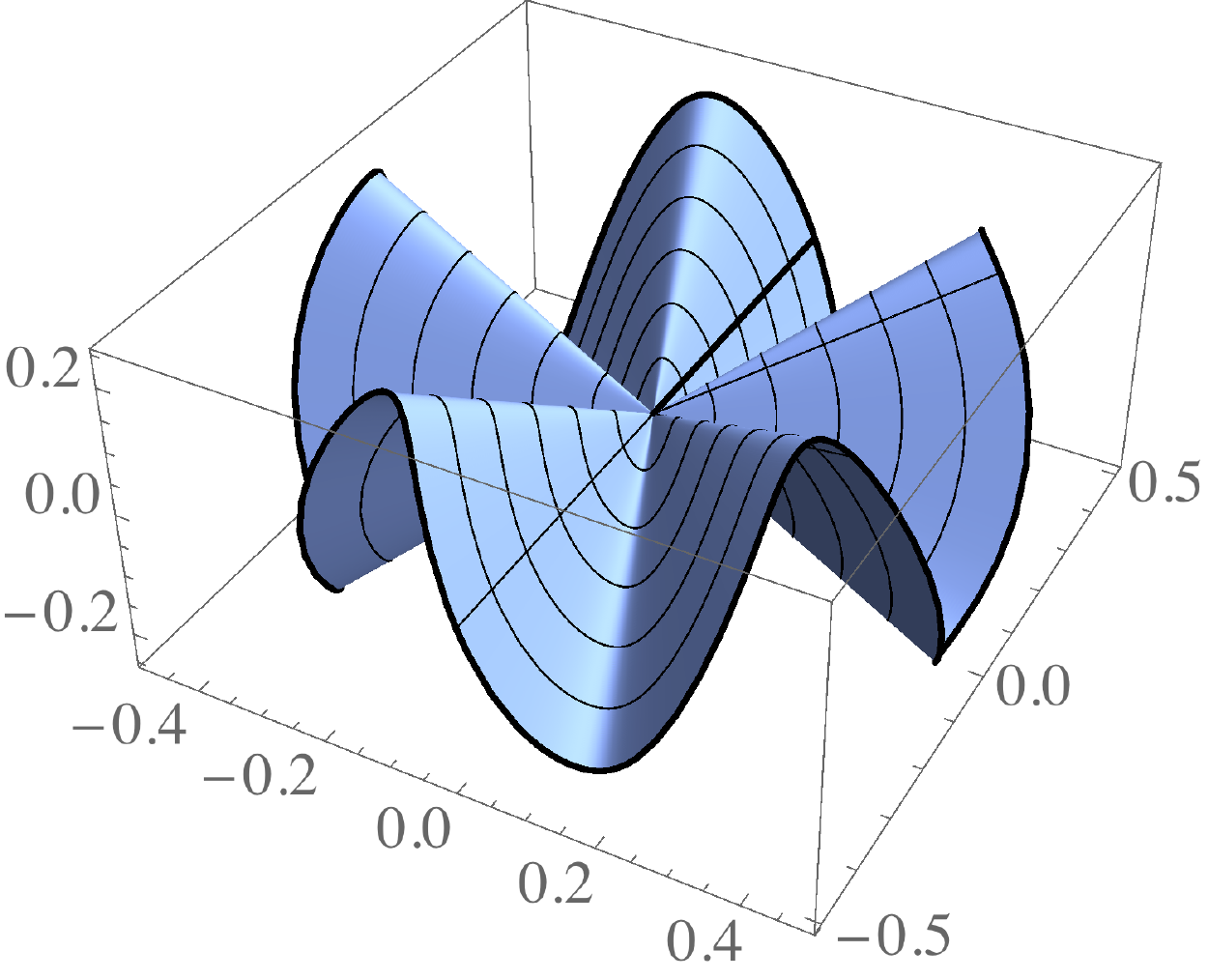}
\caption{ This figure provides a visualization of a space with an `additional angle' totaling $4\pi$  around the origin.  This suggests the spatial geometry created by a heavy degenerate operator with dimension $h_{2,1} = -\frac{c}{8}$ at large $c$.  The $h_{r,1}$ always produce a total angle equal to the integer $r$ times $2 \pi$. } 
\label{fig:Visualization4PiAngle}
\end{center}
\end{figure}

In general, one obtains an $(rs)^{\mathrm{th}}$ order differential equation for correlators of $\CO_{r,s}(z)$.
For the fairly wide range of cases of degenerate states with dimension $h_{r,1}$, the null descendant can be written in closed form as \cite{Benoit:1988aw, DiFrancesco:1997nk}
\be
\label{eq:DegenerateStateLs}
\sum_{p_i} \frac{[(r-1)!]^2 \left( b^2 \right)^{r-k}}{\prod_{i=1}^{k -1} (p_1 + \cdots + p_i) (r - p_1 - \cdots - p_i)}  L_{-p_1} \cdots L_{-p_k} | h_{r, 1} \>
\ee
where the sum is over partitions of $r$ into $k$ positive integers $p_i$.  In later sections we will use this relation to generate differential equations that must be obeyed by Virasoro conformal blocks involving degenerate operators.  

At large $c$,  the degenerate dimensions $h_{r,s}$ become
\be
h_{r,s} 
 \overset{c \to \infty}{\approx} 
\frac{c}{24} (1 - r^2) +
\frac{1-s}{2} + \frac{(r-1)(13 + 13r - 12s)}{24} + \frac{3 \left(r^2-s^2\right)}{2c} + \cdots 
\ee
so the $h_{1,s}$ approach a negative half-integer value at large $c$, while the $h_{r,s}$ with $r>1$ are proportional to $(-c)$.  In other words, the $h_{1,s}$ are light operators, with dimensions that do not scale with $c$, while generically $h_{r,s}$ are heavy, and have a non-trivial effect on the AdS$_3$ geometry even as $c \to \infty$.  These heavy operators lead to `additional angle' in AdS$_3$, as pictured in figure \ref{fig:Visualization4PiAngle}.

\subsubsection{Comments On Analytic Continuation and Unitarity}
\label{sec:ArgumentForAnalyticity}

States in unitary theories must all have positive norm.  In the case of CFT$_2$, this requires that both $c > 0$ and $h \geq 0$ for all states.  This means that in the limit of large positive $c$, correlators of operators with dimension $h_{r,s}$ will not be unitary.\footnote{We can study unitary values $h_{r,s} > 0$ if we take $c \to -\infty$.   This may apply to correlators in dS/CFT \cite{Strominger:2001pn}.}  An immediate consequence is that states with dimension $h_{r,1}$ correspond to large negative mass sources in AdS.  Gravitational solutions incorporating these sources will have $r_+ = i r + O (\frac{1}{c})$ in the geometry of equation \ref{eq:BTZGeometry}, which means that they have an angular surplus, for a total of $2 \pi r$ radians.  This contrasts with positive mass sources, which always create angular deficits.

At this point the reader may be wondering how we can use non-unitary conformal blocks to study information loss.  The answer is analytic continuation.  As a function of $c$ and of intermediate operator dimensions, the Virasoro blocks are meromorphic functions with only simple poles.  The well-known Zamolodchikov recursion relations \cite{ZamolodchikovRecursion, Zamolodchikovq} for the $z$ and $q$-series expansions of the blocks are based on this property.  More importantly, as a function of the  external dimensions $h_L$ and $h_H$, the Virasoro blocks are completely analytic.   This follows because the $q$-expansion of the blocks converges absolutely away from OPE limits \cite{Maldacena:2015iua}, and the coefficients in the $q$-expansion are rational functions of $c$ and polynomials in $h_L$ and $h_H$. Note that formulas like equation (\ref{eq:IntroExample}) appear to have square roots, but this only occurs because of the relation between external dimensions and $c$ for degenerate operators, which follows from equation (\ref{eq:GeneralDegenerateDimension}).  There are no branch cuts or singularities as a function of $h_H$, as can be seen by explicitly expanding equation (\ref{eq:IntroExample}) in $z$ or $t$.

We will see in section \ref{sec:ConnectingDegenerateAndLargeC} that the vacuum blocks for degenerate correlators exactly match the large $c$ blocks (including perturbative $1/c$ corrections) once we analytically continue our large $c$ results to reach external dimensions $h_{r,s}$.  In particular, in the large $c$ limit, the degenerate blocks have forbidden singularities, which are related by analytic continuation to the forbidden singularities that arise from Euclidean time periodicity.  We believe that this provides very strong support for the conjecture that the degenerate blocks `know' about the physics that resolves information loss.  We will also provide further  evidence based on the behavior of $1/c$ corrections to the general heavy-light Virasoro vacuum block in section \ref{sec:ResolutionofForbiddenSingularities}.   

\subsubsection{A Simple Example:  the $\mathcal{O}_{(2,1)}$ Degenerate State}
\label{sec:QuickHeavyDegenerateExample}

In this subsection, we provide a simple example to illustrate how forbidden singularities appear at large $c$, and why their removal at finite $c$ relies on a non-perturbative effect. 
We will study the holomorphic vacuum block $\mathcal{V}_{2,1}$ in the 4-point function $\langle\mathcal{O}_{(2,1)}(\infty)\mathcal{O}_{(2,1)}(1)\mathcal{O}_{L}(z,\bar{z})\mathcal{O}_{L}(0)\rangle$. $\mathcal{O}_{(2,1)}$ becomes heavy in the large $c$ limit since 
\begin{equation}
h_{2,1}=-\frac{1}{4}(3b^{2}+2),
\end{equation}
tends to negative infinity as we take $c \propto b^2 \rightarrow+\infty$. In this limit, the heavy operator induces an additional angle of $2\pi$. This additional angle geometry in AdS$_3$ leads to a forbidden singularity at $z=2$ in the CFT vacuum block.

\begin{figure}[t!]
\begin{center}
\includegraphics[width=0.7\textwidth]{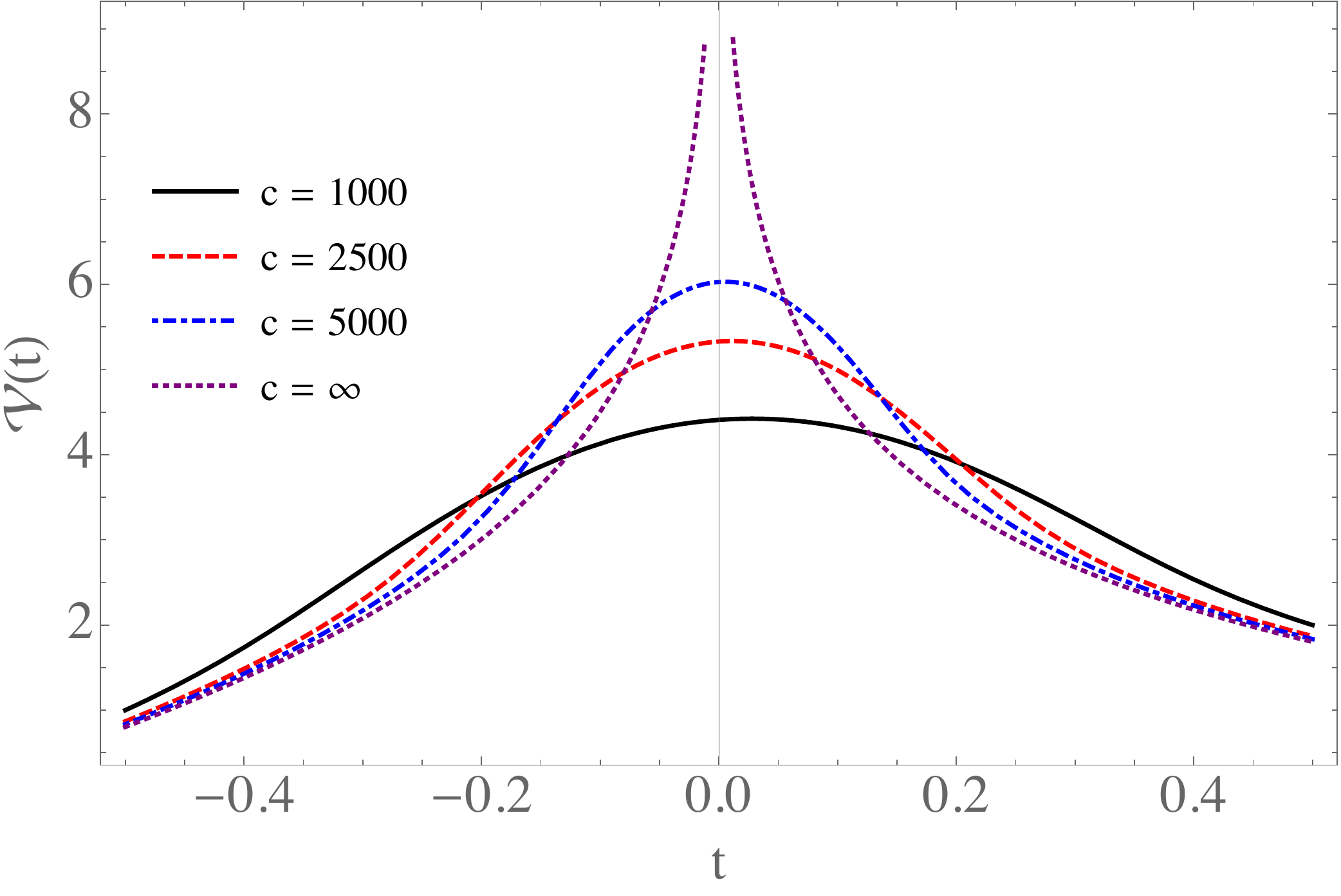}
\caption{ This figure shows the behavior of a degenerate Virasoro vacuum block  near a forbidden singularity for various values of the central charge $c$.  We have specifically plotted $\log | \CV_{2,1} |$ with $h_L = 1$ as a function of the variable $\log(z-1)$ in the vicinity of $z=2$. }
\label{fig:NearForbiddenSingularity21}
\end{center}
\end{figure}

We will take the light operator to have dimension $h_{L}=1$ for simplicity. The null condition on $\mathcal{O}_{2,1}$ implies the following 2nd order differential equation: 
\begin{equation}
\left[\frac{1}{b^{2}}\partial_{z}^{2} + \frac{1}{(1-z)^{2}}+\left(\frac{2}{b^{2}z} + \frac{(2-z)}{z(1-z)}\right)\partial_{z}\right]\tilde{\mathcal{V}}_{2,1}(b,z)=0,
\label{eq:O21equation}
\end{equation}
where we denote $\tilde{\mathcal{V}}_{2,1}\equiv z^{2h_{2,1}}\mathcal{V}_{2,1}$. 
The solution corresponding to the vacuum block is: 
\begin{equation}
\tilde{\mathcal{V}}_{2,1}(b,z)=(1-z)_{2}F_{1}(2,b^{2}+1,2b^{2}+2,z).
\end{equation}
This block is finite at $z=2$. However, $\tilde{\mathcal{V}}_{2,1}(b,2)$ is proportional to $b^{2}$ and becomes singular as $b\rightarrow+\infty$. This behavior is illustrated in figure \ref{fig:NearForbiddenSingularity21}.

Another way to see the emergence of the forbidden singularity is to take the $b\rightarrow\infty$ limit directly in (\ref{eq:O21equation}). Then we find
\begin{equation}
\left[1+\frac{(2-z)(1-z)}{z}\partial_{z}\right]\tilde{\mathcal{V}}_{2,1}(\infty,z)=0,
\label{eq:O21equation-Largeb}
\end{equation}
with the solution 
\be
\tilde{\mathcal{V}}_{2,1}(\infty,z) = \frac{1-z}{(2-z)^2},
\ee
As expected, this forbidden singularity has the same property as the $z^{-2}$ OPE singularity of $V_{2,1}$. It cannot be resolved at any order in the large $c$ (or large $b^2$) perturbation expansion.  In fact, we will show in section~\ref{sec:Borelof21} that:  
\be
\CV_{2,1}(b,z) = \sum_{k} \frac{1}{b^{2k}}\frac{p_k(z)}{(z(2-z))^{2(k+1)}},
\ee
where $p_k$ is a polynomial that is non-zero at $z=0,2$. So the forbidden singularity becomes even more singular at higher orders in $1/c$ perturbation theory. This signals the break down of the large $c$ asymptotic expansion around $z=2$, and implies that the removal of this forbidden singularity is necessarily a non-perturbative effect with the schematic form $e^{-c f(z)}$. We will characterize this non-perturbative effect in detail in section~\ref{sec:NonPerturbativeUnitarityRestoration}. 

Comparing (\ref{eq:O21equation}) and (\ref{eq:O21equation-Largeb}), we see that the crucial non-perturbative corrections to the vacuum block actually originate from a `perturbative' correction to the differential equation that the block obeys.  We will show in~\ref{sec:ResolutionofForbiddenSingularities} that this is in fact an universal mechanism that removes all forbidden singularities in the vacuum blocks involving $\phi_{r,1}$ heavy operators. We will then analytically continue in $r$ to study the late Lorentzian time behavior of the correlator induced by this type of non-perturbative effect.

\subsection{Connecting Degenerate and Large $c$ Virasoro Blocks }
\label{sec:ConnectingDegenerateAndLargeC}

In this section we will discuss the connection between correlators with degenerate operators and the more general, but much less precise results on Virasoro blocks in the heavy-light large central charge limit \cite{Fitzpatrick:2014vua, Fitzpatrick:2015zha, Fitzpatrick:2015foa, Alkalaev:2015wia, Alkalaev:2015lca, KrausBlocks, Beccaria, Fitzpatrick:2015dlt}.  In all cases we will find an exact match, but the details are interesting and lead to a useful new computational method to be explored elsewhere \cite{Hongbin}.

\subsubsection{Light Degenerate States and `Hawking from Catalan'}

In this subsection we will study the case where the degenerate state is light, in the sense that it has a fixed dimension $h_{1,s} = \frac{1-s}{2}$ at large $c$.  So for this section we regard the degenerate state as a light object probing the background created by a heavy state with dimension $h_H \propto c$.

The Virasoro blocks in the heavy-light large $c$ limit have been obtained via a number of seemingly unrelated methods \cite{Fitzpatrick:2014vua, Fitzpatrick:2015zha, Fitzpatrick:2015foa, Alkalaev:2015wia, Alkalaev:2015lca, KrausBlocks, Beccaria}.  A recent derivation was based on a brute force evaluation of Virasoro matrix elements.  This led to a suprising new expression for the blocks as a sum over diagrams composed of propagators and trivalent vertices.  It was then possible to compute the sum over all diagrams by observing it obeys a recursion relation closely related to that of the Catalan numbers.  
The final result was a second order differential equation for the heavy-light vacuum block \cite{Fitzpatrick:2015foa}.  

If we write $\mathcal{V} = (1- z)^{-h_L} W(z)^{-2h_L}$ and then use the variable $t \equiv -\log(1-z)$, this equation reduces to the simple form \cite{Fitzpatrick:2015foa}
\be
\label{eq:HawkingfromCatalan}
\partial_t^2 W(t) + \frac{1}{4} r_+^2 W(t) = 0
\ee 
where we we recall that $r_+ \equiv i \sqrt{1- 24 \frac{h_H}{c} }$.  
  This coincides precisely with the leading large $c$ limit of the $h_{1,2}$ null state equation (\ref{eq:NullDiffEqh21}).  Thus the null descendant differential equation for $h_{1,2}$ coincides with the `Hawking from Catalan' differential equation in the large $c$ limit, and of course this also implies that the blocks themselves must be identical at large $c$.  

We can also examine the cases  $h_{1,s}$ at leading order in the large $c$ expansion.  In fact,  the results of  \cite{Bauer:1991ai,  DiFrancesco:1997nk} imply that in the large $c$, 
the differential equation takes the form
\be
\left[ \prod_{k=-(s-1)+2j\atop j=0, \dots, s-1}\left( \partial_t - \frac{i k r_+}{2} \right)  \right] e^{\frac{s-1}{2}t}\mathcal{V}_s(t) = 0 .
\label{eq:lightnullhighcsimp}
\ee 
The details are reviewed in appendix \ref{app:DerivationLightNullStateEquations}.
 Substituting $\CV_s =e^{\frac{1-s}{2}t} [W(t)]^{s-1}$ we see that  any $W$ satisfying equation (\ref{eq:HawkingfromCatalan}) will automatically satisfy these differential equations.  Thus these equations all have the large $c$ heavy-light vacuum block as solutions.  

These results can be extended to obtain information about perturbative $1/c$ corrections to the heavy-light vacuum blocks.  The idea is to assume that the general heavy-light vacuum block $\mathcal{V}$ can be written as the ansatz\footnote{Until recently it was not clear whether such an ansatz would be valid, but \cite{Fitzpatrick:2015zha} provides a derivation for the case of the vacuum block.  However, a similar expansion of general Virasoro blocks in the intermediate operator dimension $\frac{h_I}{c}$ would not be valid, as the  large $c$ limit with $h_I$ fixed is not equivalent to the large $c$ limit with $h_I/c$ fixed.}
\be
\CV = \exp \left[ h_L \, \sum_{n, m=0}^\infty \left(\frac{1}{c}\right)^m \left( \frac{h_L}{c} \right)^n f_{mn} \left( \frac{h_H}{c}, z \right)\right]
\ee
Then the functions $f_{mn}$ can be determined by expanding the exact results for degenerate external operators and matching \cite{Hongbin}.  We have used this method to verify that the degenerate states match onto results for the vacuum block \cite{Beccaria:2015shq, Fitzpatrick:2015dlt} to first order in $1/c$ perturbation theory.

\subsubsection{Heavy Degenerate States}
\label{sec:HeavyDegenerateStates}

We can also study the limit where the light operator dimension $h_L$ is a free variable, while the heavy operators are degenerate states with dimension $h_{r,1}$.  In fact, this case will be of greater interest in the sections to follow, because the associated vacuum blocks have forbidden singularities at $c=\infty$ and interesting non-perturbative structure in the $1/c$ expansion.  For now we will focus on the connection between these correlators and the general heavy-light large $c$ Virasoro blocks.

When the $h_H = h_{r,1}$, we find that $r_+ = 2 \pi r$ with positive integer $r$, and so the heavy-light large $c$ vacuum block becomes
\be
\tilde{\CV}(t) = \frac{e^{h_L t} (1- e^{-t})^{2 h_L} }{\left[ \sinh \left( \frac{r}{2} t \right) \right]^{2h_L}},
\ee
where we recall $t = -\log(1-z)$.  This has $r$ singularities at $t = \frac{\pi i k}{r}$ for $k=0,1, \cdots, r-1$, where the case $k=0$ is the OPE limit and the other singularities are forbidden.  

This result can also be obtained from the large $c$ limit of the $r^{\mathrm{th}}$ order null state differential equation obtained from the operator of equation (\ref{eq:DegenerateStateLs}), as we now show. In fact, when expanded at large $c$, we find that the differential equations become first order, with the universal form
\be
\left( \partial_t  - h_L g_r(t) \right) \tilde{\CV}(t) = 0,
\label{eq:heavynullfirstorder}
\ee
 where
\be
g_r(t) = \coth \left( \frac{t}{2} \right) - r \coth \left( \frac{r t}{2} \right)
\ee
This equation has the heavy-light vacuum block  with $h_H = h_{r,1}$ as its unique solution.  For instance, we have already discussed the exact differential equation in the case $h_{2,1}$, valid for general $c$ and $h_L$. In the current variables,  it reads
\be
\label{eq:HeavyDegenerateEqn2}
 \left[  \partial_t  - g_2(t) \frac{h_L + b^{-2} \partial_t^2}{1+b^{-2}} \right] \tilde{\CV}(t) &=& 0,
 \ee
 which approaches (\ref{eq:heavynullfirstorder}) in the limit $b\rightarrow \infty$.

To derive (\ref{eq:heavynullfirstorder}) more generally, note that in the limit of large $c$, the states $\CO_{r,1}$ with dimension $h_{r,1}$ are approximately annihilated by $L_{-r}$.  More precisely, 
\be
0 &=& \left( L_{-r} + \frac{1}{c} \sum_{p_i} b_{p_i} L_{-p_1} \dots L_{-p_k} \right) |h_{r,1}\> 
\ee
where the coefficients $b_{p_i}$ are $\CO(1)$ or smaller.  To process the resulting differential equation on the four-point function in such a way that the $1/c$-suppressed terms do not produce additional powers of $h_{r,1}$ (and therefore powers of $c$) upstairs, we write it as
\be
0 &= & \< h_{r,1}|  L_r \CO_{r,1}(0) \CO_L(x) \CO_L(y)\> + \sum_{p_i}\frac{b_{p_i}}{c}  \< h_{r,1} | L_{p_k} \dots L_{p_1}\CO_{r,1}(0) \CO_L(x) \CO_L(y)\> .
\label{eq:heavynullsetup}
\ee
  Now, all $L$'s can be commuted to the right until they annihilate the vacuum.  They all commute with $\CO_{r,1}(0)$ since this is a primary operator inserted at the origin, and the commutators with $\CO_L$ just produce factors of $h_L \sim \CO(1)$.  Consequently, only $L_r$ contributes at leading order in $1/c$.  The four-point function in the above configuration is related to $\CV(z)$ by
  \be
  \< h_{r,1}| \CO_{r,1}(0) \CO_L(x) \CO_L(y) \> &=& \frac{1}{(x-y)^{2h_L}} \tilde{\CV}\left( 1- \frac{x}{y} \right). 
  \label{eq:VinXYCoord}
  \ee
The action of $L_r$ in (\ref{eq:heavynullsetup}) therefore takes the form
  \be
  0 &=& x (x-y) \left(x^r-y^r\right) \tilde{\CV}'\left(1-\frac{x}{y}\right)+y h_{\epsilon } \tilde{\CV}\left(1-\frac{x}{y}\right)
   \left(x^r (-r x+r y+x+y)-y^r (r x-r y+x+y)\right). \nn\\
   \ee
   Setting $y=1$ and $x=e^{-t}$ (which involves multiplying the correlator by the Jacobian factor $e^{-h_L t}$), this reduces to eq. (\ref{eq:heavynullfirstorder}).

Note that since the relations we obtain from external degenerate operators are exact, we also have the ability to study `heavy-heavy' correlators, where all external operators have dimensions scaling with $h \propto c$ at large $c$.
But for this paper we will only focus on the heavy-light limit, where we have concrete expectations from AdS$_3$ and from prior CFT$_2$ calculations.

\subsubsection{Light Degenerate States and Quasi-Normal Modes}

In the heavy-light limit for heavy operators above the BTZ threshold, crossing symmetry implies that the OPE of a heavy and a light operator contains a dense spectrum of states.  The spectral function  has poles at the locations of the quasi-normal modes of the corresponding BTZ metric \cite{Birmingham:2001pj}:\footnote{Such modes are unstable and have corresponding imaginary components in the frequencies, so do not correspond to primary operators in the CFT (which are necessarily stable eigenstates).  However, in this they are not much different from unstable particles in scattering amplitudes and their corresponding poles in the complex plane.}
\be
h_n &=&  h_H+ 2 \pi i T_H(h_L+n), \qquad (n \in \mathbb{N}).
\ee
It is interesting to ask how close light degenerate states can come to reproducing this aspect of the spectrum.  At first sight, degenerate operators would seem to be qualitatively different: light degenerate states have only a finite number of operators in their OPE with any other state, and thus cannot reproduce the spectrum of quasi-normal modes.  However, we will see shortly that they come extremely close, and in the limit $r\rightarrow \infty$ they reproduce the full quasi-normal mode spectrum.

This is easiest to see in the Coulomb gas expressions for the degenerate state weights, which refer to the charge $\alpha$:
\be
h(\alpha) &=& \alpha(Q - \alpha), \qquad Q \equiv b+\frac{1}{b}.
\ee
The charges $\alpha$ and $Q-\alpha$ correspond to the same weight and in fact to the same operator.
The charges of degenerate operators are
\be
\alpha_{r,s} &=& - \frac{1}{2} \left( (r-1)b + (s-1)b^{-1} \right).
\ee
When the degenerate operator $\CO_{r,s}$ fuses with an operator $\CO_H$ of charge $\alpha_H$, the only states it can  make have charge $\alpha_b$ satisfying \cite{DiFrancesco:1997nk}
\be
\alpha_b &=& \alpha_H + p \frac{b}{2} + q \frac{b^{-1}}{2}, 
\ee
for the following allowed values of $p,q$: 
\be
p &=& -(r-1) , -(r-3), \dots, (r-3), (r-1), \nn\\
q &=&-(s-1), -(s-3), \dots, (s-3), (s-1).
\ee

In the case where the degenerate operator is a light probe, one has $r=1$.  
At large $b$ with $h_H/c$ fixed, 
we have $\alpha_H \approx \frac{b}{2} \left( 1 \pm 2 \pi i T_H \right) +\frac{1}{2b} \left( 1\mp \frac{1}{24 i \pi T_H} \pm \frac{13 i \pi T_H}{6} \right)$.
It follows that the spectrum of operators in the $\CO_{1,s} \times \CO_H$ OPE at large $c$ is\footnote{While this paper was in preparation, \cite{Turiaci:2016cvo} appeared which also demonstrates this point.} 
\be
h_b  &=& h_H  + 2 \pi i T_H \left( h_{1,s} + n \right), \qquad n=0, \dots, (r-1),
\ee
which is exactly the spectrum of quasi-normal modes in a black hole background, truncated at $n=(r-1)$.

\subsection{Universal Resolution of Forbidden Singularities}
\label{sec:ResolutionofForbiddenSingularities}

In section \ref{sec:ConnectingDegenerateAndLargeC} we explained how the vacuum Virasoro blocks involving a pair of degenerate external operators agree with recent results on more general vacuum blocks in the heavy-light, large central charge limit.  At large $c$, heavy-light blocks have forbidden singularities, as discussed in section \ref{sec:ForbiddenSingularitiesVirasoroVacuum}, and these persist to all orders in the perturbative $1/c$ expansion.  Since for any finite value of $c$ the vacuum block only has OPE singularities, the  forbidden singularities must be resolved by non-perturbative or `$e^{-c}$' effects.  

In this section we provide an explicit characterization of the way that these forbidden singularities are resolved at finite $c$ by the degenerate Virasoro vacuum blocks.  We begin by providing some sample data concerning these singularities.  However, our most interesting finding is that for all heavy degenerate operators with dimensions $h_{r,1}$, forbidden singularities are resolved in a universal way.  In the vicinity of a forbidden singularity at $x=0$, the vacuum block always behaves like
\be
\label{eq:UniversalFunctionForSingularityResolution}
S(x,c) \approx \int_0^\infty dp \, p^{2h_L - 1} e^{-p x - \frac{\sigma^2}{2 c} p^2}
\ee
at large $c$, up to some order one coefficient $\sigma^2$ in the exponent, which we explicitly compute.  This means in particular that the singularities have a characteristic `width' of order $\frac{1}{\sqrt{ h_L c}}$ in the $z$ or $t$ coordinates.  We conjecture that this function also characterizes the general Virasoro vacuum block near forbidden singularities at large but finite $c$.  As we discuss in section \ref{sec:NearAForbiddenSingularity} and appendix \ref{app:ForbiddenSingularitiesinCorrectionGeneralV}, a study of the $1/c$ corrections to the general heavy-light vacuum block provides strong evidence in support of this conjecture.

\subsubsection{Growth of OPE Coefficients at Finite vs Infinite $c$}

\begin{figure}[t!]
\begin{center}
\includegraphics[width=0.46\textwidth]{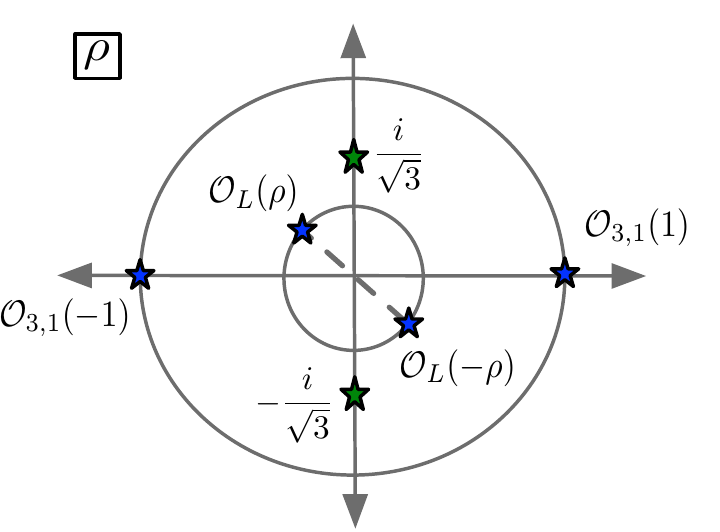}
\includegraphics[width=0.49\textwidth]{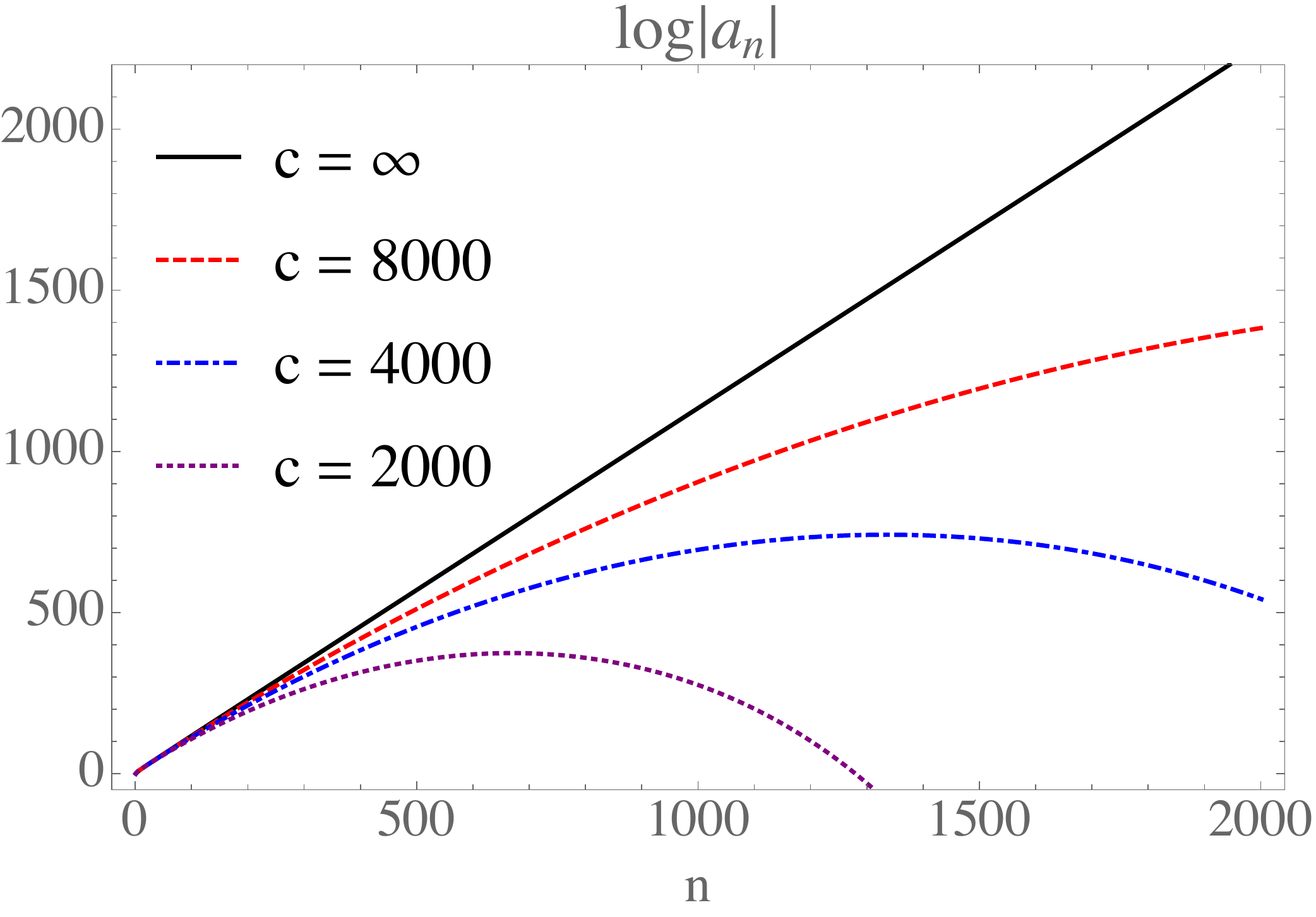}
\caption{ The figure on the left indicates the positions of the operators and the forbidden singularities (green stars) associated with the $\rho$-expansion of $\CV_{3,1}$.  The figure on the right displays the logarithm of the $2n^{\mathrm{th}}$ coefficient from equation (\ref{eq:RhoExpansionDefinition31}) for various values of $c$.  At $c=\infty$ the coefficients grow exponentially for all $n$, leading to a forbidden singularity at $|\rho_\pm | = \frac{1}{\sqrt{3}}$.  At finite $c$ the coefficients initially grow exponentially, but then fall back to sub-exponential behavior at large $n$.}
\label{fig:SeriesCoefficients31Block}
\end{center}
\end{figure}

Both Virasoro and global conformal blocks are expected to converge in the region $|z| < 1$.  We can greatly extend the region of convergence by switching to the $\rho$ coordinates, related to $z$ by
\be
\rho \equiv \frac{z}{(1 + \sqrt{1-z})^2}
\ee
Using $\rho$ is equivalent to performing radial quantization of the correlator $\< \CO(1) \CO(-1) \CO(\rho) \CO(-\rho) \>$, as pictured in figure \ref{fig:SeriesCoefficients31Block}, leading to convergence for $| \rho | < 1$.  This corresponds to the entire $z$-plane minus a branch cut from $[1,\infty)$.
In fact, for CFT$_2$ we can obtain an even greater range of convergence using the uniformizing coordinate $q$ \cite{Zamolodchikovq, Maldacena:2015iua}, but for this pragmatic exercise  the simpler $\rho$ coordinates will be sufficient.

The convergence of $\CV$ will be curtailed in the presence of forbidden singularities.  Let us study the concrete example of the $h_{1,3}$ degenerate Virasoro block.  As $c \to \infty$, it develops  forbidden singularities at $z_\pm = 1 - e^{\pm \frac{2 \pi i}{3}}$, which correspond with $\rho_\pm = \mp \frac{i}{\sqrt{3}}$.     If we write it as
\be
\label{eq:RhoExpansionDefinition31}
\mathcal{V}_{3,1}(\rho) = \sum_n a_n \rho^{n}
\ee
then the presence of the forbidden singularities would lead to exponential growth of $a_n \propto 3^{n/2}$.  Thus at large $c$, we expect to see this growth in the low order coefficients, but eventually $a_n$ must transition to a polynomial behavior in $n$ at large orders.  To study this behavior, we can use the 3rd order null state differential equation to obtain a recursion relation for the coefficients $a_n$:
\be
a_n &=& 
\frac{\left((n-4) \left(28 b^4+b^2 (8 n+2)-3
   (n-6) (n-3)\right)-64 b^2 \left(4 b^2-n+6\right)
   h_L\right)}{n \left(2 b^2+n+1\right) \left(6
   b^2+n+2\right)} a_{n-4} 
\nn \\
   && +
   \frac{ \left((n-2) \left(-28
   b^4+b^2 (8 n-50)+3 (n-3) n\right)-64 b^2 \left(4
   b^2+n\right) h_L\right)}{n \left(2 b^2+n+1\right)
   \left(6 b^2+n+2\right)} a_{n-2}
\ee
where we note that only even powers of $\rho$ appear.  The initial conditions for the recursion relation are $a_0 = 1$ and $a_2 = -\frac{32 (1+2 b^2) b^2 h_L}{6 b^2 +13 b^2+6}$.  We have plotted the behavior of the $a_n$ for various values of $c$ in figure \ref{fig:SeriesCoefficients31Block}.  Experimentally, we have observed that the exact coeffiicents at finite $c$ begin to diverge from the large $c$ coefficients by an order one factor at $n \propto \sqrt{c}$.

\subsubsection{Behavior Near a Forbidden Singularity}
\label{sec:NearAForbiddenSingularity}

To explore how the singularities are resolved in the degenerate state differential equation (as shown in figure \ref{fig:NearForbiddenSingularity21}) we need to work beyond the leading order in large $c$ equation (\ref{eq:heavynullfirstorder}).  From (\ref{eq:DegenerateStateLs}), we can read off that the null state at sub-leading order in $1/c$ is\footnote{It is somewhat easier to derive the coefficients in (\ref{eq:heavynullsecondorder}) directly by applying the constraints $\lim_{c\rightarrow \infty} \frac{1}{c} \< h_{r,1} | L_i L_{r-i} | \psi\> =0$.}
\be
0 &=& |\psi\> \approx \left( L_{-r} + \frac{6}{c} \sum_{j=1}^{r-1} \frac{1}{j(r-j)} L_{-r+j} L_{-j} \right) | h_{r,1} \>.
\label{eq:heavynullsecondorder}
\ee
As before, the large $c$ differential equation is most effectively extracted from the operators arranged as follows:
\be
0 &=& \< \psi | \CO_{r,1}(0) \CO_L(x_1) \CO_L(x_2) \> \nn\\
 &\approx& \left\< h_{r,1} \left| \CO_{r,1}(0) \left[ L_r + \frac{6}{c} \sum_{j=1}^{r-1} \frac{1}{j(r-j)}L_j L_{r-j}, \CO_L(x_1) \CO_L(x_2)\right] \right. \right\>
\ee
where the four-point function in this configuration is related to the function $\CV(z)$ by (\ref{eq:VinXYCoord}).  It is straightforward though tedious to work out the commutator of any individual factor $ L_j L_{r-j}$ above. However, our main interest is in the behavior near the forbidden singularities, at $z=1- e^{\frac{2\pi i n}{r}}$.  To explore the behavior around this singularity, we take $x_1 = 1, x_2=1-z$ in a scaling limit 
\be
1-z = e^{- \frac{2 \pi i n}{r} -\frac{x}{b}},
\ee 
where $b \rightarrow \infty$, and $b$ is defined conventionally by $c = 1 + 6(b+1/b)^2$.  At fixed $x$ and large $|b|$, this scaling limit therefore zooms in on the singularity and allows us to see explicitly how the divergence is cut off by finite $c$ effects.  The correction terms in (\ref{eq:heavynullsecondorder}) survive in this large $b$ limit, and the new resulting leading order differential equation is 
\be
0 &=&  2h_L \CV(x) + x \CV'(x) -  \CV''(x) \sigma^2_n(r),  
\label{eq:heavynullsecondorderscaling}
\ee
where 
\be
\sigma^2_n(r) &\equiv& 4\sum_{j=1}^{r-1} \frac{\sin^2\left( \frac{ j n \pi}{r} \right)}{r j(r-j)} .
\ee
This differential equation is solved by the function ${}_1F_1(h_L, \frac{1}{2}, \frac{x^2}{\sigma^2_n(r)})$.  It has an integral representation of the form of equation (\ref{eq:UniversalFunctionForSingularityResolution}), to be further  discussed in section \ref{sec:NotToyModel}, and so the function $\sigma_n(r)$ sets the width of the correlator around the saddle point $x=0$.  While this differential equation is derived for $r$ a positive integer, $\sigma^2_n(r)$ can be analytically continued as a function of $r$, and it is tempting to guess that this generalizes (\ref{eq:heavynullsecondorderscaling}) beyond the case of degenerate operators to a universal rule for how the forbidden singularities are resolved in the conformal blocks.  At large $r$, $\sigma_n(r)$ is particularly simple:
\be
\label{eq:DegenerateU}
\sigma^2_n(r) \stackrel{r\gg 1}{\approx} \frac{4}{r^2} \int_0^{2\pi n} \frac{dt}{t} \sin^2\left( \frac{t}{2} \right),
\ee
suggesting that $\sigma^2_n(h_H) \approx -\frac{c}{6h_H} \int_0^{2\pi n} \frac{dt}{t} \sin^2\left( \frac{t}{2} \right)$ in the limit of large $c$ and $h_H/c$.

Now we will provide a piece of evidence that the forbidden singularities in general heavy-light Virasoro vacuum blocks at large $c$ are resolved in the same way.  Let us assume for a moment that the blocks are well approximated by the following solution to (\ref{eq:heavynullsecondorderscaling}):
\be
S(x,c) \approx \int_0^\infty dp \, p^{2h_L - 1} e^{-p x - \frac{\sigma^2}{2 b^2} p^2}
\label{eq:blockintegralappx}
\ee
in the vicinity of their forbidden singularities.   Then the $1/c$ corrections to the leading large $c$ limit near the singularity must take the form
\be
\label{eq:ExpectationNearGeneralSingularity}
\frac{1}{x^{2h_L}} - \frac{6 \sigma^2 h_L (2h_L+1)  }{c} \frac{1}{x^{2 h_L + 2}} + \cdots
\ee
Notice that this makes a precise prediction for the strength of the singularity in $x$, for the relationship between the $h_L^2$ and $h_L$ terms, and for the overall coefficient.
Since we have an explicit expression for the leading and $1/c$ corrected heavy-light blocks \cite{Fitzpatrick:2015dlt, Beccaria}, we can search for the $x^{-2 h_L-2}$ term in the vicinity of forbidden singularities, and extract the coeffiicent $\sigma^2$.  
In appendix \ref{app:ForbiddenSingularitiesinCorrectionGeneralV} we show that the general Virasoro blocks match precisely to the prediction from this analysis and from equation (\ref{eq:DegenerateU}).

\subsection{Large Lorentzian Time Behavior from an Interesting Approximation}
\label{sec:LateLorentzianBehaviorDiffEq}

AdS correlators in a black hole background decay exponentially at late times, signaling loss of information  concerning initial perturbations.  As we discussed in section \ref{sec:BehaviorLargeLorentzianTime}, the heavy-light Virasoro blocks with $h_H > \frac{c}{24}$ (above the BTZ black hole threshold) exhibit the same behavior as $c \to \infty$.  Thus it would be very interesting to be able to compute the exact heavy-light blocks at late Lorentzian times.
We do not have an exact relation for these blocks, but we can make a very interesting approximation that incorporates the non-perturbative physics that resolves the forbidden singularities. 

We showed in section \ref{sec:HeavyDegenerateStates} that the blocks with heavy degenerate operators obey a 1st order differential equation to leading order at large central charge.
Furthermore, a universal 2nd order differential equation seems to resolve all forbidden singularites, as explained in section section \ref{sec:NearAForbiddenSingularity}.  In fact, all of these differential equations can be obtained from limits of a single, 2nd order master equation.   It can be written as
\be
 -h_L g_r(t) \CV(t)  + \CV'(t) + \frac{\Sigma_r(t) + \Sigma_{-r}(t)}{b^2}  \CV''(t) = 0,
\label{eq:MasterEquation}
\ee
 where 
\be
g_r(t) &=& \coth \left( \frac{t}{2} \right) - r \coth \left( \frac{r t}{2} \right)
\\ 
\Sigma_r(t) &=&
-\frac{1}{r \sinh \left( \frac{r t}{2} \right)} \left( e^{- \frac{ r t}{2}} \tilde{B}_r(t) + e^{\frac{r t}{2}} \tilde{B}_r(-t) -2 \cosh \left( \frac{r t}{2} \right) \tilde{B}_r(0) \right).
\ee
We have introduced the function $\tilde{B}_r(t)$ which can be represented in a few different ways that each have different advantages.  First, it arises directly from the sum over the different terms in (\ref{eq:heavynullsecondorder}) as the following sum:
\be
\tilde{B}_r(t) &=& \sum_{j=1}^{r-1} \frac{e^{t j}}{j}.
\ee
This finite sum can be written as the difference of two infinite sums when $|e^t|<1$:
\be
\tilde{B}_r(t) &=& -\log(1-e^{t}) - e^{r t} \sum_{k=0}^\infty \frac{e^{k t}}{k+r} \nn\\
 &=& -\log(1-e^{t}) - \frac{e^{ r t} {}_2F_1(1,r,1+r,e^t)}{r} = -\log(1-e^t) - B(e^t;r,0),
 \label{eq:UpsHyp}
\ee
where $B(z;a,b)$ is the incomplete beta function. This second form is more useful since we are interested in analytically continuing the function to imaginary $r$.  More precisely, the main reason for our interest in equation (\ref{eq:MasterEquation}) is that we can analytically continue $r \to 2 \pi i T_H$ to study physical correlators associated with BTZ black hole physics.  We already saw in section \ref{sec:NearAForbiddenSingularity} and appendix \ref{app:ForbiddenSingularitiesinCorrectionGeneralV} that this procedure appears to produce sensible results.\footnote{  In particular, we invert the relation $\frac{c}{24} (1-r^2) = h_H$ and analytically continue as a function of $h_H$.  Because the inverse $r=\pm \sqrt{1-24 \frac{h_H}{c}}$ passes through a branch cut when $h_H/c$ becomes positive, we have to make a choice about how to treat the two different roots.  In (\ref{eq:MasterEquation}), we have taken both roots and added them together.  Our motivation in doing this is that this prescription passes a highly non-trivial check in appendix \ref{app:ForbiddenSingularitiesinCorrectionGeneralV}, and seems very reasonable given the analytic structure of the blocks themselves.}  Here, we will use it to study the large Lorentzian time behavior of the vacuum Virasoro block.  

\begin{figure}[t!]
\begin{center}
\includegraphics[width=0.48\textwidth]{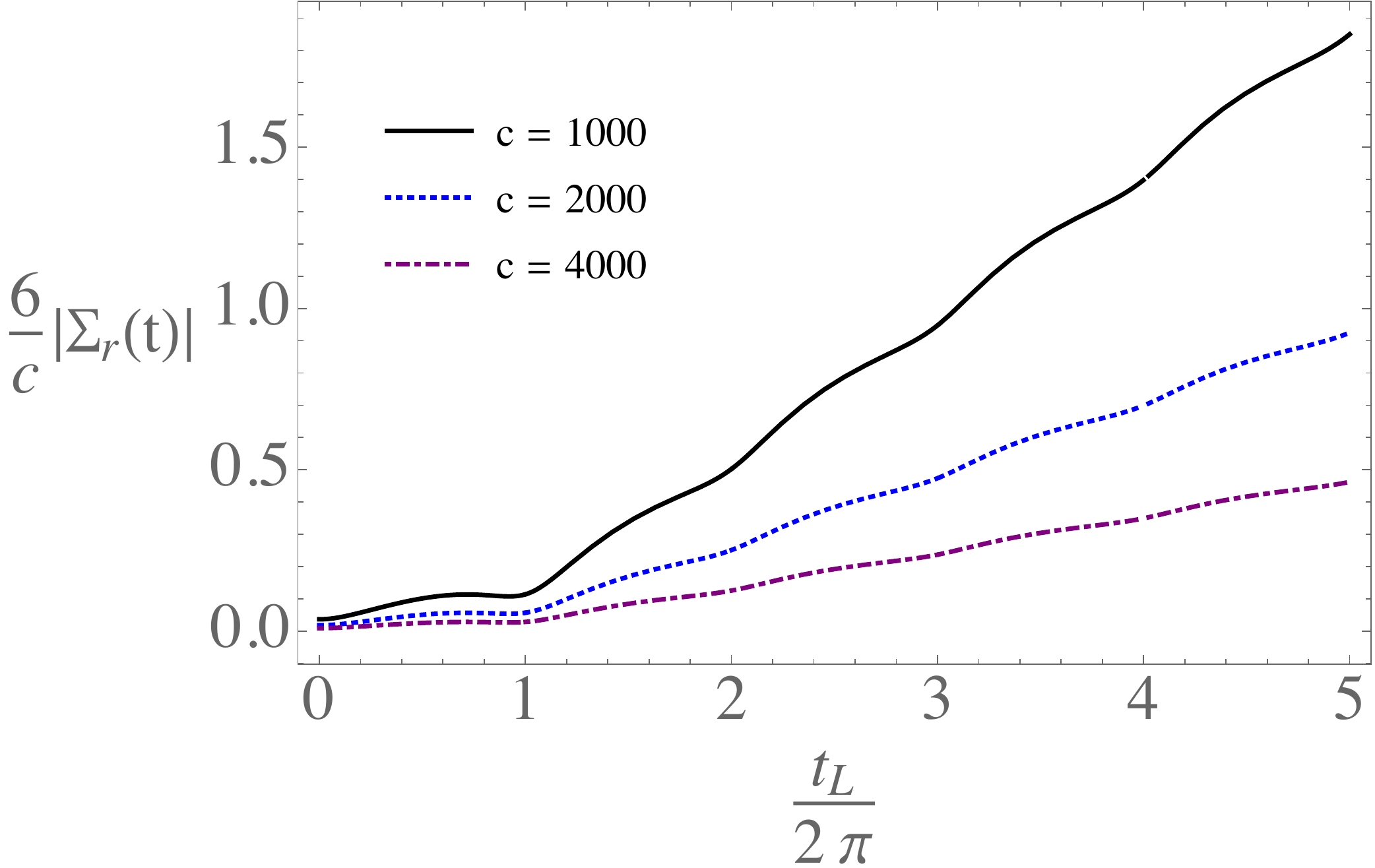}
\includegraphics[width=0.48\textwidth]{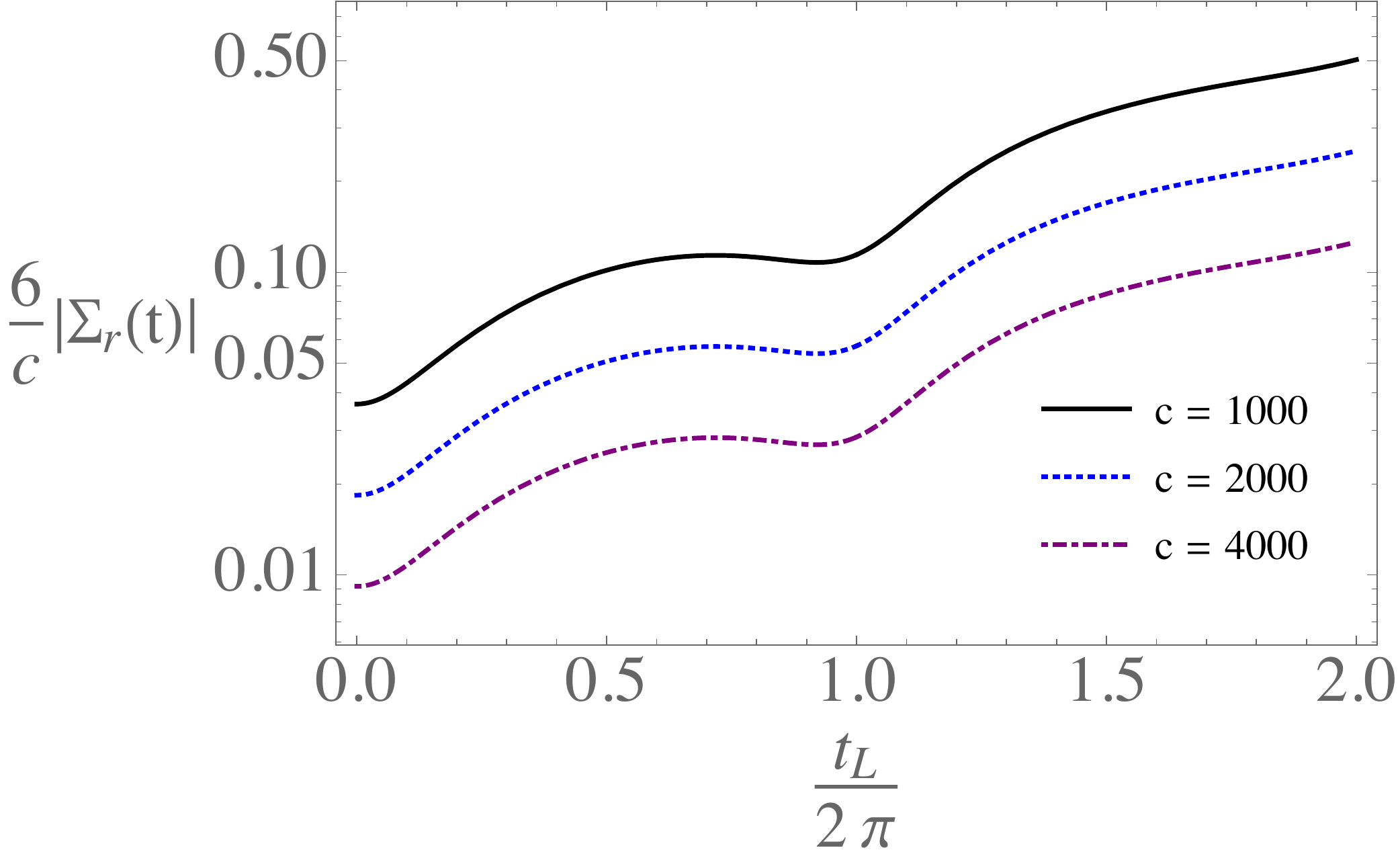}

\vspace{0.4cm}

\includegraphics[width=0.55\textwidth]{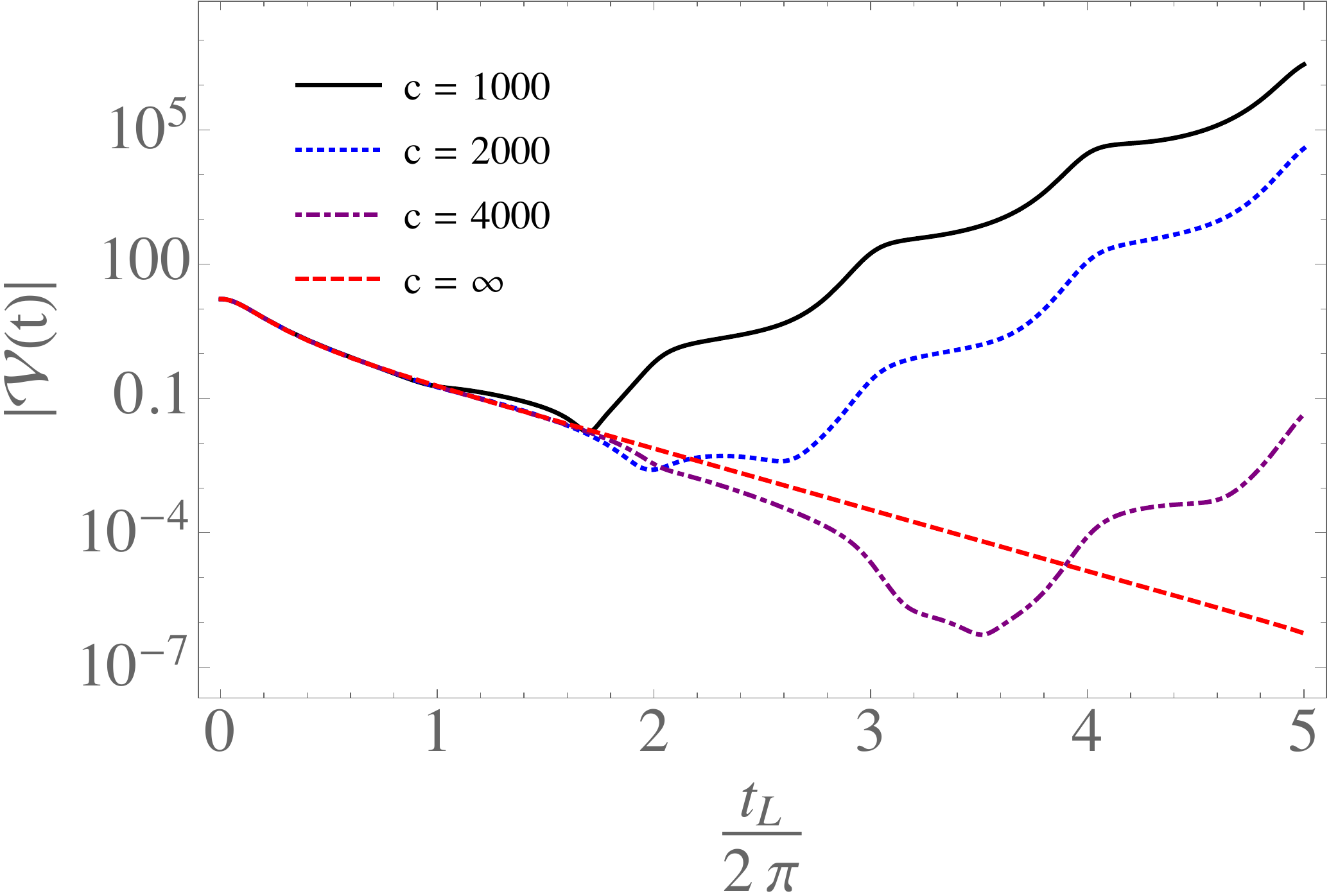}
\caption{In all cases, we take $r = 2 \pi i T_H = \frac{i}{2}$ and $t=t_E + i t_L$ with fixed $t_E=-i$.  {\it Left, top:} The  magnitude of the coefficient term $\frac{6 }{c}\Sigma_r(t)$ for various values of $c$ as a function of Lorentzian time.  At late Lorentzian time, it grows approximately linearly with a slope of order $1/ S_{BH}$.  {\it Right, top}: Same as {\it left, top}, but on a log scale to show the absolute size more clearly.  {\it Bottom}: Plots of the numerical solution to (\ref{eq:MasterEquation}) for various values of $c$. The solutions track the $c=\infty$ solution at early times, until the correction term becomes important at times of order $t_L \sim  S_{BH}$, as can be seen in the plot. The large time behavior should not be taken literally, as our approximations break down for $t_L \gtrsim S_{BH}$. }
\label{fig:lorentz}
\end{center}
\end{figure}

A final form for $\tilde{B}_r(t)$ that is useful for understanding its analytic continuation in Lorentzian time is its derivative
\be
e^{-t} \frac{d}{dt} \tilde{B}_r(t)= \frac{1-e^{t(r-1)}}{1-e^{t}}.
\label{eq:UpsInt}
\ee
As one increases ${\cal I}m(t)$, $e^t$ winds around 1 in the complex plane, picking up a contribution each time from the pole at $t=0$.  This allows the function to `remember' how much Lorentzian time has passed. We discuss this effect in more detail below.

In the large $c \propto b^2$ limit we can drop the entire $\CV''$ term to obtain equation (\ref{eq:heavynullfirstorder}), while equation (\ref{eq:heavynullsecondorderscaling}) can be obtained by scaling equation (\ref{eq:MasterEquation}) towards a forbidden singularity.  This master differential equation can be derived by repeating the analysis from section \ref{sec:NearAForbiddenSingularity} without taking the large $b$ limit with fixed $x= bz$.  For $r=2$ this equation is exact, but for degenerate operators $\phi_{(r,1)}$ with $r > 2$ it neglects effects of order $\frac{1}{c^2}$ through $\frac{1}{c^{r-1}}$.  
We have also neglected effects $\sim \frac{1}{c}$ in the coefficients of the $\CV(t)$ and $\CV'(t)$ because they are sub-dominant to the leading order terms when this $1/c$ expansion is controlled, and because unlike the $\CV''(t)$ term, they are not necessary to regulate the forbidden singularities.

 The master equation appears perturbative, but in fact its solutions incorporate both perturbative and non-perturbative effects in $1/c$, as can be seen by (\ref{eq:blockintegralappx}), with non-perturbative effects becoming important in the vicinity of forbidden singularities. One way to understand this is to perturbatively expand $\CV$ in powers of $1/c$.  The leading term in $\CV$ just solves (\ref{eq:MasterEquation}) without the $1/c$ correction terms, and produces a source term for the subsequent higher orders.  When $\CV$ is $\CO(1)$, then source term it produces is $\CO(1/c)$, and consequently (\ref{eq:MasterEquation}) just produces perturbative $1/c$ effects.\footnote{In fact, solving (\ref{eq:MasterEquation}) (plus the $1/c$ coefficients of $\CV$ and $\CV'$ terms that we have neglected) at next-to-leading order in a formal $1/c$ is one way of deriving the perturbative $1/c$ corrections to the block.  Since the analytic continuation from the non-unitary region to the unitary region appears to be justified, the differential equation (\ref{eq:MasterEquation}) may provide an easier method for deriving $1/c$ corrections to Virasoro conformal blocks in the heavy-light limit than that adopted in \cite{Fitzpatrick:2015dlt}.}  However, when $\CV$ is $\CO(c)$ or larger, as it is in the vicinity of forbidden singularities, the source term is large and (\ref{eq:MasterEquation}) captures some non-perturbative effects as well.  

The correction term can of course also become important when it becomes large through its time-dependence.  In the Lorentzian regime, increasing $t_L = {\cal I}m(t)$ causes  $\tilde{B}_r(t)$ to pick up a shift by an exponential function every $2 \pi$, as can be seen from the integral expression (\ref{eq:UpsInt}) or from its expression (\ref{eq:UpsHyp}) in terms of hypergeometric functions.  This produces linear times exponential growth, which becomes approximately linear growth in $\Sigma_r$ at late Lorentzian times due to the factor of $\sinh(\frac{rt}{2})$ in the denominator.  This is shown explicitly in Figure \ref{fig:lorentz}, where we plot the magnitude of $\Sigma$ for a range of $c$ as a function of $t_L$; in all cases, we choose $r=\frac{i}{2}$ and $t= t_E + i t_L$ with $t_E = -i$.  We also show the behavior of the solutions $\CV$ to (\ref{eq:MasterEquation}) for these choices of parameters. At early times, these track the $c=\infty$ solution, and begin to deviate significantly after $\Sigma(t)$ grows sufficiently large.  Once $\Sigma(t) \sim c$, keeping just $1/c$ terms in the differential equation (\ref{eq:MasterEquation}) is no longer justified.  So we do not have a controlled approximation for the conformal blocks at very late Lorentzian times.  We have checked that the $1/c$ suppressed coefficients of $\CV$ and $\CV'$ grow at the same rate as $\Sigma(t)$, so we cannot neglect their effects at late times either. 

However, we can still determine when these $1/c$ correction terms start to become important.  Parametrically, because of the linear growth in $t_L$ and the factor of $r= 2 \pi i T_H$ in the denominator, we have
\be
\frac{1}{c}\Sigma_r(t) \sim {\cal O}\left( \frac{ t_L}{c T_H}\right) \sim {\cal O}\left( \frac{t_L}{S_{BH}(h_H)} \right),
\ee
where $S_{BH}(h_H) = \frac{\pi^2}{3} c T_H$ is the corresponding black hole entropy.  Thus we see that deviations from the exponential decay at late Lorentzian time appear to arise at times of order $t_L \sim S_{BH}$.  More precisely, we find that the third term in equation (\ref{eq:MasterEquation}) becomes equal in magnitude to the first two terms exactly when the parametric relation $\log  \CV  \approx -S_{BH}$ obtains.   In general, by analytically continuing in $r$, we will obtain  from equation (\ref{eq:DegenerateStateLs}) an infinite series of other corrections at order $1/c^n$ with increasingly complicated time-dependent coefficients.  It would be interesting to understand the Lorentzian time-dependence of further sub-leading terms, and to see if the $1/c$ approximation entirely breaks down at $t_L \sim S_{BH}$.

\section{Information Restoration as a Non-Perturbative Effect}
\label{sec:NonPerturbativeUnitarityRestoration}

In section \ref{sec:AdSCFTInformationLossLargeC} we discussed forbidden singularities as a signature of information loss at large central charge.  Then in section \ref{sec:ExactVirasoroBlocks} we connected the exact correlators of degenerate operators to more general results about the heavy-light Virasoro vacuum block, obtaining some explicit theoretical `data' about the resolution of forbidden singularities at finite $c$.  We also obtained some results on the late Lorentzian time behavior of Virasoro blocks. 

In this section we will try to characterize the resolution of information loss as an explicit non-perturbative effect.  We consider two closely related approaches.  As reviewed in section \ref{sec:BorelResummationReview}, singularities in the Borel resummation of a perturbation series are intimately connected to classical solutions of the field equations.  So our first approach will be to study the Borel resummation of the $1/c$ perturbation series.  With our second approach, we will represent a degenerate conformal block as a contour integral and study its large $c$ asymptotics.  In both cases, the aspiration is to connect the behavior of the Virasoro vacuum block to the saddle points and the contour of integration for the gravitational (or Chern-Simons \cite{Witten:2010cx, Gaiotto:2011nm}) path integral in AdS$_3$ in future work.  In other words, we would eventually like to express the exact CFT$_2$ conformal block as a specific sum over AdS$_3$ geometries, including both the perturbative vacuum and other solutions corresponding to the exchange of heavy states.  Understanding the large $c$ saddle points of the Virasoro blocks themselves is a natural first step.

\subsection{Borel Resummation and a Vacuum Block}
\label{sec:BorelofVacuumBlock}

We will study the Borel resummation of the $1/c$ expansion of Virasoro vacuum blocks, focusing on  heavy degenerate external states.  First we will study a simple model that seems to describe the universal behavior of the blocks in the vicinity of a forbidden singularity, and then we will study the full $h_{2,1}$ block.

\subsubsection{Not Just a Toy Model}
\label{sec:NotToyModel}

Forbidden singularities take the form of power-laws $x^{-2h}$.  We would like a model that regulates this singularity in such a way that at $c=\infty$ we recover the power-law form, but at finite $c$ we are left with an entire function of $x$.  It would not be sufficient to simply move the singularity from $x=0$ to some other point(s) in the complex plane; we must eliminate the singularity completely at finite $c$.
We argued in section \ref{sec:NearAForbiddenSingularity} that in fact the forbidden singularities are resolved by a simple universal function given in equation (\ref{eq:UniversalFunctionForSingularityResolution}), which is actually one of simplest models one might imagine with the desired properties.   Let us consider the Borel resummation of this function in $1/c$ perturbation theory.

We can write $S$ as the formal series in $1/c$ by expanding the integrand
\be
S(x, c) &=& \int_0^\infty dp \, p^{2h-1} e^{-x p} \sum_{n=0}^\infty \frac{1}{n!} \left( -\frac{1}{c} \right)^n p^{2n}
\nn \\
&=& \sum_{n=0}^\infty \frac{ \Gamma(2h+2n)}{n!} \left( -\frac{1}{c} \right)^n  x^{-2h-2n}
\ee
but this series expansion does not converge for any value of $x$ or $c$, due to the factorial growth of the gamma function.  Moreover, the higher order terms become ever more singular near $x = 0$.  As we discuss in section \ref{sec:NearAForbiddenSingularity} and appendix \ref{app:ForbiddenSingularitiesinCorrectionGeneralV}, the behavior of the $1/c$ term can be used to verify our conjecture that the forbidden singularities have a universal resolution.

To resum the $1/c$ perturbation series, we define the Borel function
\be
B(x,y) &=& \sum_{n=0}^\infty \frac{\Gamma(2h+2n)}{n!^2} \left( -y \right)^n   x^{-2h-2n}
\nn \\
&=& \frac{\Gamma(2h)}{ x^{2h}} \, {}_2 F_1  \left( h, h+\frac{1}{2}, 1, -\frac{4 y}{x^2} \right)
\ee
We expect that the original function can be recovered from
\be
S(x) = \int_0^\infty dy \, e^{-y} B \left(  x, \frac{y}{c} \right)
\ee
if the integral is well-defined, ie if there are no singularities on the real $y$-axis.  But this is not always the case, because the hypergeometric ${}_2 F_1$ will have a branch cut in its last argument extending from $1$ to infinity.\footnote{The function ${}_2 F_1  \left( h, h+\frac{1}{2}, 1, 1 - z \right) \sim A + B z^{\frac{1}{2} - 2h}$ near $z=0$, so for special values of $h$ such as the $h_L=1$ example that we will consider in the next section, the branch cut can simplify somewhat.} This means that there is a branch cut in the Borel integrand beginning at
\be
y = -\frac{x^2 c}{4} 
\ee
and extending to infinity.  This intersects the real axis when e.g. $c > 0$ is real and $x$ is imaginary.    In applications to the Virasoro blocks we would take $x = z- z_*$, with $z_*$ the position of a forbidden singularity, so imaginary values of $x$ would be physically relevant.   
Most importantly, when $x \approx 0$ with fixed $c$ we are in the vicinity of the forbidden singularity, and in this region the Borel resummation becomes completely ambiguous, signaling the importance of non-perturbative effects near the forbidden singularities.

\subsubsection{The Vacuum Block with a Heavy Degenerate State}
\label{sec:Borelof21}

We would like to study the vacuum block involving a pair of heavy degenerate states with dimension $h_{2,1}$ and a light state with dimension $h_L$.  This example was discussed in section \ref{sec:QuickHeavyDegenerateExample}, and its vacuum block was plotted in the vicinity of its forbidden singularity in figure \ref{fig:NearForbiddenSingularity21}.  It is easy to write this block in closed form; for example for $h_L = 1$ it takes the particularly simple form 
\be
\tilde \CV_{2,1}(z) = z^{2h_{2,1}} \CV_{2,1} = (1-z) {}_2 F_1(2, b^2+1, 2b^2+2, z),
\ee
where we wrote $\tilde \CV$ because we factored out an overall $\frac{1}{z^2}$ for simplicity later on.
Our goal in this section will be to study its behavior in a $1/c \propto 1/b^2$ perturbation expansion, and then to Borel resum the resulting asymptotic series.  

The idea is to write this degenerate vacuum block as a series in $\frac{1}{c} \approx \frac{6}{b^2}$ with functions of the kinematic variable $t = -\log(1-z)$ as coefficients.  The 2nd order differential equation that the block obeys provides a recursion relation for these functions.  It turns out that for the particular value $h_L = 1$, the recursion relation takes an especially simple and useful form.  We define a new variable\footnote{The reason for choosing this $s$ variable is the hypergeometric function identity: $\tilde{V}_{2,1}(z)|_{h_{L}=1}={}_{2}F_{1}(1,b^{2},b^{2}+\frac{3}{2};\frac{z^{2}}{4(z-1)})$, where $e^{s}=-\frac{z^{2}}{4(z-1)}$.}
\be
\label{eq:DefinitionsVar}
s \equiv 2 \log \left( \sinh \left( \frac{t}{2} \right) \right),
\ee
noting that the forbidden singularity at $z=2$ corresponds to $s=\pi i$.
Now if we write
\be
\tilde \CV_{2,1}(s) =  
\sqrt{1 + e^{-s} } 
 \sum_{k} \frac{1}{\left(b^2 \right)^{k}} G_k(s) ,
\ee
we find that the coefficient functions are remarkably simple\footnote{We used the identity $\partial_{w}\left[\left(1-w\right)^{a+b-c}w^{c-a}F\left(a,b,c,w\right)\right]=\left(c-a\right)w^{c-a-1}\left(1-w\right)^{a+b-c-1}F\left(a-1,b,c,w\right)$ in deriving this result.}
\be
G_{k}(s) = \left( -\partial_s+\frac{1}{2} \right) (-\partial_s)^{k-1} G_0(s), 
\ee
where the leading coefficient is:  
\be
G_0(s) = \frac{e^{\frac{s}{2}}}{\left( 1+e^s \right)^{\frac{3}{2}}} .
\ee
The simple form of $G_k$ implies that the Borel function can be obtained from translations $s \to s + y$ of $G_0(s)$ and $\int G_0(s)$.  Explicitly 
\be
B(s,y) = \sum_{k=0}^\infty \frac{y^k}{k!} \left( -\partial_s+\frac{1}{2} \right) (-\partial_s)^{k-1} G_0(s) =-\frac{e^{\frac{3}{2}(s-y)}}{\left(1+e^{s-y}\right)^{\frac{3}{2}}}+\frac{1}{(1+e^{-s})^{\frac{1}{2}}} .
\ee
Therefore, $\tilde \CV_{2,1}$ can be written as the Borel transform 
\be
\label{eq:Borelofh21Block}
\tilde \CV_{2,1}(b,s) 
&=& 1-\sqrt{1 + e^{-s} }  
\int_0^\infty dy \frac{e^{-y-\frac{3y}{2b^{2}}+\frac{3s}{2}}}{\left(1+e^{s-\frac{y}{b^{2}}}\right)^{\frac{3}{2}}}.
\ee
One can directly verify that this Borel integral reproduces $\tilde \CV_{2,1}$.

We are interested in the behavior of the integrand of equation (\ref{eq:Borelofh21Block}), and especially in its singularities as a function of $y$ for various values of $b$ and $s$ (which depends on our usual kinematic variable $t$ through equation (\ref{eq:DefinitionsVar})).   The simple denominator has singularities at
\be
y_n =   b^2 \left( s -  \pi i  (1+2n)   \right)
\ee
for integers $n$.  This is interesting because when $s \approx \pi i$, the Borel integrand has a singularity at $y = 0$, the very beginning of the integration contour.  This signals the complete breakdown of $1/c$ perturbation theory about the vacuum, which is exactly what we expect in the vicinity of a forbidden singularity.  Note that if we expand $s(t)$ about the forbidden singularity at $t = \pi i$, we find
\be
s(t) \approx i \pi  + \frac{1}{4} (t - i \pi)^2 + \cdots
\ee
We see that to keep the singularities $y_n$ in the Borel plane fixed as we take the semi-classical limit $b \to \infty$, we must keep the quantity $b (t - i \pi)$ constant.  This is the scaling we discovered in section \ref{sec:NearAForbiddenSingularity} and it is also appropriate for the Gaussian example from the previous section, recalling that $c \propto b^2$.

We explained in section \ref{sec:BorelResummationReview} that singularities of the Borel integrand should correspond with classical solutions of the relevant field equations, which in this case would be Einstein's equations in AdS$_3$.  We expect that these classical solutions or `solitons' correspond to heavy, non-perturbative states in the theory.  So we would like to  determine which physical state(s) are associated with the non-perturbative effect that we have discovered.  Since  $\CV_{2,1}$ represents a correlator of degenerate CFT operators, the physical states that are exchanged follow from the fusion rule
\be
\CO_{(2,1)} \times \CO_{(2,1)} = {\bf 1} + \CO_{(3,1)} .
\ee
So the non-perturbative effect must come from the exchange of a `heavy' $\CO_{3,1}$ state.

We can argue for this conclusion more explicitly by noting that  $\CV_{(2,1)}$ obeys the second order differential equation (\ref{eq:HeavyDegenerateEqn2}).  So the contribution of a contour wrapped around  the branch cut in equation (\ref{eq:Borelofh21Block}) must also obey this differential equation.  The two solutions to that equation correspond to the vacuum or `${\bf 1}$' Virasoro block and to the $\CO_{(3,1)}$ Virasoro block.  Thus we see that when we expand the exact vacuum block in  $1/c$ perturbation theory, there is a non-perturbative contribution in the Borel plane from the `solitonic' $\CO_{(3,1)}$ state.  

We have not found an explicit expression for the Borel resummation of the $1/c$ perturbation expansion of more general degenerate vacuum blocks.  However, based on the universality of the forbidden singularities, we expect that the general features from the $(2,1)$ example will continue to hold.  In particular, we expect that only states allowed in the OPE of $\CO_{H} \times \CO_{H}$ will appear as branch cuts in the Borel plane of the resummed vacuum block.  For the examples of interest with identical heavy degenerate operators, the fusion rules are \cite{DiFrancesco:1997nk}
\be
\label{eq:fusionrules}
\CO_{(r,s)} \times \CO_{(r,s)} = \sum_{k=1}^{r} \sum_{k'=1}^{s}  \CO_{(2k-1,2k'-1)}  
\ee
with $r \geq 2$.  We do not expect every possible state in the OPE to contribute as large $c$ non-perturbative contributions to the vacuum block. For example, we do not expect the light states $\CO_{(1,2k'-1)}$ appearing in the OPE to be related to the resolution of forbidden singularities.  We will study the example of $\CO_{(3,1)} \times \CO_{(3,1)}$ and $\CO_{(2,2)} \times \CO_{(2,2)}$ in appendix \ref{app:Asymptotics31and22}.

Finally, notice that in both this section and the last, we found a branch cut in the Borel plane, not a set of isolated poles.  We suspect this is because we are seeing the combined contribution of a given state (e.g. $\CO_{(3,1)}$) plus all of its Virasoro descendants.  In the physically relevant case of a generic heavy-light Virasoro vacuum block, we would expect to find an infinite number of branch cuts, one for each forbidden singularity.  It will be interesting to understand whether these form a continuum of heavy states in AdS$_3$, and whether such a continuum begins at the BTZ black hole threshold.

\subsection{Asymptotic Analysis of a Degenerate Block}
\label{sec:SaddlesStokesDegenerateBlocks}

The degenerate Virasoro vacuum blocks can be written as contour integrals, known as the Coulomb gas representation \cite{Dotsenko:1984nm, Dotsenko:1984ad, DiFrancesco:1997nk}.  This means that we can study these Virasoro blocks at large central charge using the methods of asymptotic analysis.  In particular, we can re-write the Coulomb gas integrals in the form
\be
\CV = \int_{\CC} d w \, e^{\CI(b, z; w)}
\ee
for some contour $\CC$, and study the saddle points\footnote{For a relevant review see section 3 of \cite{Witten:2010cx}.} of the `action' $\CI$ at large but finite $c \propto b^2$.  
As compared to the Borel resummation approach of the previous section, these methods  are not as intimately connected to $1/c$ perturbation theory, but they might have a more direct relationship with the semi-classical gravitational path integral in AdS$_3$.  For example, we might hope to uncover a relationship between the saddle points of the Coulomb gas integrals and classical solutions of AdS$_3$ gravity, Chern-Simons theory, or Liouville theory.  

As we will discuss, the physical states exchanged in a conformal block do not correspond with a single saddle point of $\CI$.  In the cases we examine, a single saddle point can be associated with the vacuum block, but non-vacuum blocks arise as linear combinations of saddle points.   At this stage it is unclear whether we should focus on the saddle points or CFT primary states, so we will comment on both.  In section \ref{sec:ContourIntegral21} below we review the methodology and discuss the simplest example; we relegate more complicated examples to appendix  \ref{app:Asymptotics31and22}, providing only a brief summary in section \ref{sec:Summary31and22}.

\subsubsection{Virasoro Blocks with External $\CO_{(2,1)}$}
\label{sec:ContourIntegral21}

In this  subsection we revisit the simplest heavy degenerate operator $\CO_{(2,1)}$.   We will take the simplifying limit  $h_L = 1$, in which case the degenerate four point function can be written as 
\begin{align}
\label{eq:HypergeometricAsIntegral}
 \<  \CO_L(\infty) \CO_L(1) \CO_{(2,1)}(z)  \CO_{(2,1)}(0)   \>
& \ \sim \ 
z^{-2h_{(2,1)}}(1-z)\! \int_{\cal C} \!d w \exp{\cal I},
\end{align}
where we refer to the exponent
\be\label{eq:Action21}
{\cal I}_{(2,1)}\equiv 
b^2 \log \big[w(1-w)\big]-2\log(1-w z)
\ee
as the ``action'', and we have written the central charge as $c = 1 + 6 \left(  b + \frac{1}{b} \right)^2$. 
  We use ``$\sim$'' instead of an equality because we will not keep track of the normalization constant, and because different contours of integration can produce different conformal blocks or correlators.  The integrand $\exp \, \CI_{(2,1)}(w)$ has a singularity at $w= \frac{1}{z}$.  This singularity will play a crucial role in this section, since integration contours must be deformed to avoid it as we analytically continue the kinematic variable $z$.
  
If our goal is to pick out specific Virasoro blocks, then we can choose the integration contour $\cal C$  to be either $[0,1]$ or $[\frac{1}{z}, +\infty)$.\footnote{In general, by a contour on $[x,y]$ we are really referring to a Pochhammer contour \cite{DiFrancesco:1997nk} associated with the points  $x$ and $y$.}  To connect a given contour to a specific conformal block (or linear combination of blocks) we can study the OPE limit $z \to 0$. At small $z$, the integral on $[0,1]$ becomes
\be
z^{-2h_{(2,1)}}(1-z)
\int_0^1 d w\, e^{{\cal I}_{(2,1)}}
&\propto& 
z^{-2h_{(2,1)}},
\ee
which means that this contour of integration produces the Virasoro vacuum block.  The integral over $[\frac{1}{z},\infty]$ at small $z$ is
\begin{align}
z^{-2h_{(2,1)}}(1-z)
\int_{\frac{1}{z}}^\infty d w\, e^{{\cal I}_{(2,1)}}
\ \propto \
z^{-2h_{(2,1)}}\times z^{-2b^2-1}
\ = \ 
 z^{-2h_{(2,1)} + 2h_{(3,1)}}.
\end{align}
This is the Virasoro block corresponding to the exchange of the primary state created by $\CO_{(3,1)}$  and its Virasoro descendants.

As will be familiar from the study of path integrals, critical points (or saddle points) occur when the action is stationary with respect to the integration variable $w$. Each critical point $w=p_i$ is associated with steepest descent contours (of the real part of the action $\CI$) passing through it. Following \cite{Witten:2010cx}, we refer to the union of steepest descent contours  passing through a critical point $p_i$ as ${\cal J}_i$, which are known as `Lefschetz thimbles'.  In general, the $\CJ_i$ are not in one-to-one correspondence with CFT states.

\begin{figure}[t!]
\begin{center}
\includegraphics[width=0.45\textwidth]{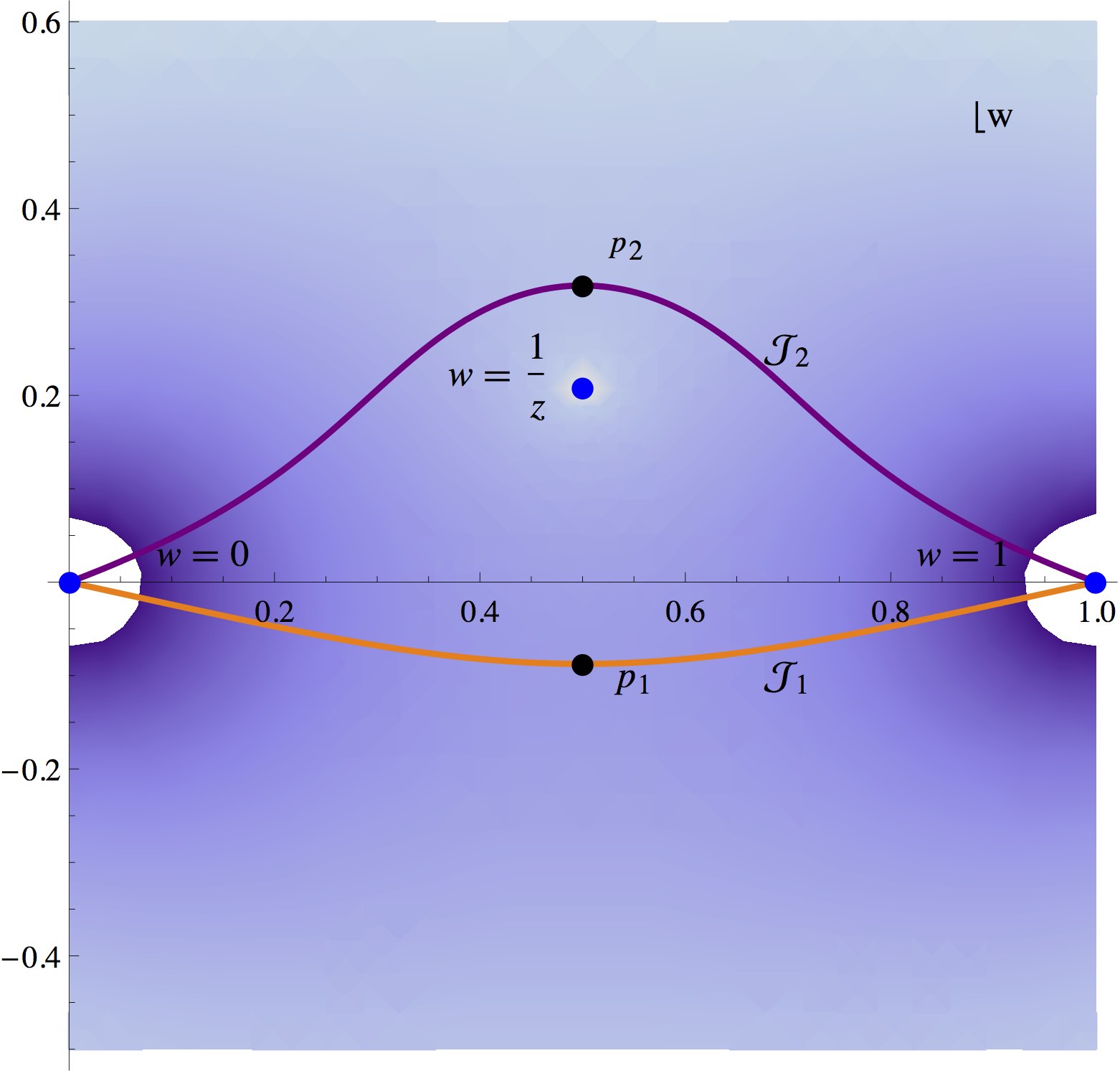}
\includegraphics[width=0.45\textwidth]{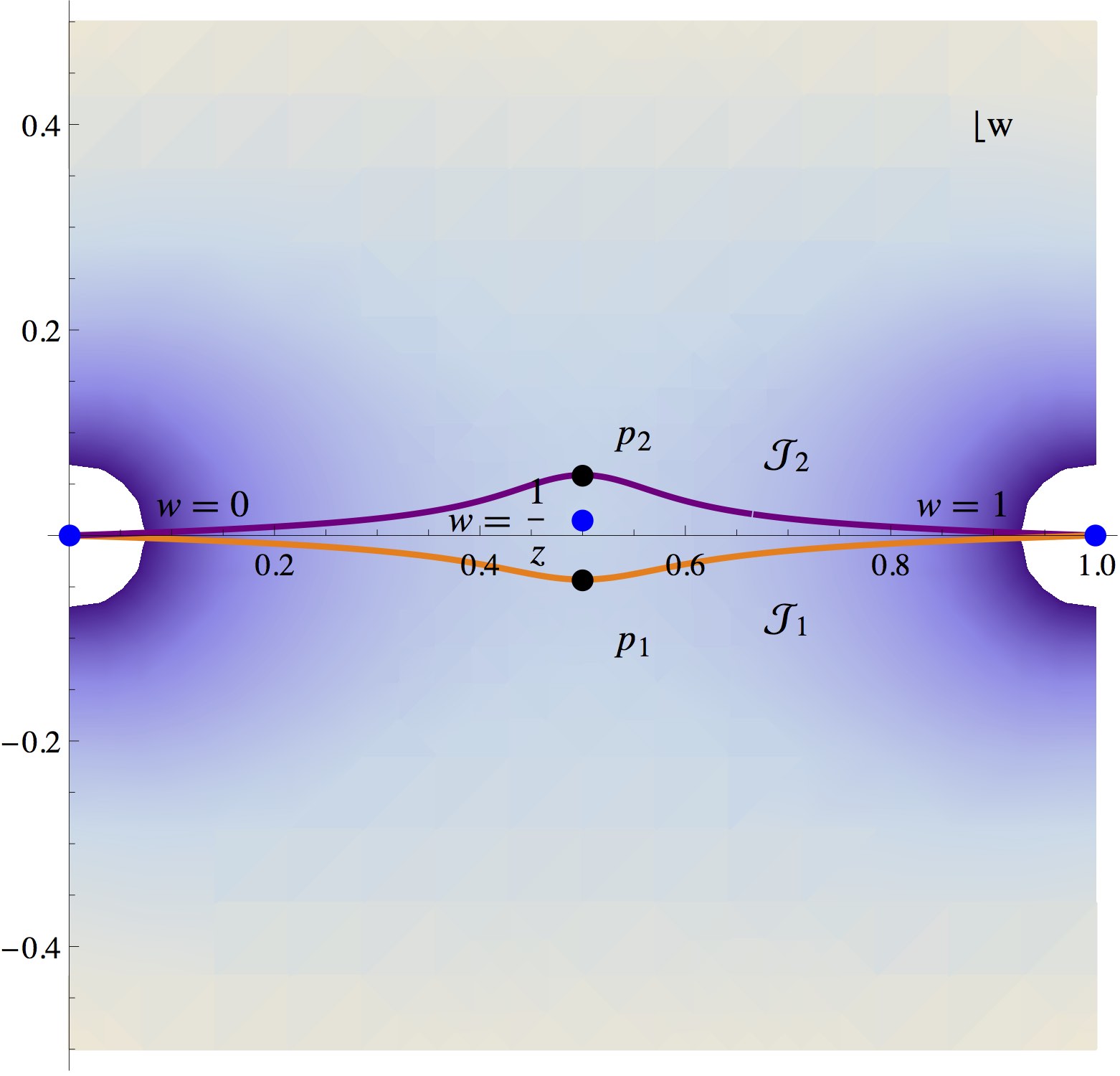}
\caption{These figures show the positions of the critical points (black dots) of the action ${\cal I}_{(2,1)}$ and the associated steepest descent contours.  Darker regions indicate smaller values of Re $\CI_{(2,1)}$, and the steepest descent contours can only end at $w=0$ and/or $w=1$, where $\CI_{(2,1)} \to -\infty$.  The integrand has a singularity at the branch point $w= \frac{1}{z}$, and so the contours of integration cannot cross this point without changing the value of the integral; this is why $\CJ_1 \neq \CJ_2$.  In the left figure, we chose $b^2=10$ and $z=1-e^{\frac{3 i \pi}{4}}$, away from all singularities. In the right figure, we chose $b^2=100$ and $z=1-e^{0.98 i \pi}$ to show how the contours  approach each other in the vicinity of the forbidden singularity at $z=2$.  Note that as $z$ revolves around $1$, the branch point singularity at $1/z = \frac{1}{2} + \frac{1}{2} i \cot(\theta/2)$ parallels the imaginary axis, forcing contour deformations.  This leads to a monodromy for the vacuum block.
}
\label{fig:phi21contour}
\end{center}
\end{figure}

The steepest descent contours are curves in the complex $w$ plane parameterized by a real number $t$, so that $w(t)$ satisfies a flow equation
\be\label{eq:flow equation}
\frac{d w}{d t}=-\frac{\partial {\bar{\cal I}}}{\partial {\bar w}}\;,\quad
\frac{d {\bar w}}{d t}
=-\frac{\partial \cal I}{\partial  w}\;,
\ee
Famously, Im $\cal I$ is constant along a steepest descent contour,
\be
\frac{1}{2i}\frac{d ({\cal I}-{\bar {\cal I}})}{d t}=\frac{1}{2 i}
\left[
\frac{\partial {\cal I}}{\partial w}\frac{d w}{d t}
-\frac{\partial {\bar {\cal I}}}{\partial {\bar w}}\frac{d {\bar w}}{d t}
\right]=0,
\ee
so steepest descent contours can be determined from the algebraic equation  Im ${\cal I}_{(2,1)} =$ constant.  In the present case, the action ${\cal I}_{(2,1)}(w)$ vanishes at $w=0$ and $w=1$, blows up as $w \to e^{i \phi} \infty$  for any $\phi$, and also blows up at $w = \frac{1}{z}$, so all steepest descent contours end at $w=0$ or $w=1$.


The action ${\cal I}_{(2,1)}$ of equation \eqref{eq:Action21} has two saddle points at $w = p_1$ and $p_2$, which depend on $b$ and $z$. We will focus on the regime of real $b$ with $b^2 \gg 1$ to simplify the analysis.   For the purpose of studying the forbidden singularity, it is convenient to write $z=1-e^{i\theta}$, since the singularities of degenerate blocks all lie on the unit circle around $1$.  Notice that in this parameterization
$\frac{1}{z} = \frac{1}{2} + \frac{1}{2} i \cot \left( \frac{\theta}{2} \right)$, so the pole of the integrand in equation (\ref{eq:HypergeometricAsIntegral}) cuts across the $[0,1]$ $w$-contour each time $\theta \to \theta + 2 \pi$.  This produces a monodromy in the Virasoro vacuum block.
We write the two critical points  as $ p_{1}=\frac{1}{2}+i q_-$ and $p_2 = \frac{1}{2} + i q_+$ with
\begin{align}\label{eq:qpm}
q_\mp=\frac{2 b^2 \cos
   \left(\frac{\theta }{2}\right)
\mp
   \sqrt{2\left(b^2-2\right)^2 \cos\theta+2b^4+8b^2-8}}{8 \left(b^2-1\right)\sin\left(\frac{\theta }{2}\right)}\;.
\end{align}
The critical points and steepest descent contours are pictured in figure \ref{fig:phi21contour}.
Notice that $q_+(\pi) = q_-(-\pi)$, so the two critical points coincide at the forbidden singularity as $b \to \infty$, which lies on a Stokes line.  We discuss the Stokes phenomena further and provide a more general parameterization of the critical points in  appendix \ref{app:OtherAnalysis21}.  

The steepest descent contours, critical points, and branch points are displayed for two choices of $z$  in figure \ref{fig:phi21contour}.  Note that the $\CJ_1$ contour can be continuously deformed into a line segment $[0,1]$, so evaluating the  integral along $\CJ_1$ produces the Virasoro vacuum block.  In the more involved examples in appendix \ref{app:Asymptotics31and22} we also use $\CJ_1$ to denote a contour that produces the vacuum block.  The critical points and the branch point are also pictured in figure \ref{fig:CriticalPointForbiddenSingularity21}.

\begin{figure}[t!]
\begin{center}
\includegraphics[width=0.48\textwidth]{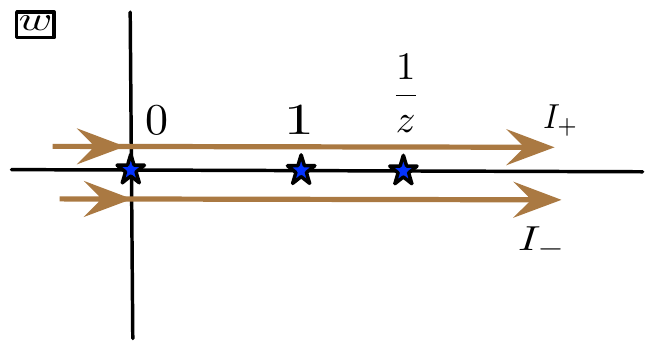}
\caption{This figure illustrates the integration contours $I_\pm$ from equation (\ref{eq:IPMContours}).  By studying the equation $I_+ - I_- = 0$ we can relate the difference between the $\CJ_i$ contours to integration on $[\frac{1}{z}, \infty)$.
}
\label{fig:IPlusIMinus}
\end{center}
\end{figure}

The interpretation of the $\CJ_2$ contour is not as simple -- it corresponds to a linear combination of the vacuum block and the $\CO_{(3,1)}$ block (see figure 9.3 of \cite{DiFrancesco:1997nk}).  There is a general connection between the steepest descent contours we are finding and other contours relevant for conformal blocks, such as $[\frac{1}{z}, \infty)$.  With real $z \in (0,1)$, consider the integrals
\be
\label{eq:IPMContours}
I_{\pm} &=& \int_{- \infty \pm i \epsilon }^{ \infty \pm i \epsilon } 
dw \, (-w)^A (1- w)^B \left( \frac{1}{z}  - w\right)^C 
 \\
&\propto& I(-\infty,0) + e^{\pm i \pi A} I(0,1) + e^{\pm i \pi(A+B)} I \left( 1, \frac{1}{z} \right) + e^{\pm i \pi(A+B+C)} I \left( \frac{1}{z}, \infty \right)
\nn
\ee
which are pictured in figure \ref{fig:IPlusIMinus}.  The contour for $I_{\pm}$ is either directly above or below the real axis, and $I(x,y)$ involve piecewise integrals  along the real axis with the same integrand, which gain phase factors when the monomials in the integrand change sign. With $A+B+C < -1$ we can close both contours $I_{\pm}$ at infinity to find that $I_+ - I_- = 0$, so we can relate the integral on $[\frac{1}{z}, \infty)$ to a contour integral enclosing the points $0,1,$ and $\frac{1}{z}$.    This means that up to an overall factor, the integral on $[\frac{1}{z}, \infty)$ is equal to an integral on $\CJ_2 - \CJ_1$.  For more general $A,B,C$ we can use this method to relate $\CJ_2 - \CJ_1$ to a Pochhammer contour \cite{DiFrancesco:1997nk}.  This procedure is also useful for interpreting the higher dimensional integrals discussed in  appendix \ref{app:Asymptotics31and22}.

As we take the large $c \propto b^2$ limit and approach the forbidden singularity at $z=2$ (ie $\theta=\pi$), the branch point $w=\frac{1}{z}$ moves towards the real $w$ axis, as do the steepest descent contours $\CJ_2$ and $\CJ_1$; see figure \ref{fig:CriticalPointForbiddenSingularity21} and the right panel of figure \ref{fig:phi21contour} for illustrations. The vacuum block contour integral \eqref{eq:HypergeometricAsIntegral} diverges in the limit 
\be\label{FS limit}
b\to +\infty, \quad \theta\to \pi\;,
\ee
since the integration contour $\CJ_1$ is pulled closer and closer to the branch point $w=\frac{1}{z}$. From the contour integral perspective, this is the origin of the forbidden singularity.  We find a similar pattern  for the more general forbidden singularities discussed in appendix \ref{app:Asymptotics31and22}.

\begin{figure}[t!]
\begin{center}
\includegraphics[width=0.56\textwidth]{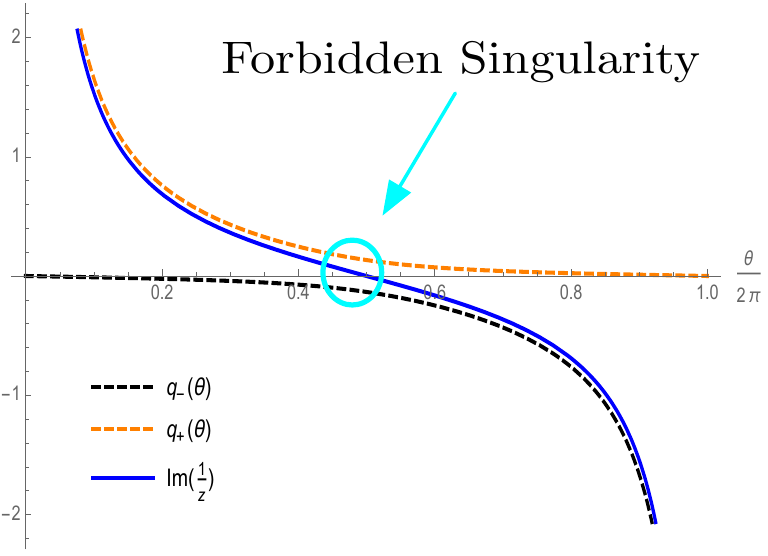}
\caption{This figure illustrates the positions of the critical points of $\CI_{(2,1)}$  as $z = 1 - e^{i\theta}$ varies over $\theta \in [0, 2\pi]$, with $c \approx 90$.  The critical points are located at $w = \frac{1}{2} + i q_\pm$.  The solid line indicates the  branch point $\frac{1}{z}$, where $\exp \CI_{(2,1)}$ has a pole, while the various dashed lines are critical points.  In the vicinity of the forbidden singularity, both critical points approach the branch point in the $w$ coordinate, intersecting it as $c \propto b^2 \to \infty$.  A single critical point approaches the branch point in the OPE limit $z \to 0$.
}
\label{fig:CriticalPointForbiddenSingularity21}
\end{center}
\end{figure}

The action of equation (\ref{eq:Action21}) has a singularity at $w = \frac{1}{z}$, so integration contours must be deformed to avoid this point.  The difference between a contour above and below $w = \frac{1}{z}$ is simply the combined contour $\CJ_1 - \CJ_2$,  which is a contour that encloses all the branch points $0, 1$ and $\frac{1}{z}$.  As we explained above, the  contour $\CJ_1 - \CJ_2$ is equivalent to the $\left[ \frac{1}{z},\infty \right)$ contour \cite{DiFrancesco:1997nk}.  When we cross the forbidden singularity we also cross a Stokes line (discussed further in appendix \ref{app:OtherAnalysis21}), and the vacuum block shifts by the Virasoro block associated with $\CO_{(3,1)}$:
\be
\int_{\CJ_1 - \CJ_2}  \!d w \exp{\cal I}_{(2,1)}  \ \propto \ \frac{(z-2) (z-1)^{b^2-1} }{z^{2 b^2 + 1}}
\ee
In other words, the contour integral along $\CJ_1 -\CJ_2$ can be viewed as a non-perturbative `one instanton' correction to the asymptotic series for the vacuum block.  This is the contour integral manifestation of what we found using Borel resummation in section \ref{sec:Borelof21}.

\subsubsection{More General Examples}
\label{sec:Summary31and22}

We examine more complicated examples with external heavy degenerate operators $\CO_{(3,1)}$ and $\CO_{(2,2)}$ in appendix \ref{app:Asymptotics31and22}.  The motivation is to gather data on the relationship between Virasoro block, critical points in the Coulomb gas integrals, and heavy states, with the hope of connecting these phenomena to the gravitational path integral in AdS$_3$ \cite{Witten:2010cx, Gaiotto:2011nm} in future work.  We find that in all cases, forbidden singularities occur when two or more critical points approach a branch point  (a pole of the integrand $\exp {\CI}$).   The first case has a fusion rule
\be
\CO_{(3,1)}\times \CO_{(3,1)}={\bf 1} \oplus \CO_{(3,1)}\oplus \CO_{(5,1)}
\ee
so we can connect the behavior of the critical points to two different heavy states.  We find that when the forbidden singularity is approached by analytic continuation from the OPE region $z \sim 0$, there is sense in which  the $\CO_{(3,1)}$ state and the vacuum make the dominant contributions.  However, we also obtain contributions from the $\CO_{(5,1)}$ state as $z$ fully encircles $1$.  The other case has a fusion rule
\be
\CO_{(2,2)}\times \CO_{(2,2)}={\bf 1} \oplus \CO_{(3,1)}\oplus\CO_{(1,3)}\oplus \CO_{(3,3)}\;,
\ee 
which is of interest because in the large $c$ limit it contains both the heavy states $\CO_{(3,1)}$ and $\CO_{(3,3)}$, whose dimensions grow with $c$, and the light state $\CO_{(1,3)}$, which has a fixed dimension as $c \to \infty$.  However, in our preliminary work we have not found a clean separation between effects from the light and heavy states, as all critical points of the $\CI_{(2,2)}$ action appear to coalesce in the neighborhood of the forbidden singulairty.  In the future it would be interesting to explore the distinctions between the $\CO_{(2,1)}$ and $\CO_{(2,2)}$ cases, since these have nearly identical dimensions at large $c$.  Ultimately, we would like to understand which critical points and heavy states are associated with the resolution of forbidden singularities and information loss in the  physical case of generic $\CO_H$ and $\CO_L$.

\section{Discussion}

Information loss in AdS black hole backgrounds infects conformal field theory correlation functions with sharply defined pathologies.  We have discussed two:  forbidden singularities in Euclidean CFT correlators \cite{Fitzpatrick:2015dlt}, and late time exponential decay in the Lorentzian regime \cite{Maldacena:2001kr}.  Both pathologies appear at the level of the Virasoro conformal blocks in the large $c \propto \frac{R_{AdS}}{G_N}$ limit.  Since these are the universal building blocks of all CFT$_2$ correlators,  information loss in AdS$_3$/CFT$_2$ has a very robust character.  We can learn a great deal about black hole physics by studying the Virasoro blocks themselves, without  focusing on a specific CFT$_2$.  In this work we obtained exact information using degenerate external operators, resolving the Virasoro blocks' forbidden singularities via non-perturbative effects in $1/c$.  The same physics qualitatively alters the blocks' behavior at Lorentzian times $t \sim S_{BH}$; beyond that timescale, available approximations seem to  break down.

We expect that the Virasoro blocks can be obtained in $1/c$ perturbation theory from the gravitational or Chern-Simons path integral \cite{HartmanLargeC, KrausBlocks, Hijano:2015qja}, and so they provide a natural project for resurgence theory \cite{Dorigoni:2014hea, Cherman:2014ofa, Basar:2013eka}; some of the groundwork has already been laid \cite{Witten:2010cx, Gaiotto:2011nm}.  An important open problem is to determine which states appear as saddle points in heavy-light Virasoro blocks, and to understand their role in resolving information loss.  As a first step, it would be interesting to better understand the simplifications that occur in the case of degenerate operators \cite{Gaiotto:2011nm, Castro:2011ui}.   More generally, one would like to obtain classical solutions in Chern-Simons corresponding to heavy-light correlators \cite{Hijano:2015qja} with a heavy operator exchange, since we expect that these solutions may arise as saddle-points associated with the restoration of information.  Studies of black hole formation in AdS$_3$ \cite{Matschull:1998rv, Lindgren:2015fum, Anous:2016kss} should be useful for this purpose.   Since the Virasoro blocks isolate and encapsulate purely gravitational phenomena, there may be a well-defined interpretation for a gravitational path integral in AdS$_3$ as an object that precisely generates the Virasoro blocks without specifying a particular CFT$_2$.\footnote{A similar perspective was suggested long before the advent of AdS/CFT \cite{Witten:1988hf, Witten:1988hc, Verlinde:1989ua, Elitzur:1989nr}; for example  \cite{Witten:1988hf}: ``The basic connection that we have so far stated between general covariance in $2 + 1$ dimensions and conformally invariant theories in $1 + 1$ dimensions is that the physical Hilbert spaces obtained by quantization in $2 + 1$ dimensions can be interpreted as the spaces of conformal blocks in $1 + 1$ dimensions.''}

If information loss in AdS$_3$/CFT$_2$ stems primarily from the behavior of the Virasoro blocks as $c \to \infty$, then it would seem that string and brane states play little to no direct role in the resolution of the information paradox.  Perhaps this goes hand in hand with the possibility that the gravitational path integral in AdS$_3$ can be well-defined.   In AdS$_{d+1}$ with $d>2$ it seems exceedingly unlikely that the gravitational path integral has a sharp definition.  And in fact it is only in higher dimensions that we can study the scattering of localized gravitons, a process that does generically require the inclusion of higher spin states to avoid violations of causality in perturbation theory \cite{Camanho:2014apa, Maldacena:2015waa, Hartman:2015lfa}.  Somewhat beyond perturbation theory, the correspondence between strings and black holes \cite{Horowitz:1996nw} suggests that stringy states play a natural role in quantum gravity.  But it remains unclear whether extended objects are useful for resolving information loss in generic theories.

It seems unlikely that our results would have any direct analog in higher dimensions, where multi-stress tensor OPEs (and multi-graviton scattering amplitudes) are not rigidly determined, and where the gravitational sector is inextricably linked with the full CFT spectrum.  Nevertheless, there is a more general lesson:  thermodynamics in holographic theories must be governed by the exchange of multi-stress tensor states between heavy microstates and light probes.  This just restates black hole thermodynamics in the language of the bootstrap \cite{FerraraOriginalBootstrap1, PolyakovOriginalBootstrap2, Rattazzi:2008pe, Rychkov:2016iqz, Simmons-Duffin:2016gjk}, but the fact that it has not been derived or incorporated into the bootstrap in $d>2$ dimensions seems like a conceptual, and perhaps a technical, shortcoming.  In the future it will be interesting to see if thermodynamic constraints on CFTs can extend the power of the bootstrap approach.  In a certain sense, these constraints do greatly simplify the analysis of two-dimensional CFTs \cite{Hartman:2014oaa, Chang:2015qfa}.

Extensions of our analysis may lead to a resolution of information loss in AdS$_3$/CFT$_2$, explaining why correlators are `too thermal' at large central charge, and displaying the non-perturbative corrections that restore unitarity.  But a much larger question remains -- can we reconstruct local bulk physics across the horizon of a black hole, resolving the information \emph{paradox}?  Perhaps with sufficiently sharp technical tools supervening upon the thermodynamic limit, we may better understand proposals \cite{Papadodimas:2012aq, Papadodimas:2013wnh} for the reconstruction of the black hole interior.

\section*{Acknowledgments}

We would like to thank Chris Beem, Hongbin Chen, Aleksey Cherman, Ethan Dyer, Ilya Esterlis, Daniel Jafferis, David E. Kaplan, Ami Katz,  Juan Maldacena, Yasunori Nomura, Eric Perlmutter, David Simmons-Duffin, David Ramirez, Douglas Stanford, Andy Strominger, Mithat Unsal, Hermann Verlinde,  Matt Walters, Edward Witten, and Xi Yin for valuable discussions.  ALF is supported by the US Department of Energy Office of Science under Award Number DE-SC-0010025.  JK  is supported in part by NSF grants PHY-1316665 and PHY-1454083, and by a Sloan Foundation fellowship.

\addtocontents{toc}{\protect\setcounter{tocdepth}{1}}

\appendix

\section{Derivation of Large $c$ Light Degenerate State Equations}
\label{app:DerivationLightNullStateEquations}

This appendix reviews\footnote{See specifically \cite{DiFrancesco:1997nk} exercise 8.8; this appendix streamlines their argument  for present purposes. } results from \cite{Bauer:1991ai, DiFrancesco:1997nk} in order to derive the light null state differential equations in the large $c$ limit.
An alternate explicit closed form for the $|h_{1,s}\>$ null states is given in \cite{DiFrancesco:1997nk} (see their eqs. (8.26) and (8.28)). Let $D_{1,s}$ be the following matrix:
\be 
D_{1,s} = -J_- + \sum_{m=0}^\infty\left(\frac {J_+}{b^2}\right)^m L_{-m-1},  
\ee
where $J_\pm$ are matrix generators of the spin $(s-1)/2$ representation of $SU(2)$:
\be
(J_0)_{ij} &=& \frac{1}{2} (s-2i+ 1)\delta_{ij}, \nn\\
(J_-)_{ij} &=& \left\{ \begin{array}{cc}  \delta_{i,j+1} & (j=1,2, \dots, s-1) \\ 0 & \textrm{else}\end{array} \right. ,\qquad \qquad  \qquad \qquad \begin{array}{c} \left[J_+, J_- \right] = 2 J_0 , \\ \left[J_0, J_\pm \right] = \pm J_\pm . \end{array}  \nn\\
(J_+)_{ij} &=& \left\{ \begin{array}{cc} i(s-i)\delta_{i+1,j} & (i=1, 2, \dots, s-1), \\ 0 & \textrm{else} \end{array} \right.  . 
\ee
Then, the null state equation of motion is given by the equation $f_0 = 0$ after eliminating $f_1, \dots, f_{s-1}$ from the equation
\be
D_{1,s} \left( \begin{array}{c} f_1 \\ f_2 \\ \vdots \\ f_s \end{array} \right) &=& \left( \begin{array}{c} f_0 \\ 0 \\ \vdots \\ 0 \end{array} \right) .
\ee
Formally, this can be written
\be
0 &=& \Delta_{1,s}(b) |h_{1,s}\>,  \nn\\
\Delta_{1,s}(b)  &\equiv& \det \left[ -J_- + \sum_{m=0}^\infty\left( \frac{J_+}{b^2}\right)^m L_{-m-1} \right].
\ee
The differential operator on the four-point function can be obtained by taking 
\be
0 &=& \< h_{1,s}|\left( \Delta_{1,s} \right)^\dagger \CO_{1,s}(0) \CO_H(x) \CO_H(y)\>
\ee 
and commuting $L_{m}$s to the right.   In the large $c$ limit, it immediately follows that only $L_2$ and $L_1$ give non-vanishing contributions the the null state differential equation at infinite $c$ in the above formula.  This is because of the suppression factor of $b^{-2(m-1)}$ for $L_{m}$ when we take the limit $b\rightarrow \infty$.  For $L_{1}$, this is no suppression, and for $L_{2}$, this is suppression by a factor of $1/c$ which is compensated for by the factor of $h_H$ when $L_{2}$ hits the heavy operators.  For $L_{3}$ and higher, there is suppression by $1/c^2$ or more, and they are never compensated by more than one $h_H$ upstairs, so their contributions vanish.  

 To work out the action of $b^{-2} L_{2}$ and $L_{1}$ in the large $c$ limit, consider a generic contribution of the form
 \be
 \<h_{1,s} | L_{p_1} L_{p_2} \dots L_{p_n} \CO_{1,s}(0) \CO_H(x) \CO_H(y) \>
 \ee
 where $p_i = 1,2$. Now, commute the $L$s to the right.  Acting on the correlation function $\<h_{1,s}| \CO_{1,s}(0) \CO_H(x)\CO_H(y)\>$,  parameterized as
 \be
 \frac{1}{(y-x)^{2h_H}} \tilde{\CV}\left(1-\frac{x}{y} \right), 
 \ee
 $L_{1}$ produces 
 \be
 \frac{x}{y} \frac{1}{(y-x)^{2h_H-1}} \tilde{\CV}'(1-\frac{x}{y}) \stackrel{x=y(1-z)}{=}  \frac{1}{y^{2h_H-1}}\frac{(1-z) }{z^{2h_H-1}} \tilde{\CV}'(z),
 \ee
 whereas $-b^{-2} L_{2} = \frac{6}{c} L_{2}$ produces
 \be
 \frac{1}{4} (1+r_+^2) \frac{1}{(y-x)^{2h_H-2}} \tilde{\CV}\left(1-\frac{x}{y}\right) \stackrel{x=y(1-z)}{=} \frac{1}{y^{2h_H-2}} \frac{1}{4}(1+r_+^2) \frac{\tilde{\CV}(z)}{z^{2h_H-2}},\nn\\
 \ee
 where we have taken the leading large $c$ piece and used $h_H = \frac{c}{24} (1+r_+^2)$.  More generally, after the action of each $L_{1}$ or $L_{2}$, we will have a function of the form $y^{a} g(z)$ for some integer $a$ and function $z$.  Acting on such a function (parameterized slightly differently for later convenience), we have:
 \be
 L_{1} \cdot \left( y^{a-2h_H} z^{-a-2h_H} (1-z)^a g(z) \right) &=& y^{1+a-2h_H} z^{-1-a-2h_H}(1-z)^{a+1} \left( z^2  g'(z) \right) \nn\\
 L_{2} \cdot \left( y^{a-2h_H} z^{-a-2h_H} (1-z)^a g(z) \right) &=& y^{2+a-2h_H} z^{-2-a-2h_H} (1-z)^{a+2} h_H \frac{z^4}{(1-z)^2} g(z) + \CO(c^0) \nn\\
 \ee
which we can summarize, changing to $t$ coordinates, as
\be
L_{1} &:& \{ g(t), a\} \rightarrow \{ 4 \sinh^2\left( \frac{t}{2} \right) g'(t), a+1\}, \nn\\
-b^{-2} L_{2} &:& \{ g(t), a\} \rightarrow \{ 4 (1+r_+^2) \sinh^4 \left( \frac{t}{2} \right) g(t) ,a+2 \}.
\ee
 
To simplify further, we can change basis by redefining the $f_{s-j}$'s  by 
\be
f_{s-j} = \sinh^{1-s}\left( \frac{t}{2} \right) \sinh^{2j}\left( \frac{t}{2} \right) \tilde{f}_{s-j}.
\ee 
 This has the effect of factoring out $\sinh^{1-s}\left( \frac{t}{2} \right)$ from $f_s$,  and also of removing the factors of $\sinh^2\left( \frac{t}{2}\right)$ from $L_1$ and $\sinh^4 \left( \frac{t}{2} \right)$ from $L_{2}$ . After this change, we effectively replace the action of $L_1$ and $L_2$ with
\be
L_{1} &:& \{ g(t), a\} \rightarrow \{ 4( g'(t) - \frac{(s-1-2j)}{2} \coth\left( \frac{t}{2} \right)g(t) ), a+1\}, \nn\\
-b^{-2}L_{2} &:& \{ g(t), a\} \rightarrow \{ 4 (1+r_+^2) g(t) ,a+2 \},
\ee 
so that in the basis of the $\tilde{f}_j$s,  $D_{1,s}$ takes the form
\be
\tilde{D}_{1,s} &=& -J_- + 4 \partial_t + 4\coth \left( \frac{t}{2} \right) J_0  +4 (1+r_+^2) J_+  .
\label{eq:Dr1simp} 
\ee    
 Now, we can perform a transform of the form $\tilde{D}_{1,s} \rightarrow U^\dagger \tilde{D}_{1,s} U$ where $U=e^{y(t) J_+}$ is an upper right-triangular matrix with 1s on the diagonal, and therefore the new basis manifestly preserves $\Delta_{1,s}=0$.  Using the algebra of $J_0, J_\pm$, it follows that $\tilde{D}_{1,s}$ transforms to
 \be
 \tilde{D}_{1,s} \rightarrow -J_- + 4 \partial_t + \left( 4 \coth \left( \frac{t}{2} \right) + 2 y(t) \right)J_0 + \left( 4 (1+r_+^2)+ 4y'(t) - y^2(t)\right) J_+.
 \ee
Choosing $y=-2 \coth \left(\frac{t}{2} \right)$ to eliminate the $J_0$ term, we finally arrive at
\be
\tilde{D}_{1,s} \rightarrow -J_- + 4 \partial_t + 4r_+^2 J_+.
\label{eq:simpleD1s}
\ee
All terms above now commute, so in computing the determinant, we can treat all contributions as regular numbers.  Computing the determinant of $( -J_- + x + J_+)$ as a function of $x$ is equivalent to computing the eigenvalues of the matrix
\be
-J_- + J_+ , 
\ee
which is $2J_1$ in a conventional notation.
This is related to $2J_0$ by a similarity transform and thus has the same eigenvalues: $x = -(s-1),-(s-3), \dots, (s-3), (s-1)$.\footnote{Equivalently, we can diagonalize (\ref{eq:simpleD1s}) directly by conjugating with $(2r_+)^{\frac{1}{2} J_0-\frac{s+1}{4}} e^{\frac{\pi}{4}(J_- + J_+) }$. }  Substituting $x\rightarrow \frac{2}{r_+} \partial_t$ in the resulting characteristic polynomial then gives the form of the singular operator at level $s$.  This proves (\ref{eq:lightnullhighcsimp}), reproduced here for convenience:\footnote{Note that $\tilde{f}_s = \sinh^{s-1}\left( \frac{t}{2} \right) \CV_s(t) = (1-z)^{\frac{1-s}{2}} \CV_s(t)$.}
\be
\left[ \prod_{k=-(s-1)+2j\atop j=0, \dots, s-1}\left( \partial_t - \frac{ik r_+}{2} \right)  \right]  \tilde{f}_s(t) = 0 .
\label{eq:lightnullhighcsimp2}
\ee

\section{Universality of Forbidden Singularities and General $1/c$ Corrections}
\label{app:ForbiddenSingularitiesinCorrectionGeneralV}

We showed in section \ref{fig:NearForbiddenSingularity21} that the degenerate Virasoro vacuum blocks take a universal form of equation (\ref{eq:UniversalFunctionForSingularityResolution}), as they are governed by the differential equation (\ref{eq:heavynullsecondorderscaling}) near their forbidden singularities.  In this appendix we will provide a piece of evidence that this universal regulator also governs the behavior of general heavy-light Virasoro vacuum blocks at large $c$.  The point is that if the blocks are well approximated by 
\be
S(x,c) \approx \int_0^\infty dp \, p^{2h_L - 1} e^{-p x - \frac{\sigma^2}{2 c} p^2}
\ee
in the vicinity of their forbidden singularities, then the $1/c$ corrections to the leading large $c$ limit near the singularity must take the form
\be
\label{eq:ExpectationNearGeneralSingularity}
\frac{1}{x^{2h_L}} - \frac{\sigma^2 h_L (2h_L+1)  }{c} \frac{1}{x^{2 h_L + 2}} + \cdots
\ee
where we inserted the constant $\sigma^2$ in accord with equation (\ref{eq:UniversalFunctionForSingularityResolution}).  Notice that this makes a precise prediction for the relationship between the $h_L^2$ and $h_L$ terms.
Since we have an explicit expression for the leading and $1/c$ corrected heavy-light blocks \cite{Fitzpatrick:2015dlt, Beccaria}, we can search for the $x^{-2 h_L-2}$ term in the vicinity of forbidden singularities, and extract the coeffiicent $\sigma^2$.  

The general $1/c$ corrections to the heavy-light Virasoro vacuum block are \cite{Fitzpatrick:2015dlt, Beccaria}
\begin{equation}
\begin{aligned}
\mathcal{V}(t) &= e^{h_L t}  \left( \frac{\pi T_H}{\sin(\pi T_H t)} \right)^{2h_L} \left[ 1 + \frac{h_L}{c} {\cal V}_{h_L}^{(1)} + \frac{h_L^2}{c} {\cal V}_{h_L^2}^{(1)} \right], \\ 
 {\cal V}_{h_L}^{(1)} &= \frac{\text{csch}^2\left(\frac{\alpha  t}{2}\right)}{2} \Big[3 \left(e^{-\alpha t} B\left(e^{-t},-\alpha ,0 \right)+e^{\alpha  t} B\left(e^{-t},\alpha, 0 \right)+e^{\alpha  t}
   B\left(e^t,-\alpha,0 \right)+e^{-\alpha  t} B\left(e^t,\alpha , 0\right)\right)  \\
    &  +\frac{1}{\alpha ^2}+ \cosh (\alpha  t)\left( -\frac{1}{\alpha ^2}+6 H_{-\alpha }+6 H_{\alpha }+6 i \pi -5 \right)+12 \log \left(2 \sinh \left(\frac{t}{2}\right)\right)+ 5\Big] \\
     & -t \frac{\left(13
   \alpha ^2-1\right)  \coth \left(\frac{\alpha  t}{2}\right)}{2\alpha }+12 \log \left(\frac{2 \sinh \left(\frac{\alpha 
   t}{2}\right)}{\alpha }\right),     \\
 {\cal V}_{h_L^2}^{(1)} &=  
 6 \Big( \text{csch}^2\left(  \frac{\alpha  t}{2}\right) \left[ \frac{B(e^{-t},-\alpha,0) +B(e^t,-\alpha ,0)+B(e^{-t},\alpha ,0)+B(e^t,\alpha,0)}{2} \right.  \\
    & \left. + H_{-\alpha }+H_{\alpha }+2 \log \left(2 \sinh \left(\frac{t}{2}\right)\right)+i \pi \right] 
    +2 \left(\log \left(\alpha  \sinh \left(\frac{t}{2}\right) \text{csch}\left(\frac{\alpha  t}{2}\right)\right)+1\right)\Big). 
    \label{eq:ExactResult}
    \end{aligned}
\end{equation}
where  
$B(x,\beta,0) = \frac{x^{\beta}{}_2F_1(1,\beta,1+\beta,x)}{\beta}$ is the incomplete Beta function,  $z \equiv 1-e^{-t}$,  $H_n$ is the harmonic function, and $\alpha \equiv \sqrt{1- \frac{24 h_H}{c}} \cong 2 \pi i T_H$.  

 Note that (despite naive appearances) if we expand either of these results around $t = 0$ they are non-singular.  However, after the analytic continuation $t \to t + \frac{n}{T_H}$ singularities develop.

Let us consider the forbidden singularities of the $1/c$ corrections to the general large $c$ heavy-light blocks.  This means we want to evaluate equation (\ref{eq:ExactResult}) in an expansion around $t = \frac{n}{T_H} = \frac{2 \pi i n }{\alpha}$, which gives a coefficient for the $n$th forbidden singularity  $1/(t-n/T_H)^{2h_L + 2}$
\be
\label{eq:GeneralHeavyLightSingularityCoeff}
 {\cal V}_{h_L^2}^{(1)} = 2 {\cal V}_{h_L}^{(1)} &\to& -\frac{3 }{2 \pi^2 T_H^2} \Big[  2 H_{-\alpha }+2 H_{\alpha }+2 \pi i + 4 \log \left(2  \sinh \left(\frac{ n  }{2 T_H}\right)\right) 
 \\ && +
 B\left(e^{-\frac{n}{T_H}},-\alpha ,0 \right)+ B \left(e^{-\frac{n}{T_H}},\alpha, 0 \right) +
   B\left(e^{\frac{n}{T_H}},-\alpha,0 \right)+ B\left(e^{\frac{n}{T_H}},\alpha , 0\right)   \Big] 
   \nn
\ee
Now let us compare to what we obtained from the degenerate blocks.  Using some special function identities, we find that the degenerate blocks have forbidden singularities characterized by the function  
\be
\sigma_{\mathrm{deg}}^2(n,r) = \frac{12 \left(B \left(e^{-\frac{2 \pi i  n }{r}},r,0\right)+  B \left(e^{\frac{2 \pi i n}{r}},r, 0 \right) + 2 \log \left( 2 \sinh \left( \frac{\pi i n}{r} \right) \right) + 2 H_{r-1} \right)}{r^2}
\nn
\ee
Using the fact that $H_{r-1} = H_r - \frac{1}{r}$, we see that by analytically continuing $r \to  \alpha = 2 \pi i T_H$, we find that
\be
\frac{\sigma_{\mathrm{deg}}^2(n,\alpha) + \sigma_{\mathrm{deg}}^2(n,-\alpha) }{2}
&=& -\frac{3}{2 \pi^2 T_H^2} \Big[  2 H_{-\alpha }+2 H_{\alpha }+2 \pi i + 4 \log \left(2  \sinh \left(\frac{ n  }{2 T_H}\right)\right) 
 \\ && + 
 B\left(e^{-\frac{n}{T_H}},-\alpha ,0 \right)+ B \left(e^{-\frac{n}{T_H}},\alpha, 0 \right) +
   B\left(e^{\frac{n}{T_H}},-\alpha,0 \right)+ B\left(e^{\frac{n}{T_H}},\alpha , 0\right)   \Big]  \nn
\ee
This result from the degenerate blocks exactly matches the result from $1/c$ correction to the general heavy-light blocks, equation (\ref{eq:GeneralHeavyLightSingularityCoeff}).  We cannot say for certain that it should have been necessary to average the $r = \pm \alpha$ results, although we think this is quite reasonable since the analytic continuation of $r = \sqrt{1 - \frac{24 h_H}{c}}$ to $h_H > c/24$  has a sign ambiguity, and so it makes sense that both signs should contribute.\footnote{ Furthermore, as we have discussed in section \ref{sec:ArgumentForAnalyticity}, the series expansion in $q$ of the conformal blocks, at every order there is an $\alpha \rightarrow -\alpha$ symmetry (which follows from the fact that the coefficients are rational functions of the weights).  Since the $q$ expansion converges absolutely on the infinite-sheeted covering space \cite{Zamolodchikovq,Maldacena:2015iua}, this should be a symmetry of the full block itself.}  In any case, averaging appears to be the correct procedure since in this case we find the two results match exactly. 

\section{Asymptotic Analysis of More Degenerate Blocks}
\label{app:Asymptotics31and22}

In this appendix we study the two-dimensional Coulomb gas integrals with external $\CO_{(3,1)}$ and $\CO_{(2,2)}$ operators.  The rather involved analysis demonstrates that groups of critical points coalesce as we approach forbidden singularities.  Via contour deformation, we can re-interpret the behavior of the critical points and their associated steepest descent contours in terms of linear combinations of specific CFT states.  In the future we would like to better understand the relationship between non-perturbative effects in the vacuum Virasoro block, critical points in the Coulomb gas integrals, heavy states, and AdS$_3$ geometry.

\subsection{Virasoro Blocks with External $\CO_{(3,1)}$}
\label{sec:ExternalO31}

Using the Coulomb gas formalism \cite{Dotsenko:1984nm, Dotsenko:1984ad}, the correlators of higher order degenerate operators can be written in terms of  higher dimensional contour integrals.  For example, we can write the $\CO_{(3,1)}$ degenerate four point function as  \cite{DiFrancesco:1997nk}
\be\label{eq: O31ContourIntegral}
\< \CO_L(\infty) \CO_L(1) \CO_{(3,1)}(z)   \CO_{(3,1)}(0)\> \sim 
\int \!d w_1\! \int\! d w_2 \, 
\frac{(1-z)^2}{z^{2h_{(3,1)}} }  \exp {\cal I}_{(3,1)}\;,
\ee
where for the special choice $h_L = 1$ 
the action ${\cal I}_{(3,1)}$ is given by
\begin{align}
{\cal I}_{(3,1)}&=2b^2 \log\big[w_1(1-w_1)w_2(1-w_2)\big]- 2b^2 \log(w_1-w_2)-2\log\big[(1-z w_1)(1-z w_2)\big]
\end{align}
Once again the action has a singularity at $w_i = \frac{1}{z}$, requiring the deformation of contours as we analytically continue in $z$.
The action ${\cal I}_{(3,1)}$ is defined on $\mathbb{C}^2$, with a symmetry under $w_1 \leftrightarrow w_2$.  There are three integration contours relevant to the study of specific Virasoro blocks, namely $[0,1]\times [0,1]$,\; $[0,1]\times [\frac{1}{z},\infty)$,\; and $[\frac{1}{z},\infty)\times [\frac{1}{z},\infty)$.  To connect them to the primary operators occurring in the fusion rules, we can study the integrals in the OPE limit of small $z$. We find
\be
z^{-2h_{(3,1)}}(1-z)^2
\int_0^1 d w_1\!\int_0^1 d w_2\, e^{{\cal I}_{(3,1)}}
\ \sim \  z^{-2h_{(3,1)}}
\ee
so the first integration contour $[0,1]\times [0,1]$ corresponds to the Virasoro vacuum block. Similarly, one can show that the $[0,1]\times [\frac{1}{z},\infty)$ and $[\frac{1}{z},\infty)\times [\frac{1}{z},\infty)$ integration contour gives rise to the $\CO_{(3,1)}$ and $\CO_{(5,1)}$ blocks, respectively.

\begin{figure}[t!]
\begin{center}
\includegraphics[width=0.48\textwidth]{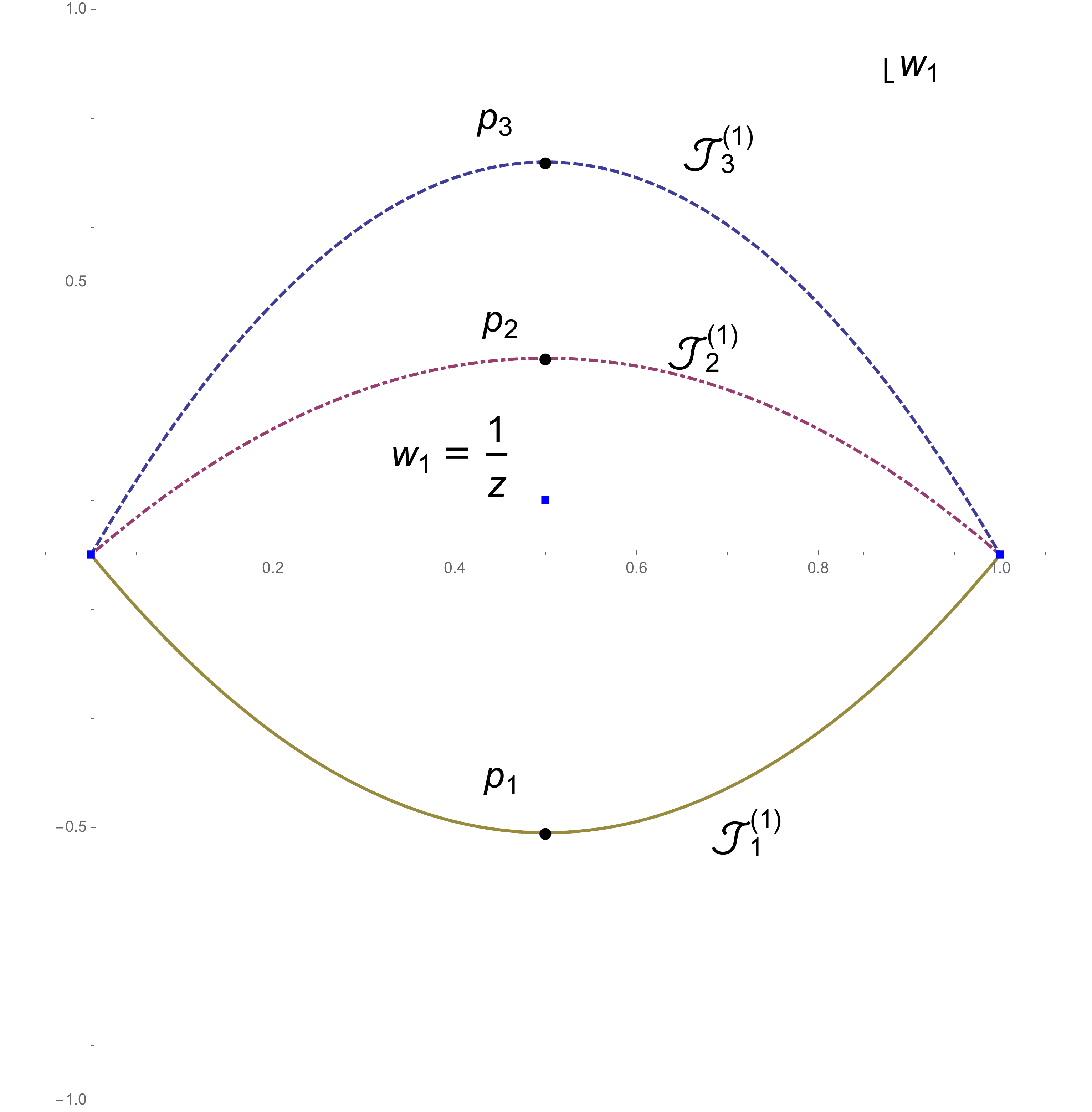}
\includegraphics[width=0.48\textwidth]{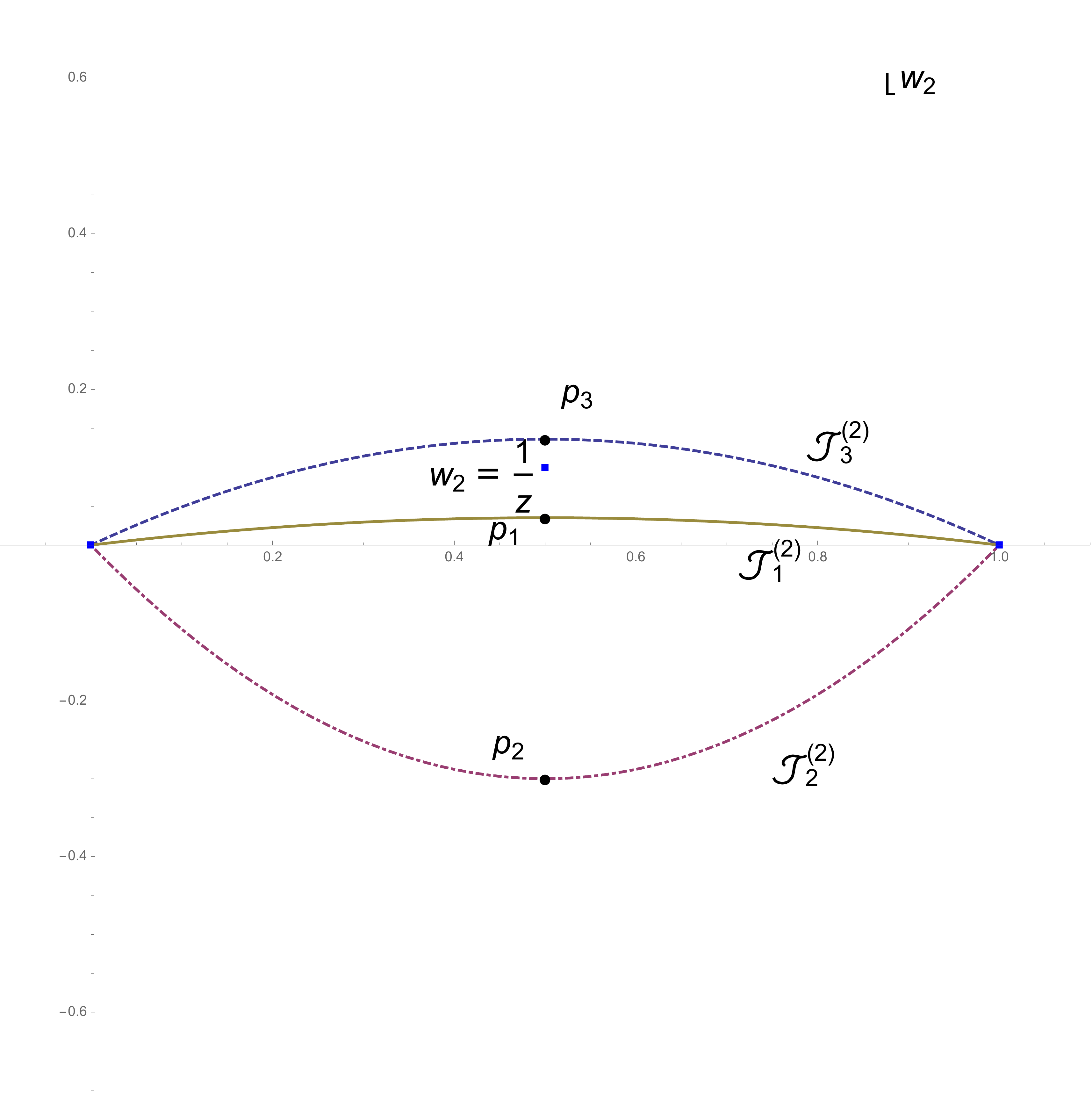}
\caption{This figure shows the projection onto the $w_1$  plane (left) and the $w_2$  plane (right) of the critical points (black points) of the action ${\cal I}_{(3,1)}$; the colored curves are sketches of the steepest descent contours attached to these critical points. At branch points (blue dots) the action diverges, and the integrand has a pole, so one cannot deform the contours across the branch points without changing the value of the integral.  In the above plots we chose $2b^2=20$ and $z$ far from all forbidden singularities. The integration contour ${\cal J}_1^{(j)}$ can be deformed to $[0,1] \times [0,1]$, and so it corresponds to the vacuum Virasoro block. Note that as $z$ revolves around $1$, the branch point singularity at $1/z = \frac{1}{2} + \frac{1}{2} i \cot(\theta/2)$ parallels the imaginary axis, forcing various contour deformations, and producing a monodromy for the vacuum block. } 
\label{fig:phi31contour1}
\end{center}
\end{figure}

We would like to understand which critical points and heavy states are involved in the resolution of forbidden singularities.  For example, in section \ref{sec:Borelof21} we saw that the Borel resummation became ill-defined in the vicinity of a forbidden singularity due to a branch cut in the Borel integrand, which we could associate with the $\CO_{(3,1)}$ state.  We want to gather information about similar phenomena using the Coulomb gas representation.  For this purpose we need to determine what happens to the integration contours as we analytically continue $z$ towards the forbidden singularities.  

Let $p_i$ be a critical point, and define the Lefschetz thimble $\CJ_i$ as the submanifold of $\mathbb{C}^2$ connected to the critical point $p_i$ by a steepest descent contour.  At critical points, the derivatives of the action with respect to $w_1$ and $w_2$ vanish.  Naively there are six critical points, but due to the symmetry $w_1 \leftrightarrow w_2$ of the action ${\cal I}_{(3,1)}$, they can be grouped into three pairs.  To simplify the analysis,  as in the previous subsection we write $z=1-e^{i \theta}$ and study the region of real $\theta \in [0,2\pi)$. In this case, the coordinates of the critical points  $(w_1, w_2)$ are $p_i = \left(\frac{1}{2}+i \eta_i,\;\frac{1}{2}+i \sigma_i\right)$ with $i=1,2,3$,  where the $\sigma_i$'s and $\eta_i$'s are solutions to algebraic equations that have been relegated to appendix \ref{app:ExplicitCriticalPoints}.  The pairs of physically identical critical points are related by the exchange $\eta_i \leftrightarrow \sigma_i$.

\begin{figure}[t!]
\begin{center}
\includegraphics[width=0.96\textwidth]{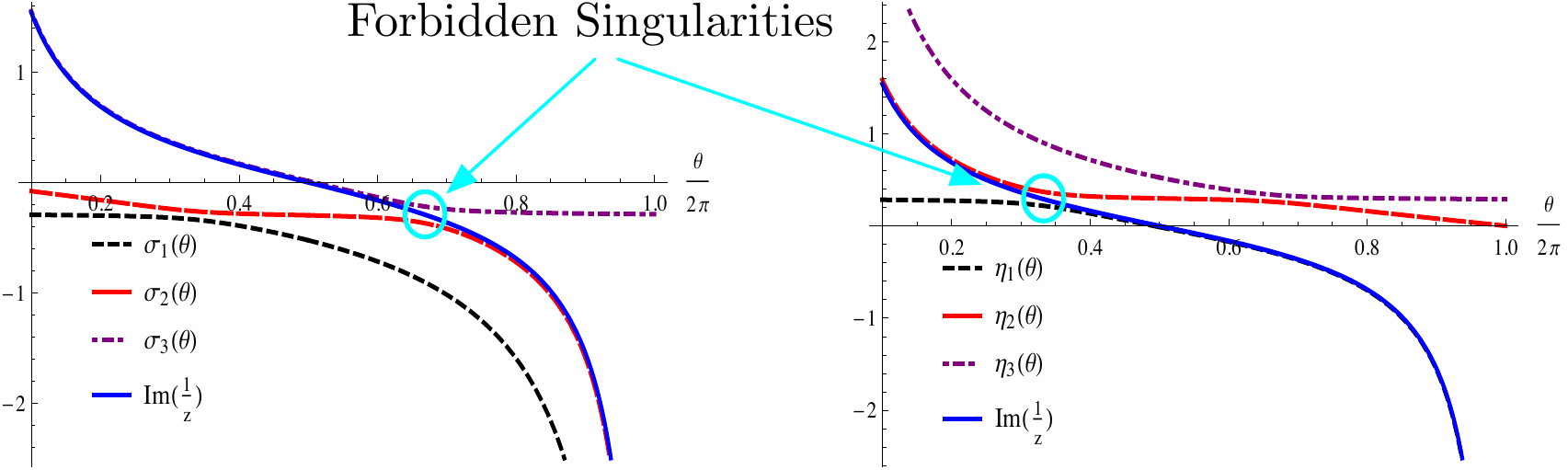}
\caption{This figure illustrates the positions of the three critical points of $\CI_{(3,1)}$  as $z = 1 - e^{i\theta}$ varies over $\theta \in [0, 2\pi]$.  The critical points have coordinates $(w_1, w_2) = \left(\frac{1}{2} + i \sigma_i, \frac{1}{2} + i \eta_i \right)$, and so the left and right plots are of the imaginary part of $w_1$ and $w_2$, respectively.  The solid line indicates the branch point $\frac{1}{z}$ where $\exp \CI_{(3,1)}$ has a pole, while the various dashed lines are critical points.  In the vicinity of the forbidden singularities, pairs of critical points approach the branch point, intersecting it as $c \propto b^2 \to \infty$.  In these plots we take $c \approx 500$, so pairs of critical points approach but do not meet.
}
\label{fig:CriticalPointsForbiddenSingularities31}
\end{center}
\end{figure}

Since the action ${\cal I}_{(3,1)}$ is defined on a four real dimensional  manifold $\mathbb{C}^2$, it is not possible to illustrate the relative positions of the branch points and the critical points in a single planar picture, as we did in the $\CO_{(2,1)}$ case. 
Instead, in figure \ref{fig:phi31contour1} we project them onto the complex $w_1$ plane (left) and the $w_2$ plane (right), respectively. The branch points of the action ${\cal I}_{(3,1)}$ are plotted in blue,  and the critical points in black.  We also schematically indicate the steepest descent contours associated with each critical point $p_i$, which we refer to as ${\cal J}_i^{(1)}$  and ${\cal J}_i^{(2)}$ when projected in the $w_1$ and $w_2$ planes, respectively.  
Note that in figures \ref{fig:phi31contour1} the integration contours passing through $p_1$ (yellow lines) can be continuously deformed into $[0,1]\times [0,1]$ without passing through any branch point. In other words, the one-dimensional integration contours $\CJ_1^{(j)}$ (yellow lines)  correspond to the vacuum Virasoro block. 

We illustrate the relationship between the positions of the critical points and the forbidden singularities in figure \ref{fig:CriticalPointsForbiddenSingularities31}.  Notice that as $z$ approaches a forbidden singularity, pairs of critical points approach the branch point of $\CI_{(3,1)}$ at $\frac{1}{z}$.  We provide analytic formulas documenting this behavior in appendix \ref{app:ExplicitCriticalPoints}.  As we discussed in the $\CO_{(2,1)}$ case, the contours of integration must be deformed to avoid the branch point at $\frac{1}{z}$, just as the Borel contour must be deformed to avoid singularities in the Borel plane. 
In the present case, we need to deform $\CJ_1^{(2)}$ (the yellow curve in $w_2$ plane in figure \ref{fig:phi31contour1}) as we approach a forbidden singularity.  As a result, there is ambiguity in the integration contour for the integral \eqref{eq: O31ContourIntegral}.  As we showed with equation (\ref{eq:IPMContours}), the ambiguity  is proportional to an integral over the contour $[0,1] \times [\frac{1}{z},\infty)$.  We encounter an identical phenomenon when we approach the second forbidden singularity at $b=\infty, z=1-e^{\frac{4 i \pi}{3}}$, as illustrated in figure \ref{fig:CriticalPointsForbiddenSingularities31}. We conclude that although the fusion rule
\be
\CO_{(3,1)}\times \CO_{(3,1)}={\bf 1} \oplus \CO_{(3,1)}\oplus \CO_{(5,1)}
\ee
contains more than one heavy state, only the $\CO_{(3,1)}$ state has an immediate connection with the behavior of the correlator near the forbidden singularities.  As we analytically continue in $z$ around $1$, we will be forced to deform the contour further and pick up contributions that can be associated with the $\CO_{(5,1)}$ state as well \cite{DiFrancesco:1997nk}.

\subsection{A Light Operator in the Fusion Rule: the $\CO_{(2,2)}$ Case}
\label{sec:ExternalO22}

Thus far we have studied heavy degenerate external  operators that can only fuse to form either the Virasoro vacuum or heavy states with dimensions proportonal to $c$.  In this subsection we will study the degenerate four point functions with external $\CO_{(2,2)}$ states. From equation \eqref{eq:fusionrules}, we see that the fusion rule of $\CO_{(2,2)}$ with itself is given by
\be
\CO_{(2,2)}\times \CO_{(2,2)}={\bf 1} \oplus \CO_{(3,1)}\oplus\CO_{(1,3)}\oplus \CO_{(3,3)}\;,
\ee 
which involves both the light operator $\CO_{(1,3)}$ and heavy operators $\CO_{(3,1)}$ and  $\CO_{(3,3)}$. Nevertheless, at large $c$ we have $h_{(2,2)} \approx h_{(2,1)}$, and so this case only has a single forbidden singularity at $z=2$, much like in the case of external $\CO_{(2,1)}$.

By studying the contour integral representation of this degenerate correlation function, we wish to shed light on the question of which saddle points and states contribute non-perturbative corrections to the vacuum block in association with forbidden singularities.  We will see that the $\CO_{(2,2)}$ case behaves rather differently from those above.  In particular, all four of the saddle points coalesce near the single forbidden singularity.

\begin{figure}[t!]
\begin{center}
\includegraphics[width=0.48\textwidth]{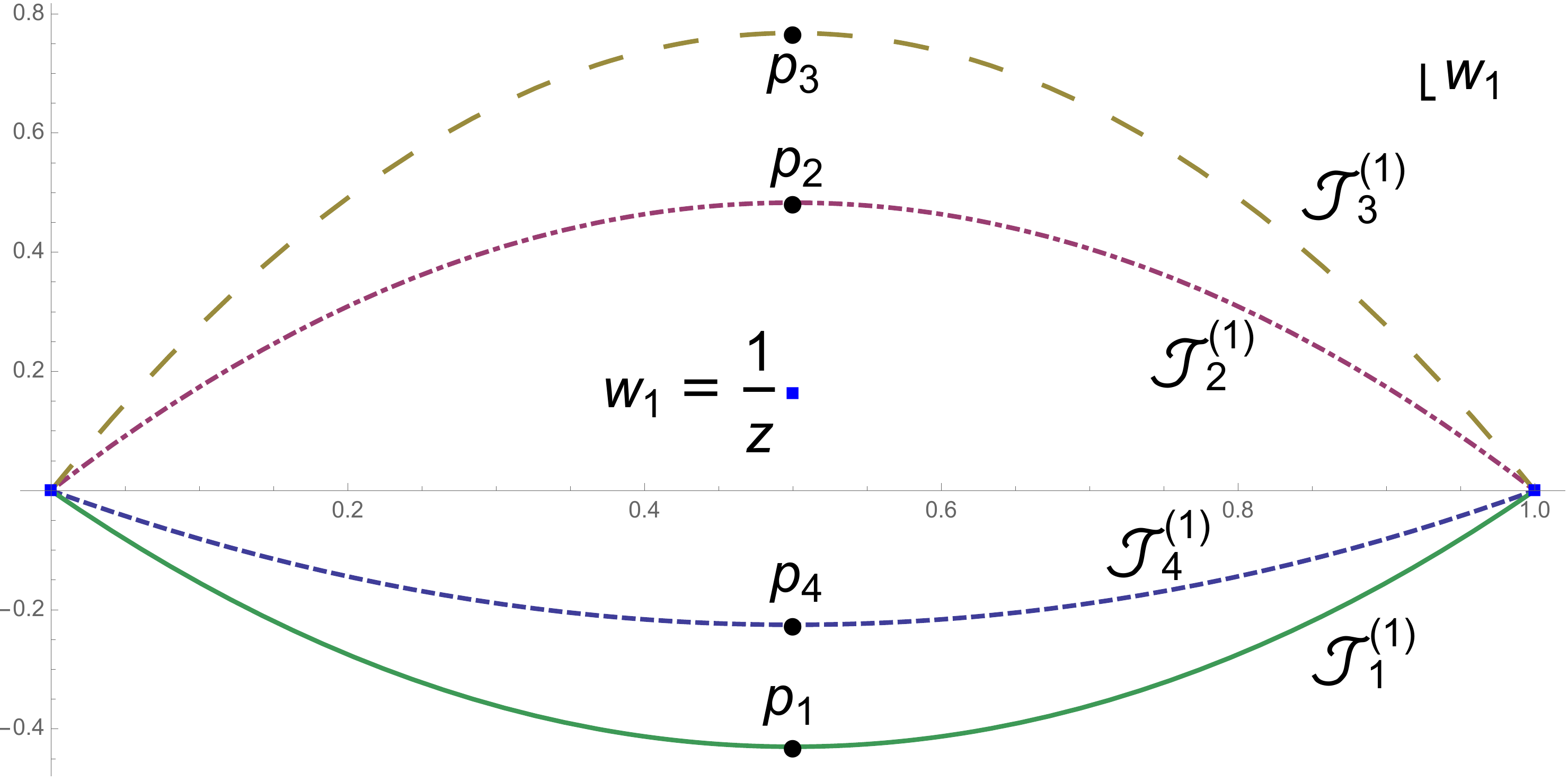}
\includegraphics[width=0.48\textwidth]{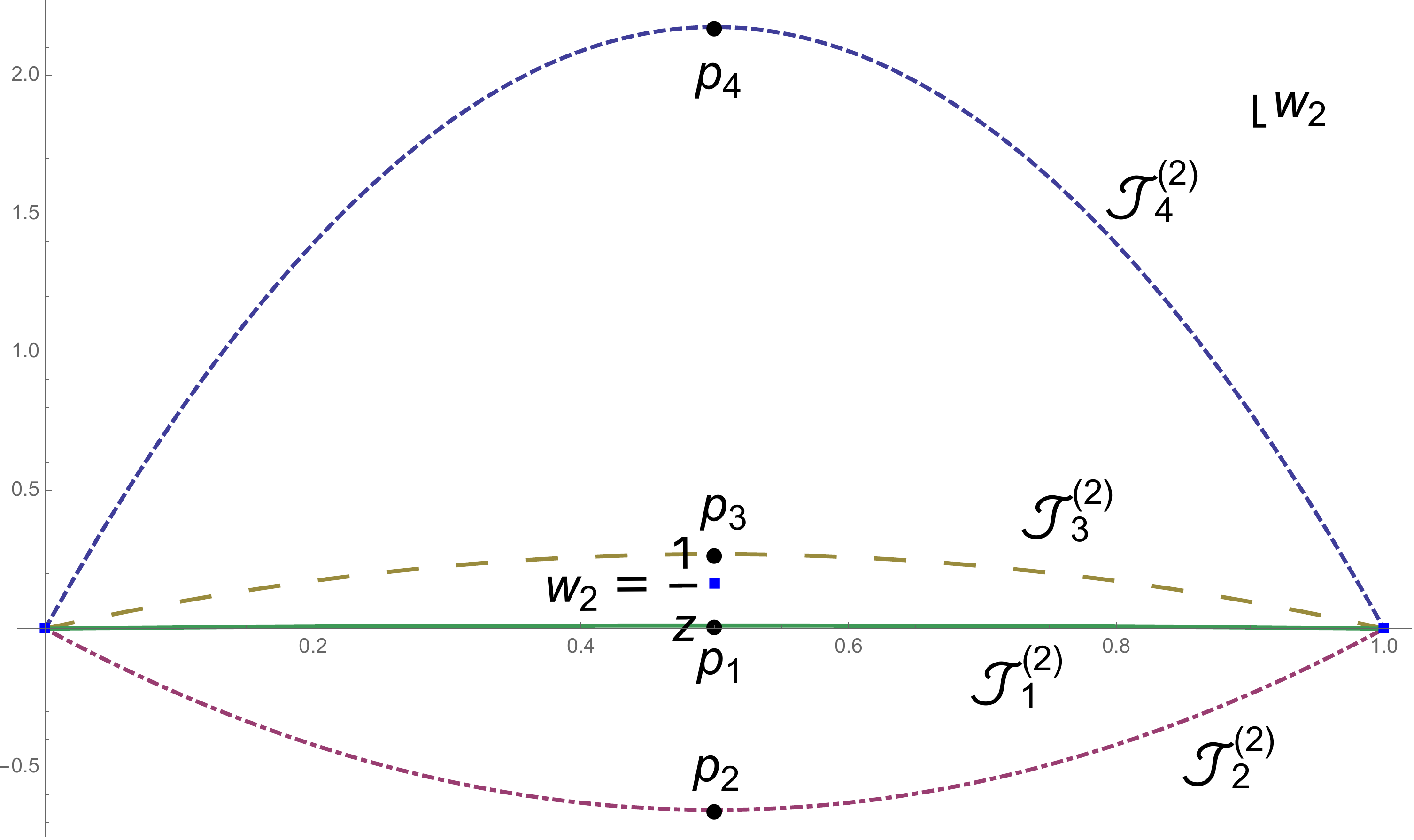}
\caption{This figure shows the projection onto the $w_1$ complex plane (left) and the $w_2$ complex plane (right) of the branch points (blue dots) and critical points (black points) of the action ${\cal I}_{(2,2)}$. Colored curves are sketches of the steepest descent contours attached to each critical points. In the above plots we chose $b^2=3$ and $z=1-e^{0.8 *i \pi}$ to indicate the generic case where one is away from forbidden singularities. The contour integral \ref{eq: O22ContourIntegral} along the curves labeled ${\cal J}_1$ is the vacuum block.  Note that as $z$ revolves around $1$, the branch point singularity parallels the imaginary axis, forcing various contour deformations, and producing a monodromy for the vacuum block.}
\label{fig:phi22contour1}
\end{center}
\end{figure}

Using the Coulomb gas formalism, we can write 
\be\label{eq: O22ContourIntegral}
\< \CO_{(2,2)}(0) \CO_{(2,2)}(z) \CO_L(1) \CO_L(\infty) \> \sim 
\int\!d w_1\! \int\! d w_2\, 
\frac{(1-z)^{1+\frac{1}{b^2}}}{z^{2h_{(2,2)}} }  \exp {\cal I}_{(2,2)}
\ee
for $h_L=1$, where the action ${\cal I}_{(2,2)}$ reads
\begin{align}
{\cal I}_{(2,2)}&=(1+b^2)\log\big[w_1(1-w_1)\big]
+\left(1+\frac{1}{b^2}\right)\log\big[w_2(1-w_2)\big]
-2\log(w_1-w_2)\nn\\
&-2 \log(1-z w_1)-\frac{2}{b^2} \log(1-z w_2)\;.
\end{align}
Once again the action has a singularity at $w_i = \frac{1}{z}$, requiring the deformation of contours as we analytically continue in $z$.
As in the ${\cal I}_{(3,1)}$ case, the contour for ${\cal I}_{(2,2)}$ is also defined on $\mathbb{C}^2$. However, notice that the point $w_1=$ finite, $w_2=\infty$ is now a regular point of the action $I_{(2,2)}$, and therefore should be considered when we search for critical points. 

Due to the absence of a symmetry swapping $w_1$ and $w_2$, there are four inequivalent integration contours relevant for the study of conformal blocks, namely $[0,1]\times [0,1]$,\; $[0,1]\times [\frac{1}{z},\infty)$,\; $[\frac{1}{z},\infty)\times [0,1]$, and $[\frac{1}{z},\infty)\times [\frac{1}{z},\infty)$.  As in the ${\cal I}_{(3,1)}$ case, it is easy to show that 
\be
z^{-2h_{(2,2)}}(1-z)^{1+\frac{1}{b^2}}\int_0^1 d w_1\! \int_0^1 d w_2\,\exp {\cal I}_{(2,2)}
\ \sim
\ z^{-2h_{(2,2)}} 
\ee
to leading order at small $z$. That is the $[0,1]\times [0,1]$ integration contour gives rise to the vacuum block.  Similarly, one finds that the $[0,1]\times [\frac{1}{z},\infty)$,\; $[\frac{1}{z},\infty)\times [0,1]$, and $[\frac{1}{z},\infty)\times [\frac{1}{z},\infty)$ contours correspond to the $\CO_{(1,3)}$, the $\CO_{(3,1)}$, and the $\CO_{(3,3)}$ blocks, respectively.

We refer to the steepest descent contours from the $i$th critical point in the $w_j$ variable as $\CJ_i^{(j)}$.
  If we include the point $(w_1, \infty)$ as part of the $(w_1, w_2)$ manifold, then there are a total of six critical points, with two at $w_2=\infty$.  However, the action $\CI_{(2,2)}$ at the points with $w_2 = \infty$ has a degenerate Hessian matrix, and so it is more difficult to define Lefschetz thimbles associated with these points.  These points do not seem to play a role in the Virasoro blocks, so we will ignore them in what follows.  For $z=1-e^{i \theta}$, the coordinates of the four finite critical points, written in the form $(w_1, w_2)$, can again be written as
$p_i = \left(\frac{1}{2}+i\sigma_i,\; \frac{1}{2}+i \eta_i\right)$ for $i=1,2,3,4$.  The explicit form of $\sigma_i$s and $\eta_i$s have been relegated to appendix \ref{app:ExplicitCriticalPoints}.

\begin{figure}[t!]
\begin{center}
\includegraphics[width=0.96\textwidth]{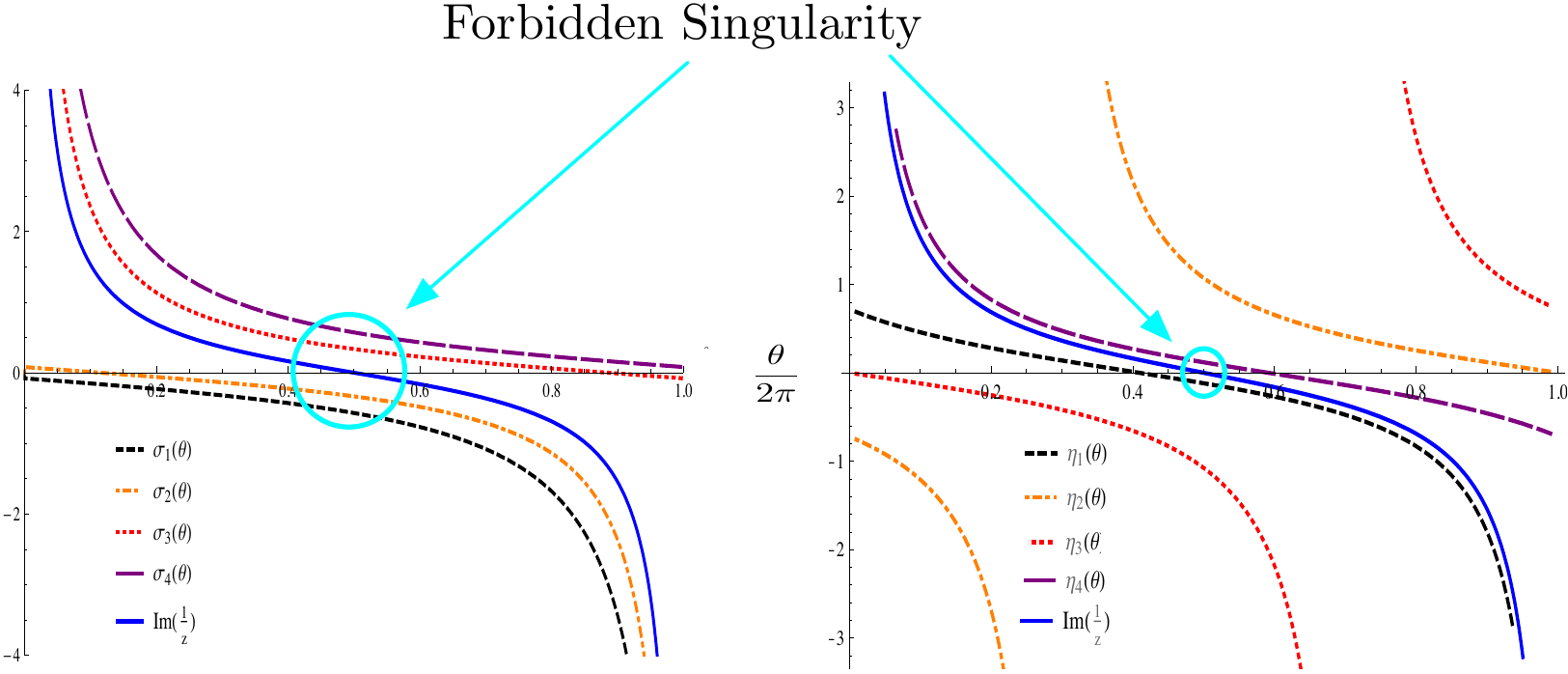}
\caption{This figure illustrates the positions of the four critical points of $\CI_{(2,2)}$  as $z = 1 - e^{i\theta}$ varies over $\theta \in [0, 2\pi]$.  The critical points have coordinates $(w_1, w_2) = \left(\frac{1}{2} + i \sigma_i, \frac{1}{2} + i \eta_i \right)$, and so the left and right plots are of the imaginary part of $w_1$ and $w_2$, respectively.  The solid line indicates the branch point $\frac{1}{z}$ where $\exp \CI_{(2,2)}$ has a pole, while the various dashed lines are critical points.  In the vicinity of the forbidden singularity, all four critical points approach the branch point in the $w_1$ coordinate, intersecting it as $c \propto b^2 \to \infty$, while two of the critical points also  approach the branch point in $w_2$.  In these plots we take $c \approx 20$, so groups of critical points approach but do not meet.
}
\label{fig:CriticalPointsForbiddenSingularities22}
\end{center}
\end{figure}

In figure \ref{fig:phi22contour1} we plot the branch points (blue dots) and the critical points (black dots) of the action ${\cal I}_{(2,2)}$, along with the steepest descent contours passing through each critical point.  These plots are of the same style as the ones in the previous subsection, projecting $\mathbb{C}^2$ onto the $w_1$ and $w_2$ complex plane. As one can see, the green curves, $\CJ_1^{(1)}$ and $\CJ_1^{(2)}$, can be deformed into $[0,1]\times [0,1]$ without crossing the branch point $w=\frac{1}{z}$, so the integral over the Lefschetz thimble $\CJ_1$ corresponds to the vacuum block.

As one approaches the forbidden singularity, all four of the $\CJ_i$  move towards the branch point $w=\frac{1}{z}$, as pictured in figure \ref{fig:CriticalPointsForbiddenSingularities22}.  We describe this phenomenon analytically in appendix \ref{app:ExplicitCriticalPoints}.  When the contours intersect the branch point, the integral $\eqref{eq: O22ContourIntegral}$ over $\CJ_1$ (the vacuum block) becomes ambiguous, as we discussed in the $\CO_{(2,1)}$ case, and so we must deform the contour to avoid the branch point.  Unlike in the case of external $\CO_{(3,1)}$, here we must deform the contour in both  $w_1$ and $w_2$ planes, picking up a linear combination of states in the $\CO_{(2,2)} \times \CO_{(2,2)}$ fusion rules.  It would be interesting to try to interpret the branch points in this $\CO_{(2,2)}$ case as a dressing of those in the simpler $\CO_{(2,1)}$  case studied in section \ref{sec:ContourIntegral21}, and to understand the behavior of the block with an intermediate $\CO_{(1,3)}$ state.

\section{Details of Critical Points and Stokes Phenomena}
\label{app:DetailsStokes}

\subsection{Stokes Phenomena in the $\CO_{(2,1)}$ Example}
\label{app:OtherAnalysis21}

This appendix provides an alternative and more general approach to section \ref{sec:ContourIntegral21}, which is useful for studying Stokes phenomena.  We will study a representation of the hypergeometric function associated with the simplest heavy degenerate operator, which obeys equation (\ref{eq:HeavyDegenerateEqn2}).   We will also take the simplifying limit  $h_L = 1$, in which case the result can be written as 
\be
\label{eq:HypergeometricAsIntegralApp}
{}_2 F_1\left( 2,b^2+1, 2b^2+2,z \right) = \frac{2^{2 b^2+1} \Gamma \left(b^2+\frac{3}{2}\right)}{\sqrt{\pi } \Gamma \left( b^2 +1 \right)} \int_0^1 w^{b^2} (1-w)^{b^2} (1-zw)^{-2} dw
\ee
when $z < 1$, where we recall that the central charge is $c = 1 + 6 \left(  b + \frac{1}{b} \right)^2$.   This function was plotted in the vicinity of its forbidden singularity in figure \ref{fig:NearForbiddenSingularity21}.  In principle, the boundaries at $0$ and $1$ can make important contributions.  However, we will be interested in the regime Re$(b^2) \gg 1$ where the integrand vanishes at these boundaries, and where the integrand blows up as $w \to e^{i \phi} \infty$  for any $\phi$.  In this case the boundary contributions vanish, and all steepest descent contours end at $w=0$ or $w=1$.  

Our integral has two saddle points at $w = p_\pm$.  The $p_\pm$  depend on $b$ and $z$, but it is more convenient to write $b$ and $z$ as functions of the $p_\pm$ via
\be
b^2=  \frac{2 p_+ p_- - p_+ - p_- + 1}{(1-2 p_+)(1-2p_-)} , \ \ \ z = \frac{2 p_+ p_- - p_+ - p_- + 1}{p_+ p_-}
\ee
The large $c \propto b^2$ limit corresponds to $p_+$ or $p_-$ near $\frac{1}{2}$.  Taking $p_- \approx \frac{1}{2}$, we have $z \approx \frac{1}{p_+}$.  The forbidden singularity is located at $z = 2$ and $b \to \infty$, which occurs precisely when the two saddlepoints coincide at $p_+ = p_- = \frac{1}{2}$.  

Stokes surfaces are locations in $b, z$ where the steepest descent contour from multiple saddles coincide.  Clearly this occurs at $p_\pm = \frac{1}{2}$.  In general, the imaginary part of the action is constant on steepest descent curves.  Therefore a Stokes line can only exist if the action has the same imaginary part at two different saddles points.  If we study the situation where $b^2$ is real and large due to $p_- \approx \frac{1}{2}$, then this can only happen if
\be
\mathrm{Im} \left[  \frac{\log \left(\left(1- p_+\right) p_+\right)}{(1 -2 p_+) } \right] = 0
\ee
which means that $0 <p_+ < 1$ with $p_-$ real and in the vicinity of $\frac{1}{2}$.  Translating this to a statement about $z$, we identify it with the region $z \in (1,\infty)$.  This coincides with the branch cut of the hypergeometric function, which is exactly what we should have expected.  Crossing the Stokes line corresponds to crossing the branch cut of the hypergeometric function.
When the parameters $b^2 \gg 1$ and $z \in (0,1)$ are real, the integrand is real for real $w$, and the contour of integration $[0,1]$ corresponds exactly with a steepest descent contour through $p_- \approx \frac{1}{2}$.  Since $p_+ > 1$ its corresponding saddle will not contribute to the hypergeometric function, and so we have ${}_2 F_1 = \CJ_1$, the contour associated with $p_1 \equiv p_-$.

\subsection{Explicit forms for the Critical Points}
\label{app:ExplicitCriticalPoints}

As we discussed in appendix \ref{app:Asymptotics31and22}, the critical points for the action ${\cal I}_{(3,1)}$ take the form of
\be
p_i=\left(\frac{1}{2}+i \sigma_i,\;\frac{1}{2}+i \eta_i\right)\;,\quad i=1,2,3\;,
\ee
for $z=1-e^{i \theta}$\;, where $\sigma_i$'s and $\eta_i$'s are solutions to the algebraic equations
\begin{align}
&-\frac{2}{b^2 \left(\cot
   \left(\frac{\theta
   }{2}\right)-2 \sigma
   \right)}+\frac{1}{\sigma
   -\eta }-\frac{8 \sigma }{4
   \sigma ^2+1}=0\;,\nn\\
&\frac{2}{b^2 \left(\cot
   \left(\frac{\theta
   }{2}\right)-2 \eta
   \right)}+\frac{1}{\sigma
   -\eta }+\frac{8 \eta }{4
   \eta ^2+1}=0\;.
\end{align}
For generic values of $b^2$ and $\theta$, it is difficult to obtain useful  analytic expressions for the solutions. However, for $b^2 \gg 1$,  which is the case of interest, we can  solve the above equations perturbatively in $b^2$.  Thus, up to $\CO(b^{-2})$, we find
\begin{align}
\sigma_1&=-\frac{1}{2\sqrt{3}}-\frac{2 \left(\sqrt{3}-9 \cot \left(\frac{\theta }{2}\right)\right)}{b^2(9-27 \cot
   ^2\left(\frac{\theta }{2}\right))}\;\quad
\eta_1=\frac{1}{2\sqrt{3}}+\frac{2 \left(9 \cot \left(\frac{\theta }{2}\right)+\sqrt{3}\right)}{b^2(9-27 \cot
   ^2\left(\frac{\theta }{2}\right))}\;,\\
\sigma_2&=-\frac{1}{2} \tan \left(\frac{\theta
   }{4}\right)+\frac{2 \sin  \left(\frac{\theta}{2}\right)-\sin (\theta   )}{2 b^2 \left(\cos \left(\frac{\theta}{2}\right)+\cos (\theta)\right)}
\;,\quad
\eta_2=\frac{1}{2} \cot \left(\frac{\theta}{2}\right)
+\frac{1}{2b^2 \left(\sin (\theta)-\sin
   \left(\frac{\theta
   }{2}\right)\right)}\;,
\nn \\
\sigma_3&=\frac{1}{2} \cot \left(\frac{\theta}{2}\right)
+\frac{1}{2b^2 \left(\sin (\theta)+\sin
   \left(\frac{\theta
   }{2}\right)\right)}\;, \quad
\eta_3=\frac{1}{2} \cot \left(\frac{\theta}{4}\right)+\frac{2\sin \left(\frac{\theta}{2}\right)+\sin (\theta)}{2b^2 \left(\cos
   \left(\frac{\theta}{2}\right)-\cos (\theta )\right)}\;. \nn 
\end{align}
although this identification will only match figure \ref{fig:CriticalPointsForbiddenSingularities31} in the regime where $\theta < \frac{2 \pi}{3}$.
This perturbative solution breaks down when $\theta=\frac{2\pi}{3}$ and $\theta=\frac{4\pi}{3}$, near the forbidden singularities. At these special values of $\theta$, we have to look for different ansatz. It turns out that at $\theta=\frac{2 \pi}{3}$, the solutions of $\sigma$ and $\eta$ are given by
\begin{align}
\sigma_1&=-\frac{1}{2\sqrt{3}}-\frac{1}{3\sqrt{2} b}\;,\quad
\eta_1=\frac{1}{2\sqrt{3}}-\frac{\sqrt 2}{3 b}\;,\\
\sigma_2&=-\frac{1}{2\sqrt{3}}+\frac{1}{3\sqrt{2}b}\;, \quad
\eta_2=\frac{1}{2\sqrt{3}}+\frac{\sqrt 2}{3 b}\;, \nn \\
\sigma_3&=\frac{1}{2\sqrt{3}}\left(1+\frac{1}{b^2}\right)\;, \quad
\eta_3=\frac{\sqrt{3}}{2}\left(1+\frac{3}{2b^2}\right)\;.\nn 
\end{align}
Notice that as $b \to \infty$, we find that $\eta_1 \to \eta_2$ at a rate set by $\frac{1}{b}$.  This effect was pictured for general $\theta$ with fixed, large $b$ in figure \ref{fig:CriticalPointsForbiddenSingularities31}.
At $\theta=\frac{4\pi}{3}$, the critical points are given by
\begin{align}
\sigma_1&=-\frac{\sqrt{3}}{2}\left(1+\frac{3}{2b^2}\right)\;,\quad 
\eta_1=-\frac{1}{2\sqrt{3}}\left(1+\frac{1}{b^2}\right)\;,\nn \\
\sigma_2&=-\frac{1}{2\sqrt{3}}-\frac{\sqrt 2}{3 b}\;,\quad
\eta_2=\frac{1}{2\sqrt{3}}-\frac{1}{3\sqrt{2}b}\;,\nn\\
\sigma_3&=-\frac{1}{2\sqrt{3}}+\frac{\sqrt 2}{3 b}\;,\quad
\eta_3=\frac{1}{2\sqrt{3}}+\frac{1}{3\sqrt{2} b}\;.\nn
\end{align}
Once again, as $b \to \infty$, we find that $\eta_1 \to \eta_2$ at a rate set by $\frac{1}{b}$, as was pictured for general $\theta$ but fixed $b$ in figure \ref{fig:CriticalPointsForbiddenSingularities31}.
As we discussed in appendix \ref{app:Asymptotics31and22}, we use notation such that the contour integrals over the steepest descendant curves $\CJ_1$ through $p_1$ correspond with the vacuum block when $z$ is near the origin.  

The critical points of the ${\cal I}_{(2,2)}$, which are relevant to section \ref{app:Asymptotics31and22}, take the form 
\be
p_i=\left(\frac{1}{2}+i \sigma_i,\;\frac{1}{2}+i \eta_i\right)\;,\quad i=1,2,3,4\;,
\ee
where the $\sigma_i$'s and $\eta_i$'s are solutions to the algebraic equations
\begin{align}
&\frac{4 \left(b^2+1\right) \sigma }{4 \sigma ^2+1}+\frac{1}{\eta
   -\sigma }+\frac{2}{\cot \left(\frac{\theta }{2}\right)-2
   \sigma }=0\;,\nn
\\
&\frac{2 \left(b^2+1\right) \eta }{b^2 \left(4 \eta
   ^2+1\right)}
+\frac{1}{2 \sigma -2 \eta }+\frac{1}{b^2 \left(\cot \left(\frac{\theta
   }{2}\right)-2 \eta \right)}=0\;.
\end{align}
Solving them perturbatively in $b^2$, we find
\begin{align}
\sigma_1&=-\frac{\tan \left(\frac{\theta
   }{2}\right)}{b^2}\;,\quad
\eta_1=\frac{1
   }{2} \cot
   \left(\frac{\theta
   }{2}\right)+\frac{1}{b^2 \left(\sin
   (\theta )-2 \tan
   \left(\frac{\theta
   }{2}\right)\right)}\;,\\
\sigma_2&=-\frac{\tan \left(\frac{\theta }{2}\right)}{2 b^2}\;,\quad 
\eta_2=-\frac{\tan (\theta )}{4}b^2-\frac{15 \cos (\theta )+2 \cos (2 \theta )+\cos (3 \theta )-2}{16  \sin (\theta ) \cos ^2(\theta )}\;,\nn \\
\sigma_3&=\frac{1}{2} \cot
   \left(\frac{\theta
   }{2}\right)+\frac{\csc (\theta
   )}{b^2}\;,\quad 
\eta_3=-\frac{1
   }{2} \tan
   \left(\frac{\theta
   }{2}\right)+\frac{2 \sin
   \left(\frac{\theta
   }{2}\right)}{b^2 \left(3
   \cos \left(\frac{\theta
   }{2}\right)+\cos
   \left(\frac{3 \theta
   }{2}\right)\right)}\;,\nn\\
\sigma_4&=\frac{1}{2} \cot
   \left(\frac{\theta
   }{2}\right)+\frac{2 \csc (\theta
   )}{b^2}\;,\quad 
   \eta_4=\frac{1}{2} \cot
   \left(\frac{\theta
   }{2}\right)+\CO(b^{-4})\;,\nn\\
\end{align}
for $\theta < \pi$.  The formulas still apply for $\theta > \pi$, but the labels must be permuted. For the special value $\theta=\pi$, ie near the forbidden singularity at $z=2$, we need to use a different ansatz. It turns out that these solutions are given by
\begin{align}
\sigma_1&=-\frac{1}{\sqrt{2}b}-\frac{3}{4\sqrt{2}b^3}\;,\quad \
\eta_1=-\frac{1}{\sqrt{2}b^3}\;,\\
\sigma_2&=-\frac{1}{2b}-\frac{1}{4b^3}\;,\quad \quad \quad \
\eta_2=\frac{b}{2}+\frac{1}{4b}\;,\nn\\
\sigma_3&=\frac{1}{2b}+\frac{1}{4b^3}\;,\quad  \quad \quad \quad
\eta_3=-\frac{b}{2}-\frac{1}{4b}\;,\nn\\
\sigma_4&=\frac{1}{\sqrt{2}b}+\frac{3}{4\sqrt{2}b^3}\;,\quad \ \ \
\eta_4=\frac{1}{\sqrt{2}b^3}\;.\nn
\end{align}
Notice that all four of the $\sigma_i \to 0$ as $b \to \infty$, so these critical points all approach the branch point at $\frac{1}{z} = \frac{1}{2} + i 0$.  We also see that $\eta_1, \eta_3 \to 0$ at a faster rate as $b \to \infty$, so this particular pair of critica points approaches the branch point in the $w_2$ plane at large $b$.  These results are plotted for fixed $b$ and general $\theta$ in figure \ref{fig:CriticalPointsForbiddenSingularities22}.

\bibliographystyle{utphys}
\bibliography{VirasoroBib}

\end{document}